\documentclass[twocolumn,showpacs,pra,aps,superscriptaddress,floatfix]{revtex4-1}
\usepackage{graphicx}
\usepackage{braket} 
\usepackage{float} 
\usepackage{natbib}
\usepackage[latin1]{inputenc}           
\usepackage{bm} 
\usepackage{amsmath}
\usepackage{amssymb}
\usepackage{mathtools}
\usepackage{bbold}
\usepackage{comment}
\usepackage{dsfont}
\usepackage{color}
\usepackage{soul}

 \definecolor{darkred}{rgb}{0.8,0.1,0.1}
 \definecolor{DARKRED}{rgb}{0.8,0.1,0.1}
 \definecolor{darkblue}{rgb}{0.1,0.1,0.7}
 \definecolor{bleudefrance}{rgb}{0.19, 0.55, 0.91}
 
\usepackage{colordvi}

\definecolor{orangevillavics}{RGB}{204,102,0}

\definecolor{redjesus}{RGB}{204,0,102}

\definecolor{DodgerBlue}{RGB}{0, 90, 156}

\definecolor{greenfernando}{RGB}{51, 102, 0}
\definecolor{violetdiego}{RGB}{153,0, 153}

\usepackage[caption=false,justification=justified]{subfig}
\usepackage[caption=false]{subfig} 
\usepackage{ragged2e} 
\DeclareCaptionJustification{justified}{\justifying}

\begin{document}
\title{Thermal Uhlmann phase in a locally driven  two-spin system}
\author{J. Villavicencio}
    \affiliation{Facultad de Ciencias, Universidad Aut\'onoma de Baja
    California, 22800 Ensenada, B.C., M\'exico}
 \author{E. Cota}
     \affiliation{Centro de Nanociencias y Nanotecnolog\'ia, Universidad
     Nacional Aut\'onoma de M\'exico, Apartado Postal 14, 22800 Ensenada, B.C., M\'exico}
  \author{F. Rojas}
    \affiliation{Centro de Nanociencias y Nanotecnolog\'ia, Universidad
    Nacional Aut\'onoma de M\'exico, Apartado Postal 14, 22800 Ensenada, B.C., M\'exico}
    \author{J.~A. Maytorena}
    \affiliation{Centro de Nanociencias y Nanotecnolog\'ia, Universidad
    Nacional Aut\'onoma de M\'exico, Apartado Postal 14, 22800 Ensenada, B.C., M\'exico}
    \author{D. Morachis Galindo}
    \affiliation{Centro de Nanociencias y Nanotecnolog\'ia, Universidad
     Nacional Aut\'onoma de M\'exico, Apartado Postal 14, 22800 Ensenada, B.C., M\'exico}
\author{F. Nieto-Guadarrama}
    \affiliation{Centro de Nanociencias y Nanotecnolog\'ia, Universidad
    Nacional Aut\'onoma de M\'exico, Apartado Postal 14, 22800 Ensenada, B.C., M\'exico}
    \email{nieto.fernando@uabc.edu.mx}
\date{\today}
\begin{abstract}
We study the geometric Uhlmann phase of mixed states at finite temperature in a system of two coupled spin-$\frac 1 2$ particles driven by a magnetic field applied to one of the spins.
In the parameter space of temperature and coupling,
we show the emergence of two topological Uhlmann phase transitions when the magnetic field evolves around the equator, where a winding number can characterize each temperature range.
%
For small couplings, the width of the temperature gap
of the non-trivial phase is roughly the critical temperature $T_c$ of one-dimensional fermion systems with two-band Hamiltonians.
%
The first phase transition in the low-temperature regime and small values of the coupling corresponds to the peak of the \textit{Schottky anomaly} of the heat capacity, typical of a two-level system in solid-state physics
involving the ground and first excited states. 
The second phase transition occurs at temperatures very close to the second maximum of the heat capacity associated with a multilevel system.
We also derive analytical expressions for the thermal Uhlmann phase for both subsystems, showing that they exhibit phase transitions.
In the driven subsystem, for minimal $g$, a topological phase transition phase appears at  $T_c$ again. However, for larger values of $g$, the transitions occur at lower temperature values, and they disappear when the coupling reaches a critical value $g_c$.
The latter is not the case for the undriven subsystem, where at low temperatures, a single phase transition occurs at $g_c$. 
Nevertheless, as the temperature rises,  we demonstrate the emergence of two phase transitions defining a coupling gap, 
where the phase is non-trivial and vanishes as the temperature reaches a critical value.
\end{abstract}
\maketitle

\section{Introduction}

Since the discovery of the quantum Hall effect, topological phases of matter have become of great interest in condensed matter physics \cite{Colloquium}. For example, the characterization of this paradigmatic effect in terms of the Chern topological invariant employs the Berry curvature as a fundamental concept \cite{NobelTopologicalQuantumMatter,zakprl89}, which has also proved to be the essential ingredient in the theoretical description of topological insulators \cite{shorty_2014,vanderbilt_2018}. The presence of topological phases of matter beyond conventional condensed matter systems has paved the way for exploring geometrical phases in optical, polaritonic \cite{TopologicalPhotonics,PerspectiveTopNanophotonics}, and superconducting systems \cite{TopologicalSuperconductors}.
Although the Berry phase has been essential to characterize the topological properties of various quantum systems by studying their ground states (pure states),  systems involving finite temperatures or out-of-equilibrium physics require a different approach since they involve statistical mixtures. 
Therefore, an extension of the Berry phase concept for pure states to the point where we have the presence of mixed states is required. The Uhlmann phase \cite{uhlmannrmp86,uhlmannlmp91}, which consists of the evolution of density matrices,  is a suitable generalization of the Berry phase to finite-temperature systems. 
The latter has acquired great relevance \cite{viyuelaprl14,Viyuela_2015} in the context of one-dimensional fermionic systems \cite{suprl79,shorty_2014,PhysRevLett.83.2636, Kitaev_2001}, where 
it defines a topological invariant that remains constant in a finite temperature interval after acquiring a null value from a specific critical temperature. A system below the critical temperature is classified as topologically protected, while the system is said to be in the topologically trivial regime above that temperature.
Geometric phases are very sensitive to thermal changes, so achieving control of their properties and a certain degree of robustness against dissipative effects is desirable for applications in quantum computing.

We have recently investigated the Uhlmann phase in mixtures generated by a noisy channel applied to a system of two spin-$\frac{1}{2}$ fermions driven by a time-dependent magnetic field \cite{pravillaeta21}. We showed how to control the phase transitions in the subsystems by manipulating the intensity of the noisy channel. 
In the latter model, we did not observe any phase transition of the 
Uhlmann and interferometric phase defined by Sj\"oqvist \textit{et al.} in  Ref.~\onlinecite{PhysRevLett.85.2845} 
for the mixed states of the composite system. Given our previous findings, an interesting research topic is to study the structure of the Uhlmann phase in a bipartite system where a different mechanism causes the mixing of the states. 
One of these mechanisms is due to the thermal effects in the system. 
Interestingly, studies of the Uhlmann topological phase for single spin-$j$ systems \cite{PhysRevA.103.042221,hou_finite-temperature_2021} show the emergence of an intermediate-thermal topological phase in different temperature regimes that can be classified using the winding numbers of the system.
Moreover, recent studies \cite{he_uhlmann_2022} show the robustness of the Uhlmann phase in a qubit under environmental and thermal effects modeled within the Lindblad equation approach \cite{rivas2011open} in topological systems like the Su-Schrieffer-Heeger (SSH) model \cite{suprl79}, Kitaev chain \cite{kitaev_unpaired_2001}, and Bernevig-Hughes-Zhang (BHZ) model \cite{bernevig_quantum_2006}.

In this work, we aim to study the effects of temperature in the Uhlmann phase for a system of two coupled fermions in the presence of a magnetic field. Using analytical expressions for the Uhlmann phase, we show the emergence of topological phase transitions in the composite system and its corresponding subsystems. We demonstrate that the composite system exhibits two phase transitions occurring at different temperatures, which define a gap $\Delta T$ where the phase is non-trivial, and that depends on the coupling value. Interestingly, we find that the first phase transition, which occurs in the low-temperature regime and small coupling values, corresponds to the maximum of the  \textit{Schottky anomaly} of the heat capacity, typical of two-level systems, suggesting a connection between the thermal Uhlmann geometric phase and a physical observable. 
We also demonstrate that the Uhlmann phase in subsystems corresponding to the driven and undriven fermions exhibit completely different phase transitions. In particular, the latter shows a peculiar \textit{double topological transition} for two critical values of the coupling, which appear at a fixed temperature value in all directions of the field with a fixed latitude at the equator of the sphere.
 
We organize our paper as follows: in Sec.~\ref{subsec:spin}, we present the model and discuss the procedure to calculate the Uhlmann phase. In Sec.~\ref{sec:theramlcompositeAB}, we explore the thermal effects and topological transitions on the composite system, and in Sec.~\ref{sec:subsysAandB}, we study the features of their corresponding subsystems. Finally, we present the conclusions in Sec.~\ref{sec:conclusions}.

\section{Model}\label{subsec:spin}

Our model involves two coupled fermions of spin-$\frac{1}{2}$ via an anisotropic Heisenberg interaction, where only one of the particles is driven by a time-dependent magnetic field. The following Hamiltonian \cite{yiprl04,pravillaeta21} determines the dynamics of the system 
\begin{equation}
    \hat{H}(\phi) = \frac{1}{2}\boldsymbol{B}(t)\cdot \boldsymbol{\hat{\sigma}}\otimes\mathbb{1}+(J/2)\,(\hat{\sigma}_x\otimes\hat{\sigma}_x-\hat{\sigma}_y\otimes\hat{\sigma}_y),
    \label{newham}
\end{equation}
where $\boldsymbol{B}(t) = B_o\boldsymbol{\hat{n}}$ is the rotating magnetic field along the direction $\boldsymbol{\hat{n}} = (\sin\theta\cos\phi,\sin\theta\sin\phi,\cos \theta)^T$, with $\theta = [0,\pi]$, and the time dependence comes from the parameter $\phi = \phi(t)$. The energy spectrum of the rescaled Hamiltonian, $\hat{H} = \hat{H}_o/(B_o/2)$, is given by,
\begin{eqnarray}
    E_1=-E_2=\sqrt{1+g^2/2+(g/2)\sqrt{g^2+4\sin^2\theta}}; \nonumber \\
    E_3=-E_4=\sqrt{1+g^2/2-(g/2)\sqrt{g^2+4\sin^2\theta}},
    \label{eigenvalues}    
\end{eqnarray}
where $g = 2J/B_o$ stands for the spin-spin coupling.
We know that unitary transformations leave invariant the spectrum of a Hamiltonian \cite{wagner_unitary_1986}, and since the eigenvalues (\ref{eigenvalues}) are $\phi$ independent, we may attempt to write (\ref{newham}) as $\hat{H}(\phi) = \hat{U}(\phi)\hat{H}(0)\hat{U}^\dagger(\phi)$. The unitary transformation that satisfies these expressions is given by
\begin{equation}
    \hat{U}(\phi) = e^{-i(\phi/2)(\hat{\sigma}_z\otimes\mathbb{1}-\mathbb{1}\otimes\hat{\sigma}_z)}.
    \label{UnitaryT}
\end{equation}
The latter transformation can be interpreted as performing a rotation about the $\boldsymbol{\hat{z}}$ axis on the first spin-$\frac{1}{2}$ particle while performing the inverse rotation on the second spin-$\frac{1}{2}$ particle. Hence the eigenvectors of our system can be expressed as $\ket{u_j} = \hat{U}\ket{u_j(0)}$, where $\ket{u_j(0)}$ are the eigenvectors of $\hat{H}(0)$. This set of eigenvectors are given by $\ket{u_j(0)} =  {\cal N}^{-1/2}_j[u_j^{(1)},u_j^{(2)},u_j^{(3)},u_j^{(4)}]$, where $u_j^{(1)} = \sin\theta$, $u_j^{(2)} = g(\cos^2\theta-E^2_j)/(1-E^2_j), u_j^{(3)} = (E_j-\cos\theta)$, and $u_j^{(4)} = g\sin\theta(\cos\theta-E_j)/(1-E^2_j)$, with ${\cal N} = \sum_i\left[u_j^{(i)}\right]^2$. We will take advantage of this unitary equivalence of eigenvectors in the computation of the Uhlmann holonomy below. 

An approach to exploring the geometric phases in composite systems is using the Uhlmann phase \cite{uhlmannrmp86,uhlmannlmp91}.  %
The Uhlmann phase, $\Phi$, introduced by Viyuela \cite{viyuelaprl14,Viyuela_2015} for exploring thermal effects in one-dimensional fermion systems is given by
\begin{equation}
    \Phi={\rm Arg}\left\{{\rm Tr}[\rho_{\lambda_0}\,V(\lambda,\lambda_0)]\right\},
    \label{uhlmana}
\end{equation}
where the Uhlmann holonomy $V(\lambda,\lambda_0)={\cal P} e^{\oint A(\lambda)}$ is a  $\lambda$  ordered integral, and $A(\lambda)$ is the Uhlmann connection. In general $A(\lambda)$ does not commute for all values of the parameter $\lambda$. 
An alternative procedure to evaluate $V(\lambda,\lambda_0)$ is by solving the differential equation for the evolution operator,
\begin{equation}
    d V(\lambda,\lambda_0)=A(\lambda) \,V(\lambda,\lambda_0),
    \label{difqeV}
\end{equation}
with the initial condition $V(\lambda_0,\lambda_0)=\mathbb{1}$, where we assume that 
$\lambda_0= 0$.
The Uhlmann connection $A(\lambda)$ is given by:
\begin{equation}
    A(\lambda)=\sum_{i,j}\ket{\psi_i}\frac{\braket{\psi_i|\left[\partial_{\lambda}\sqrt{\rho}, \sqrt{\rho}\right]|\psi_j} }{p_j+p_i} \bra{\psi_j}\,d\lambda,
    \label{uhlAU}
\end{equation}
which involves the matrix elements with respect to the eigenbasis $\{\ket{\psi_j}\}$ of the density matrix $\rho$, which we assume to be diagonalized, with eigenvalues $\{p_j\}$. In the spectral basis, $\rho=\sum_{j}p_j\ket{\psi_j}\bra{\psi_j}$.
By explicitly  computing the matrix elements of the commutator in Eq.~(\ref{uhlAU}) we write the Uhlmann connection as, 
\begin{equation}
    A(\lambda)=\sum_{i\ne j}\frac{(\sqrt{p_j}-\sqrt{p_i})^2}{p_j+p_i}\braket{\psi_i|\partial_{\lambda}\psi_j} \ket{\psi_i}\bra{\psi_j}\,d\lambda.
    \label{uhlAUb}
\end{equation}

In the following sections we investigate the general features of the Uhlmann phase in mixed entangled states for a composite system (Sec.~\ref{sec:theramlcompositeAB}), and its corresponding subsystems  (Sec.~\ref{sec:subsysAandB}).

\section{Uhlmann phase and thermal effects in a composite system} \label{sec:theramlcompositeAB}

\begin{figure}[htbp] 
    \begin{center}
        \subfloat{\includegraphics[width=4.5cm]{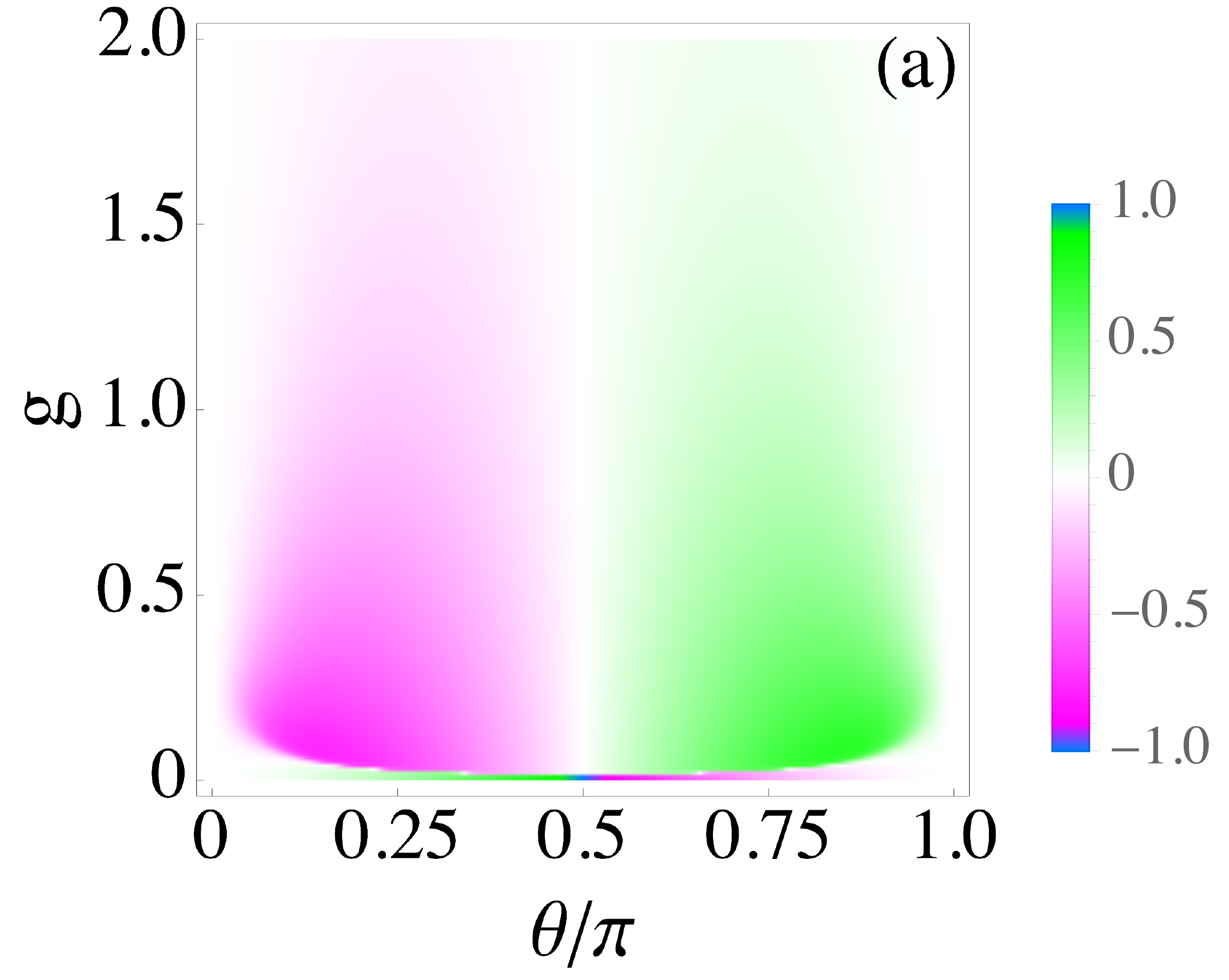}}
        \subfloat{\includegraphics[width=4.5cm]{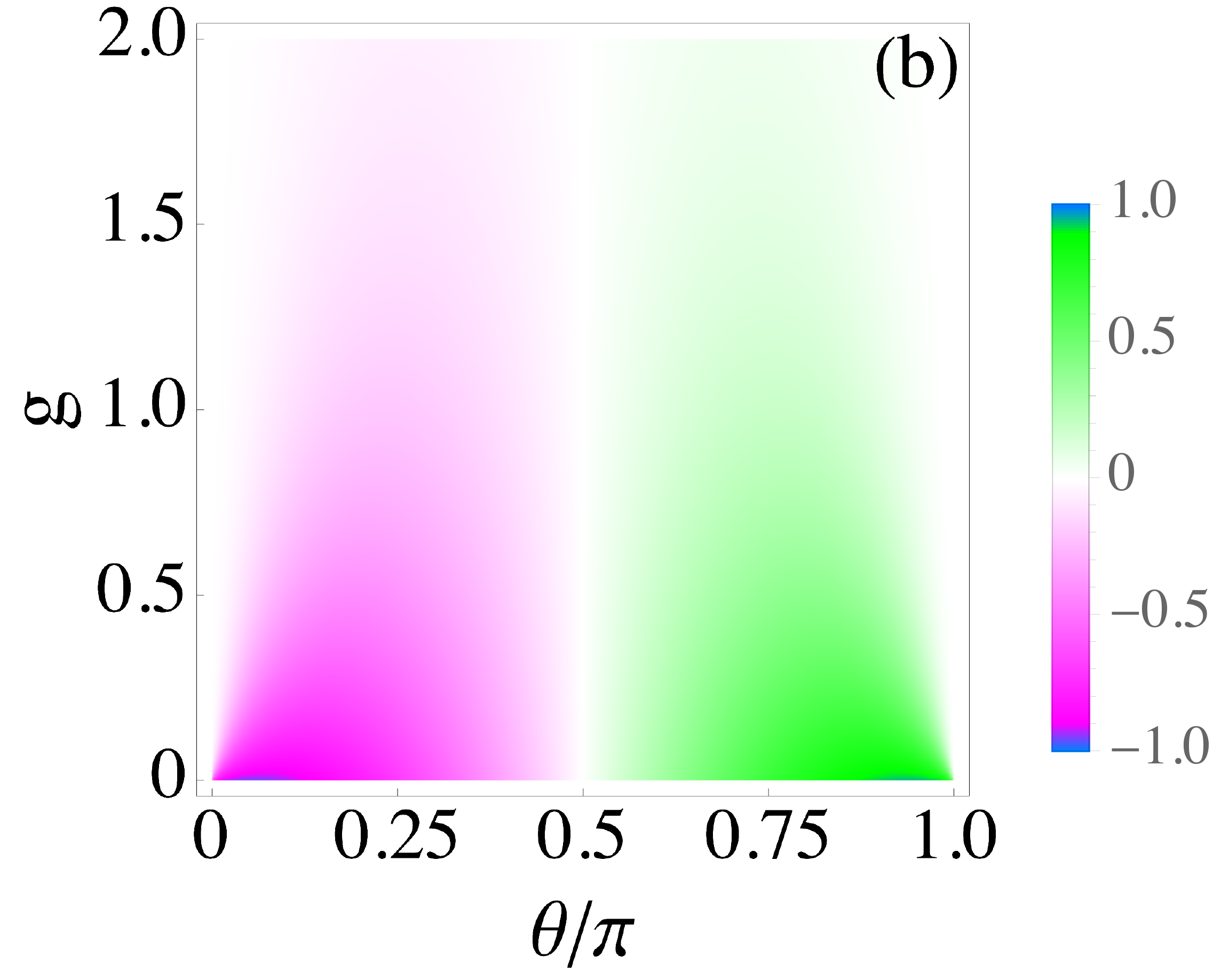}}
    \end{center}
    \caption{Color density maps of (a) the Uhlmann phase of the composite system, $\Phi^{AB}$ [Eq.~(\ref{uhlman_bis})] for $T=0.01$, and (b) the Berry phase $\gamma_2$ of the ground state, as a function of the coupling parameter $g$, and $\theta$. We show that for small temperature values, $\Phi^{AB}$ and $\gamma_2$ approximately coincide, as expected. All the phases are in units of $\pi$. }            
    \label{UhlmannvsBerrylowT}
\end{figure}

\begin{figure}[htbp] 
    \begin{center}
        \subfloat{\includegraphics[width=4.5cm]{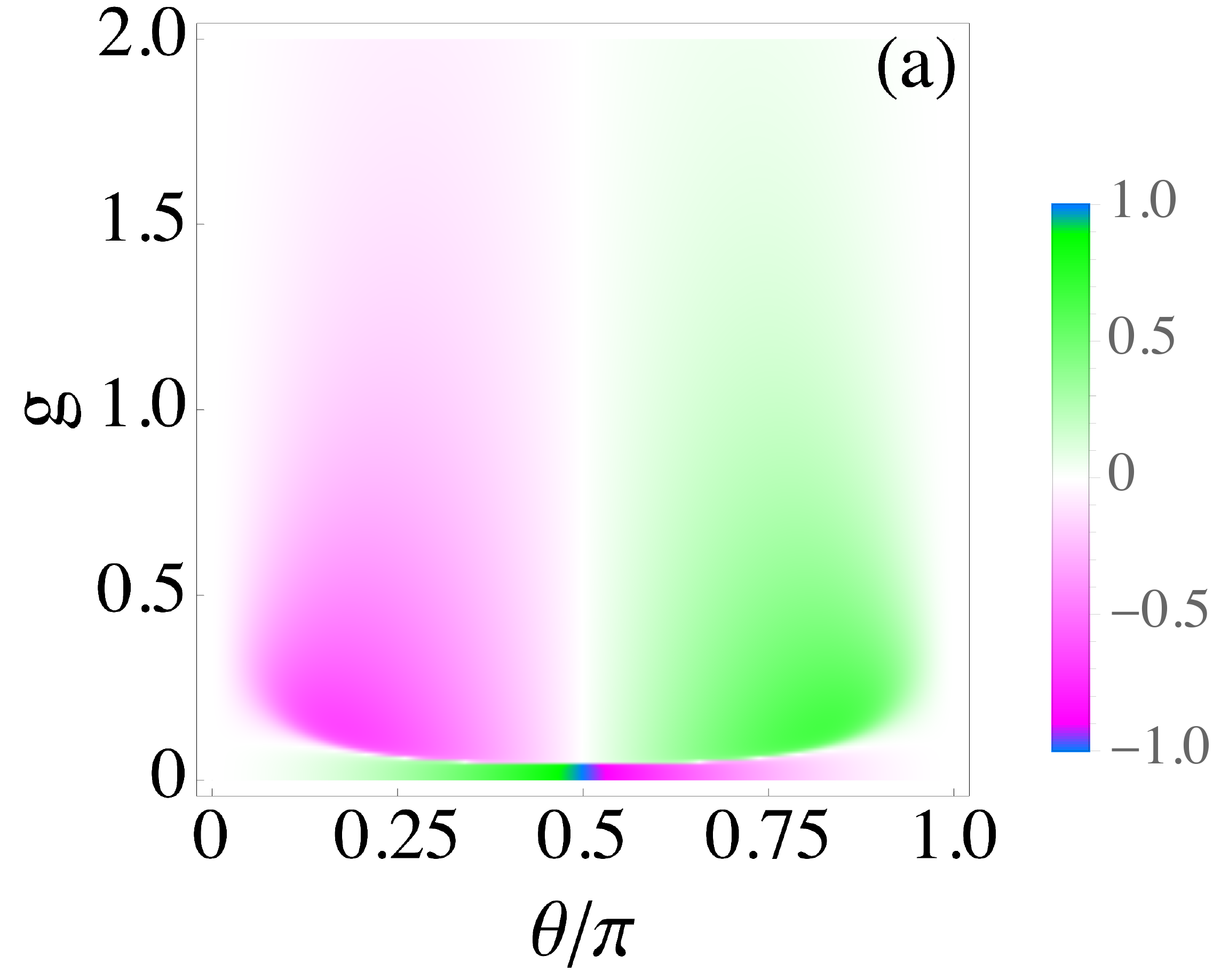}}
        \subfloat{\includegraphics[width=4.5cm]{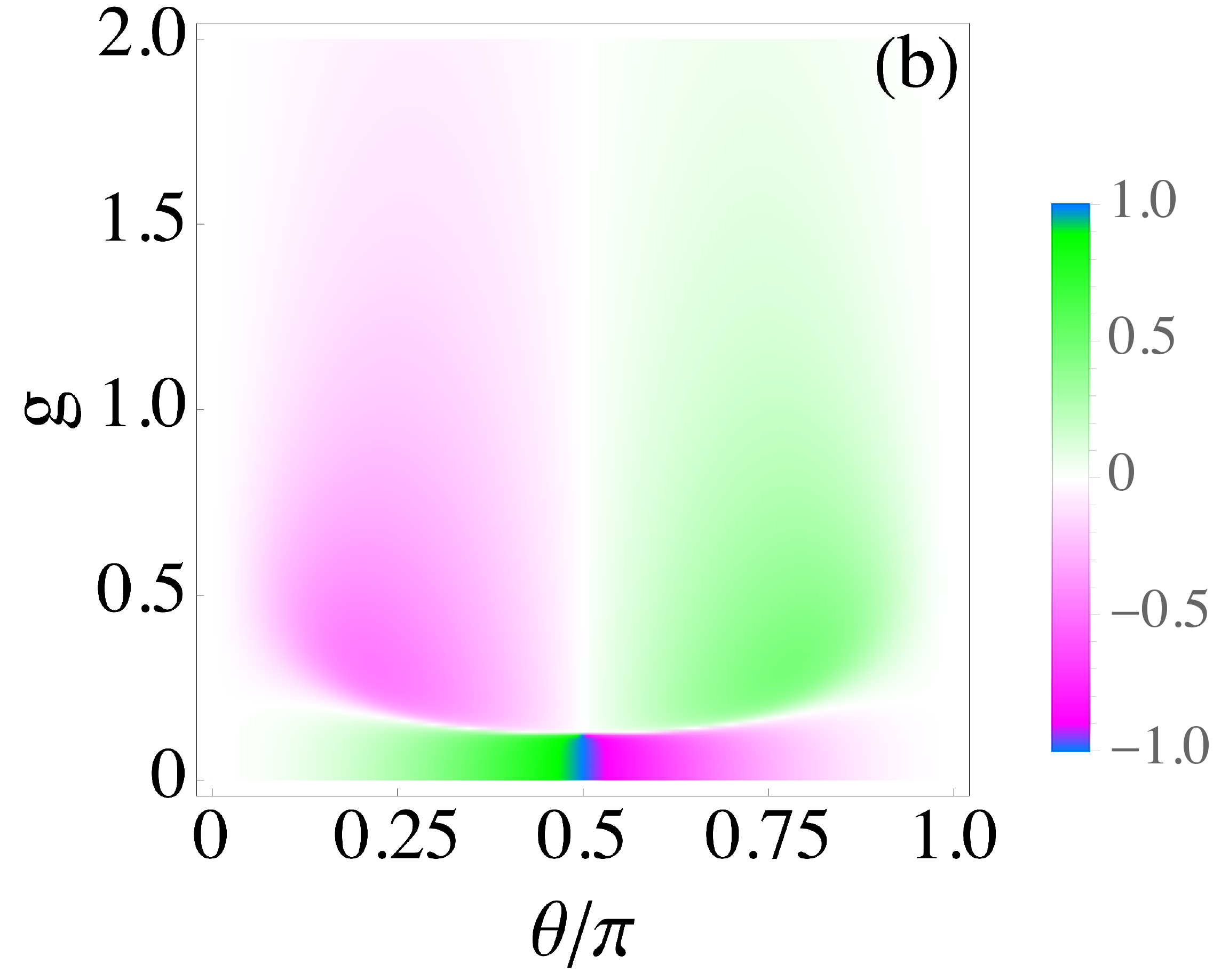}}
        \hspace{0.25cm}
        \subfloat{\includegraphics[width=4.5cm]{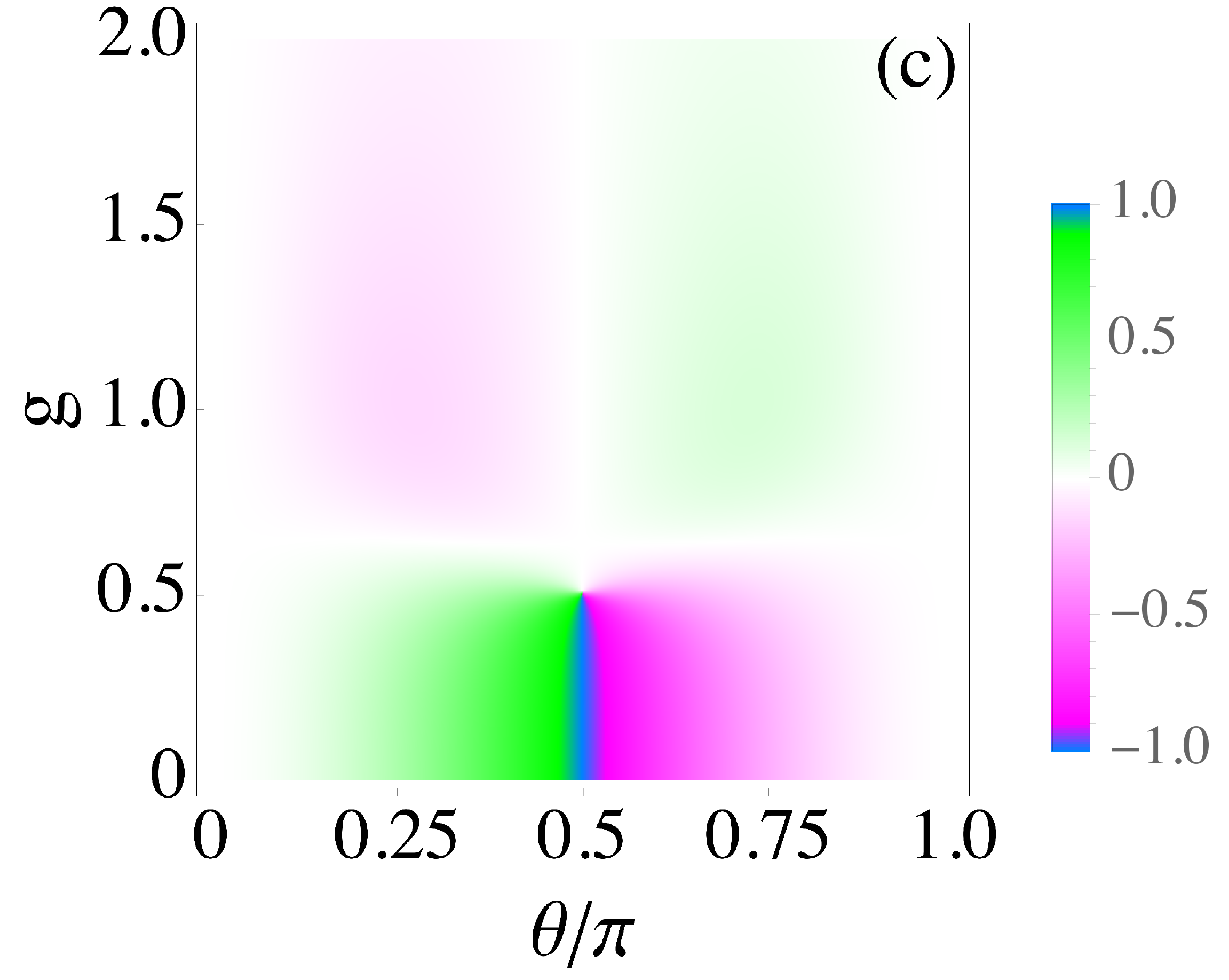}}
        \subfloat{\includegraphics[width=4.5cm]{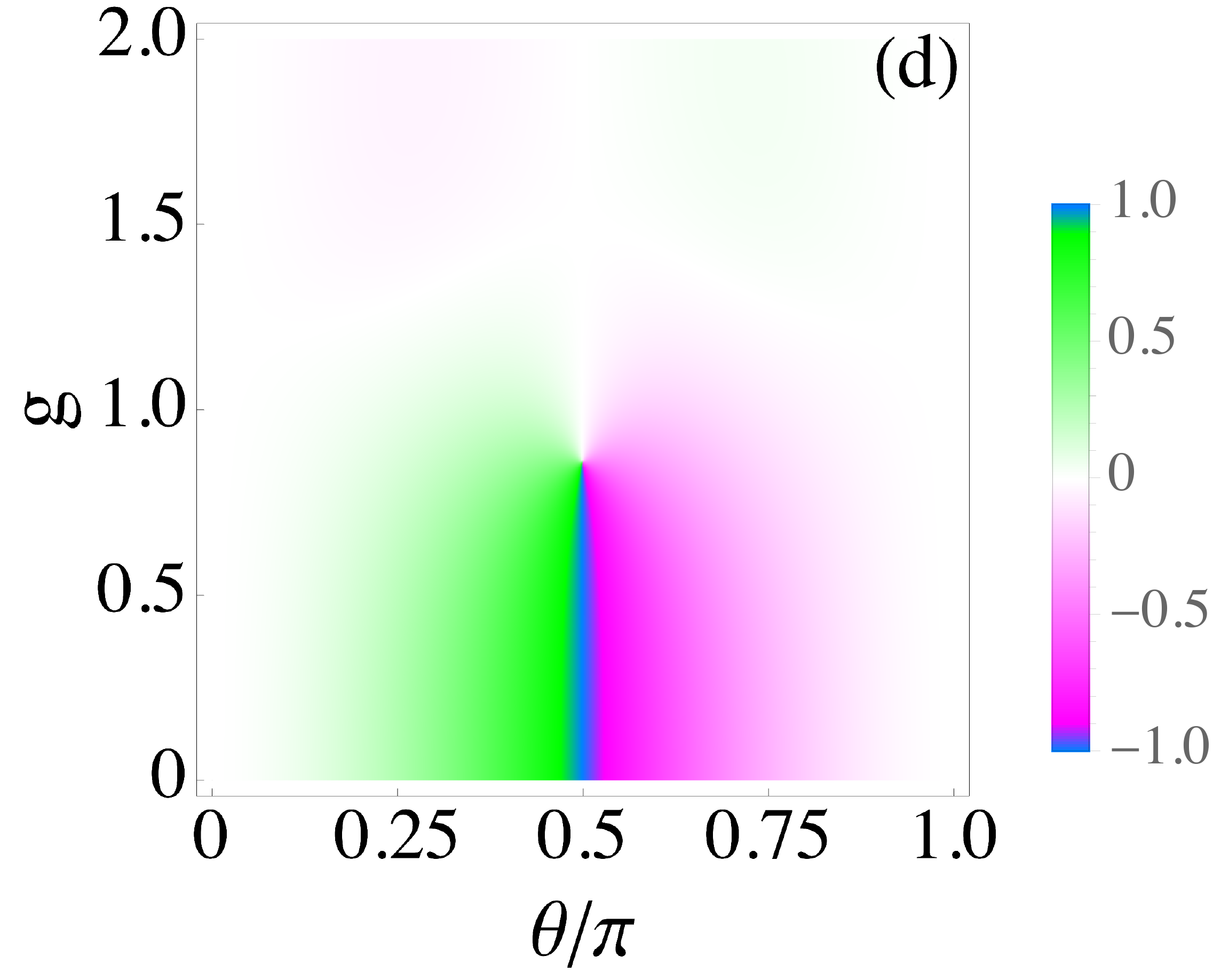}}
        \hspace{0.25cm}
        \subfloat{\includegraphics[width=4.5cm]{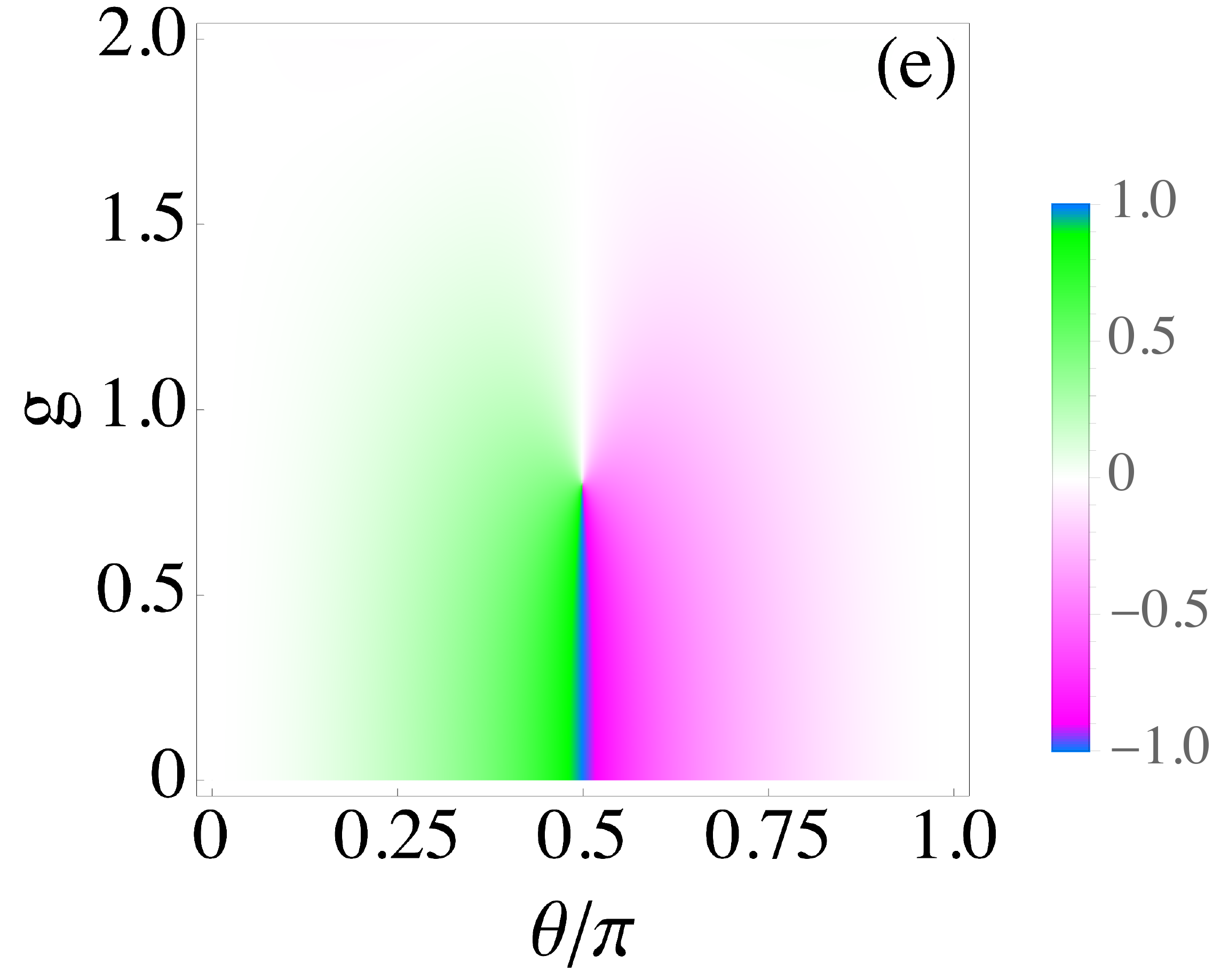}}
        \subfloat{\includegraphics[width=4.5cm]{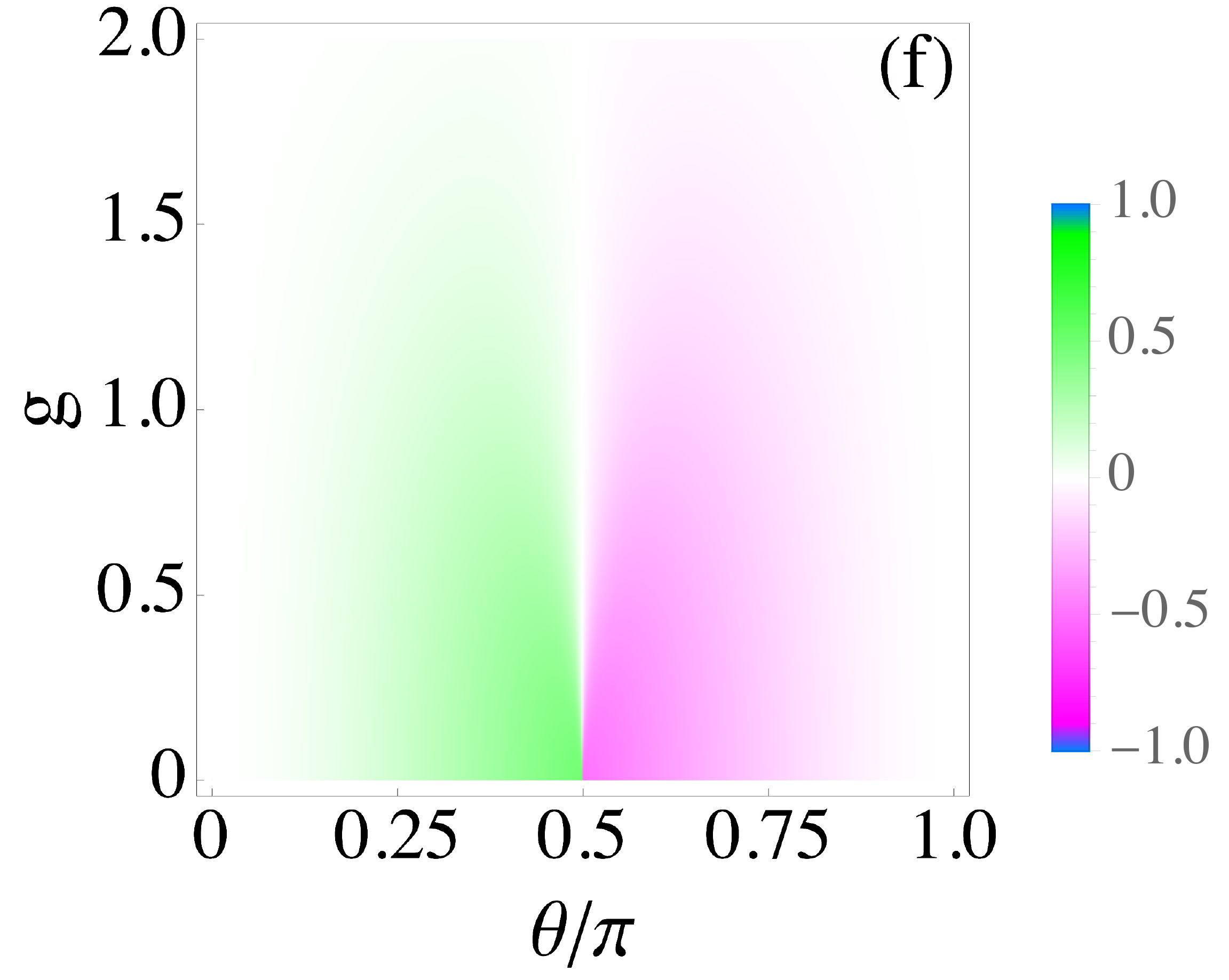}}
    \end{center}
    \caption{Color density maps of the Uhlmann phase of the composite system, $\Phi^{AB}$ [Eq.~(\ref{uhlman_bis})],  as a function of the coupling parameter $g$, and $\theta$, for different values of the temperature: (a) $T=0.02$, (b) $T=0.05$, (c) $T=0.2$, (d) $T=0.4$, (e) $T=0.6$ and (f) $T=T_c$. The vortex disappears at a critical temperature $T_c$.}
    \label{colormap1ABDepolarizing}
\end{figure}

We investigate  the mixing of states due to thermal effects of the composite system 
$\mathcal{H}^{AB} = \mathcal{H}^A\otimes\mathcal{H}^B$
of two interacting fermions with spin-$\frac{1}{2}$, by introducing the density matrix for a system in thermal equilibrium
%
%
\begin{equation}
    \rho=e^{-\beta\,\hat{H}}/\operatorname{Tr}[e^{-\beta\,\hat{H}}],
    \label{rhotermicas}    
\end{equation}
with $\beta=1/k_BT$, where $k_B$ is the Boltzmann constant (we set $k_B=1$ in our numerical calculations) and $T$ is the temperature. 
The Hamiltonian of the system $\hat{H}$ fulfills $\hat{H}\ket{u_j}=E_j\ket{u_j}$, with energy eigenvalues $E_j$ with corresponding eigenstates $\ket{u_j}$, which are also eigenfunctions of $\rho$ [Eq.~(\ref{rhotermicas})] \textit{i.e.}
$\rho\ket{u_j}=p_j\ket{u_j}$. Thus, the eigenvalues are simply given by $p_j=\operatorname{e}^{-\beta\,E_j}/Z$,
where  $Z=\sum_k \operatorname{e}^{-\beta E_k}$ is the canonical partition function. We evaluate the  Uhlmann connection $A(\lambda)$ [Eq.~(\ref{uhlAUb})], by letting the eigenstates $\ket{\psi_j}\rightarrow \ket{u_j}$, and the parameters $\lambda\rightarrow\phi$, and $\lambda_0\rightarrow\phi_0$.
By substituting in Eq.~(\ref{uhlAUb}), the analytical expressions $\braket{u_i|\partial_{\phi}u_j}=i\,(u_i^{(4)}u_j^{(4)}-u_i^{(1)}u_j^{(1)})/\sqrt{{\cal N}_i\,{\cal N}_j}$, 
the Uhlmann connection yields 
\begin{eqnarray}
    A(\phi)&=&\sum_{i\ne j}\frac{i}{\sqrt{{\cal N}_i\,{\cal N}_j}}\frac{\left(\sqrt{p_{j}}-\sqrt{p_{i}}\right)^2}  {p_{j}+p_{i}} \times \nonumber  \\
    && \left(u_i^{(4)}u_j^{(4)}-u_i^{(1)}u_j^{(1)}\right) \ket{u_i}\bra{u_j}\,d\phi.
    \label{uhlAUbbis2}
\end{eqnarray}
The Uhlmann holonomy $V(\phi,\phi_0)$ can be computed either by numerically solving the time-evolution in Eq.~(\ref{difqeV}) (with the initial condition $V(\phi_0,\phi_0)=\mathds{1}_4$) or by switching to the rotating reference frame using a unitary transformation (\ref{UnitaryT}).
In the latter case note that $\ket{u_i}\bra{u_j} = \hat{U}(\phi)\ket{u_i(0)}\bra{u_j(0)} \hat{U}^\dagger(\phi)$, and that all the coefficients in Eq.~\eqref{uhlAUbbis2} are constant. This allows us to write $\hat{A}(\phi) = \hat{U}(\phi)\hat{A}(0)\hat{U}^\dagger(\phi)$, which expresses the $\phi$-dependence of the Uhlmann holonomy by means of a unitary transformation. By solving Eq. \eqref{difqeV} in the rotating frame and transforming back to the laboratory frame, we obtain the following expression for the Uhlmann holonomy for a one-cycle evolution
\begin{equation}
    \hat{V} = e^{-2i\pi \left[ -(\hat{\sigma}_z\otimes\mathbb{1}-\mathbb{1}\otimes\hat{\sigma}_z)/2+\hat{K} \right]}.
\end{equation}
Here we have defined the Hermitian operator $\hat{K} = i\hat{A}(0)$ to clearly express the unitarity of $\hat{V}$.
Thus, the thermal Uhlmann phase of the composite system, $\Phi^{AB}(\theta,g,T)$, is given by
\begin{equation}
    \Phi^{AB}(\theta,g,T)={\rm Arg}\left\{{\rm Tr}[\rho_{\phi_0}\,\hat{V}(\phi,\phi_0)]\right\}.
    \label{uhlman_bis}
\end{equation}
We explore the Uhlmann phase for system $AB$ as a function of $g$ for all directions of the field in the low-temperature regime.
In Fig.~\ref{UhlmannvsBerrylowT}(a) we show that in the limit $T\rightarrow 0$, the Uhlmann phase resembles to the Berry phase [Fig.~\ref{UhlmannvsBerrylowT}(b)] of the ground state $\ket{u_2}$. The geometric phase was evaluated by using the general expression $\gamma_j=\int_0^{2 \pi}d\phi\,\braket{u_j|i\partial_{\phi}u_j}=(2\pi/{\cal N}_j)\{[u_j^{(1)}]^2-[u_j^{(4)}]^2\}$, corresponding to the $j$-eigenstate of the system. This result is consistent with the fact that for low temperatures, the thermal mixture of states tends to its ground state $\ket{u_2}$, governed by $E_2$.
%
%
Figure~\ref{colormap1ABDepolarizing}, shows the Uhlmann phase for system $AB$ as a function of $g$ for all directions of the field for different values of $T$. 
%
%
%
In Figs.~\ref{colormap1ABDepolarizing}(a)-(e) we show that the Uhlmann phase $\Phi^{AB}$ [Eq.~(\ref{uhlman_bis})] is 
non-trivial, with an evanescent magnitude as the temperature $T$ is increased. 
In the sequence of cases shown in Fig.~\ref{colormap1ABDepolarizing}(a)-(e), 
we note the appearance of a vortex in the $\theta=\pi/2$  direction, with a peculiar behavior for increasing values of temperature. Its position in $g$ increases as $T$ increases until reaching a particular temperature value from which its position begins to decrease. The position of the vortex continues to decrease until the temperature reaches the critical value $T_c=1/\operatorname{ln}[2+\sqrt{3}]$ 
after which the vortex disappears. Interestingly, this value coincides with the critical temperature reported by Viyuela \cite{viyuelaprl14} for two level systems.

\begin{figure}[htbp] 
    \begin{center}
        \subfloat{\includegraphics[width=4.5cm]{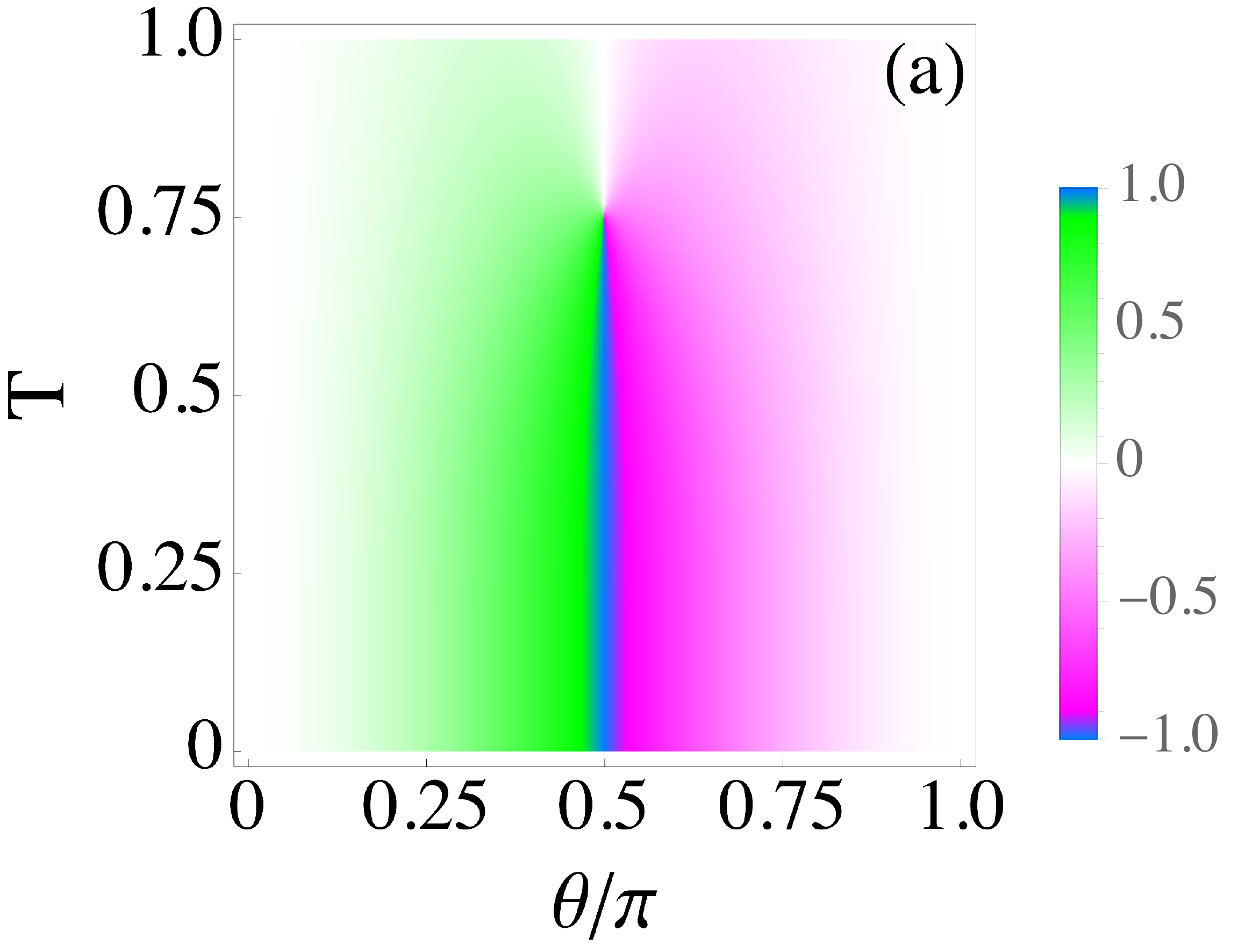}}
        \subfloat{\includegraphics[width=4.5cm]{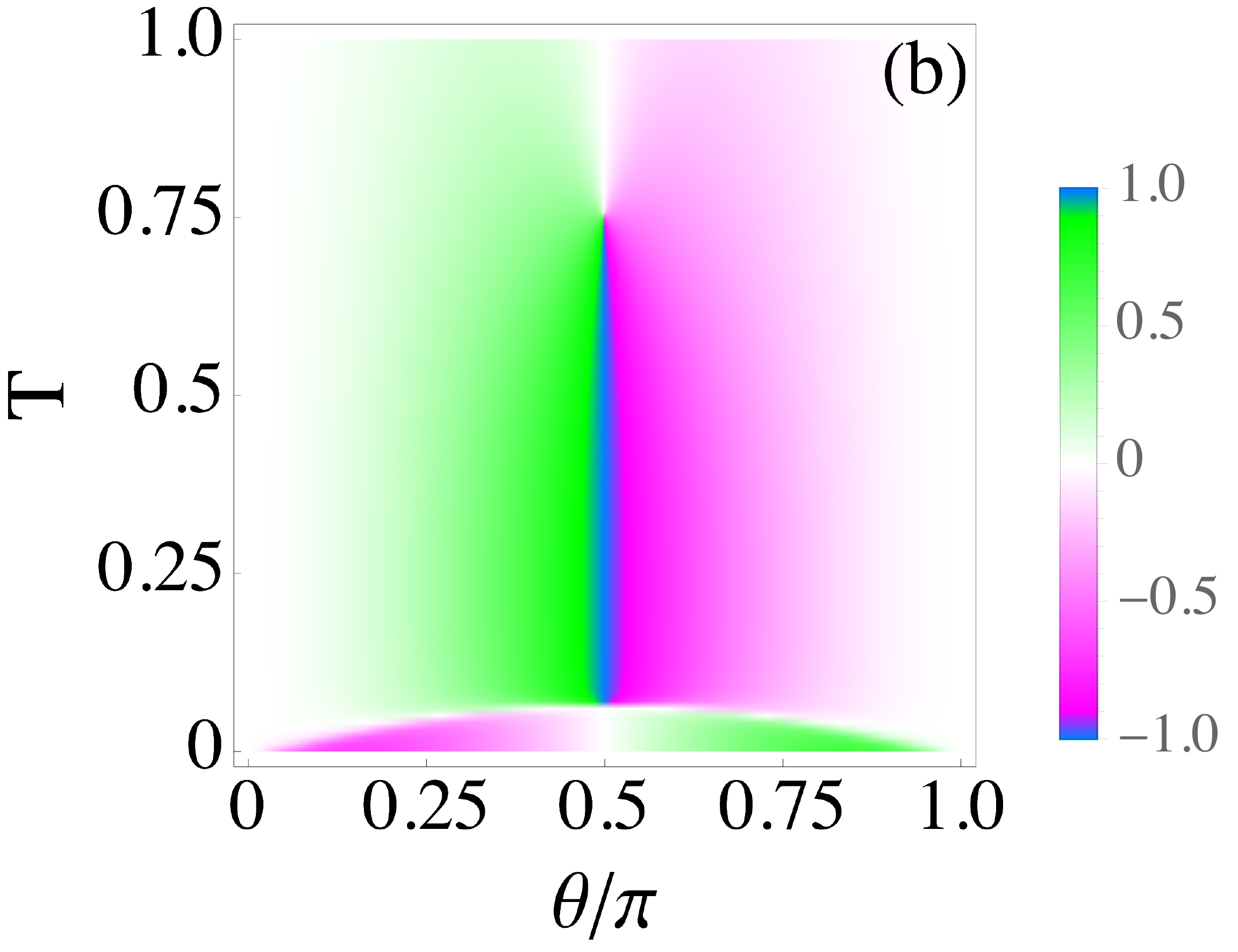}}
        \hspace{0.25cm}
        \subfloat{\includegraphics[width=4.5cm]{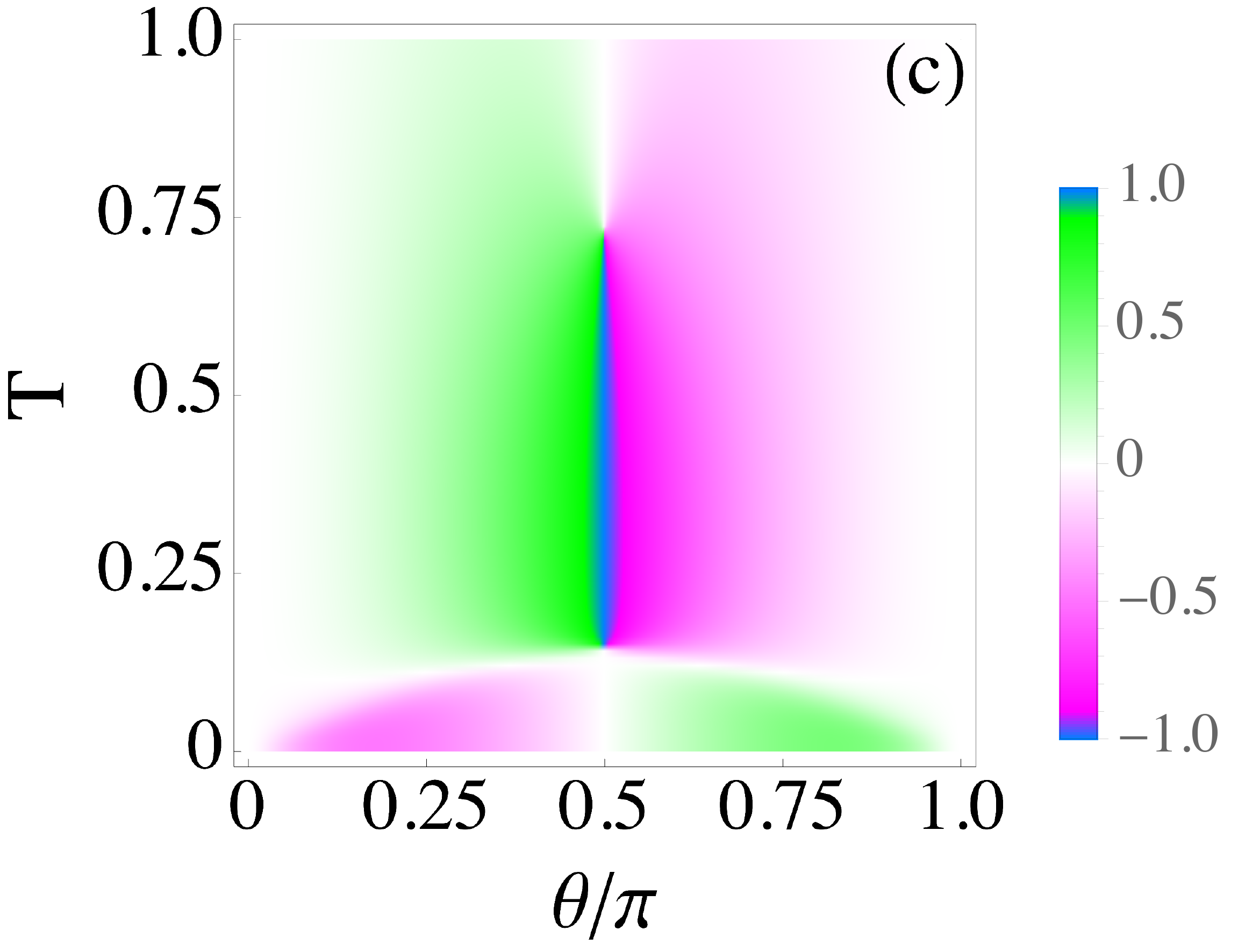}}
        \subfloat{\includegraphics[width=4.5cm]{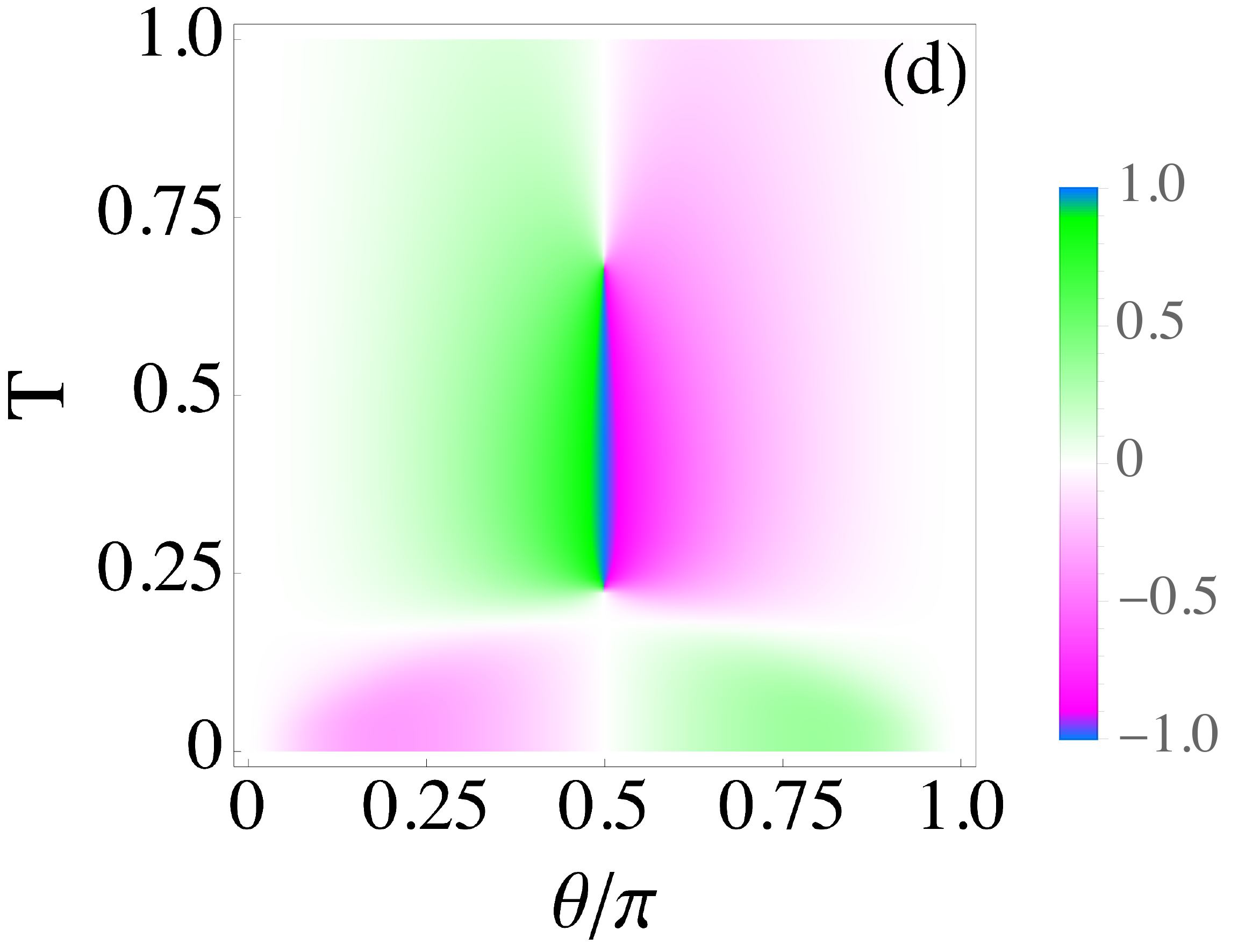}}
        \hspace{0.25cm}
        \subfloat{\includegraphics[width=4.5cm]{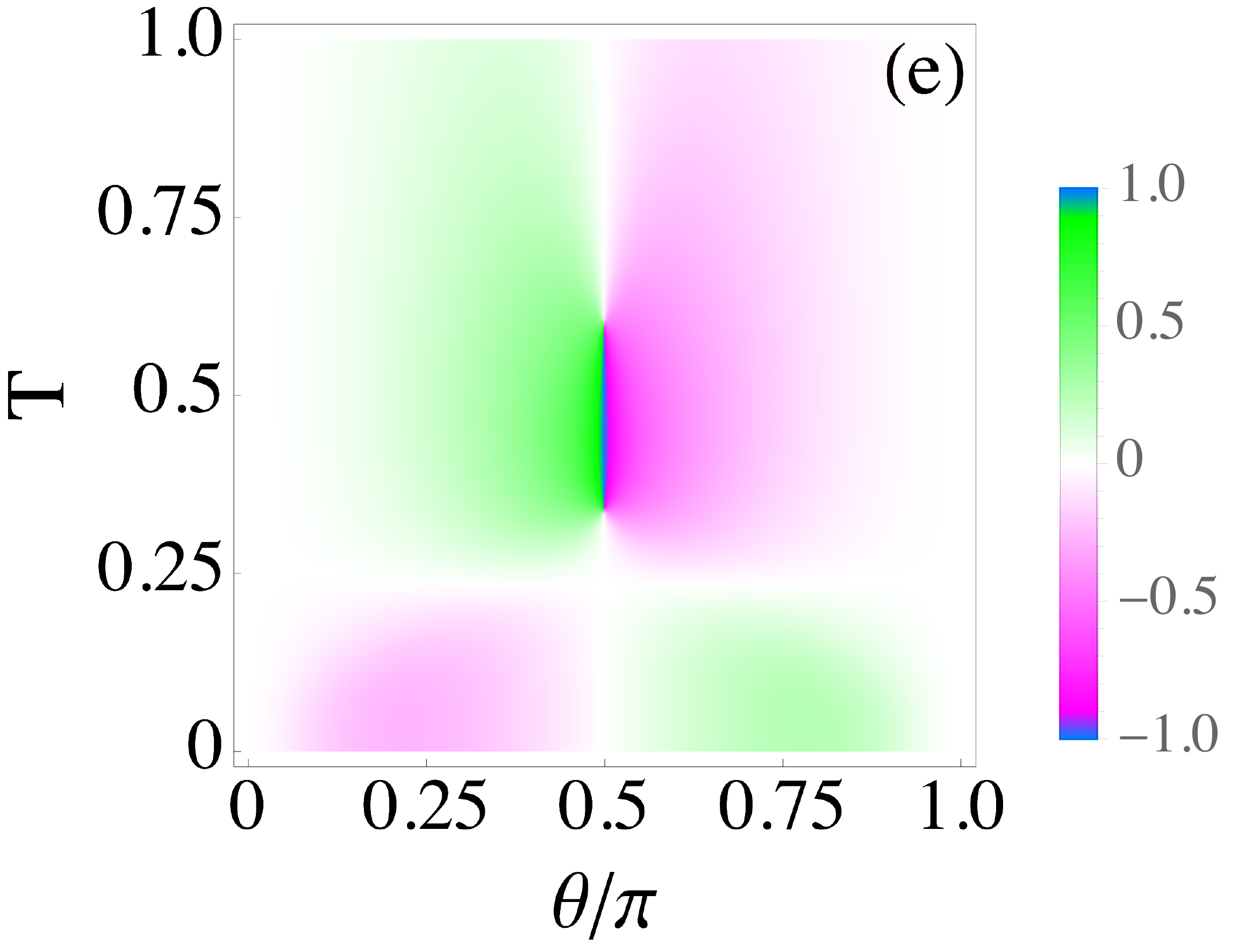}}
        \subfloat{\includegraphics[width=4.5cm]{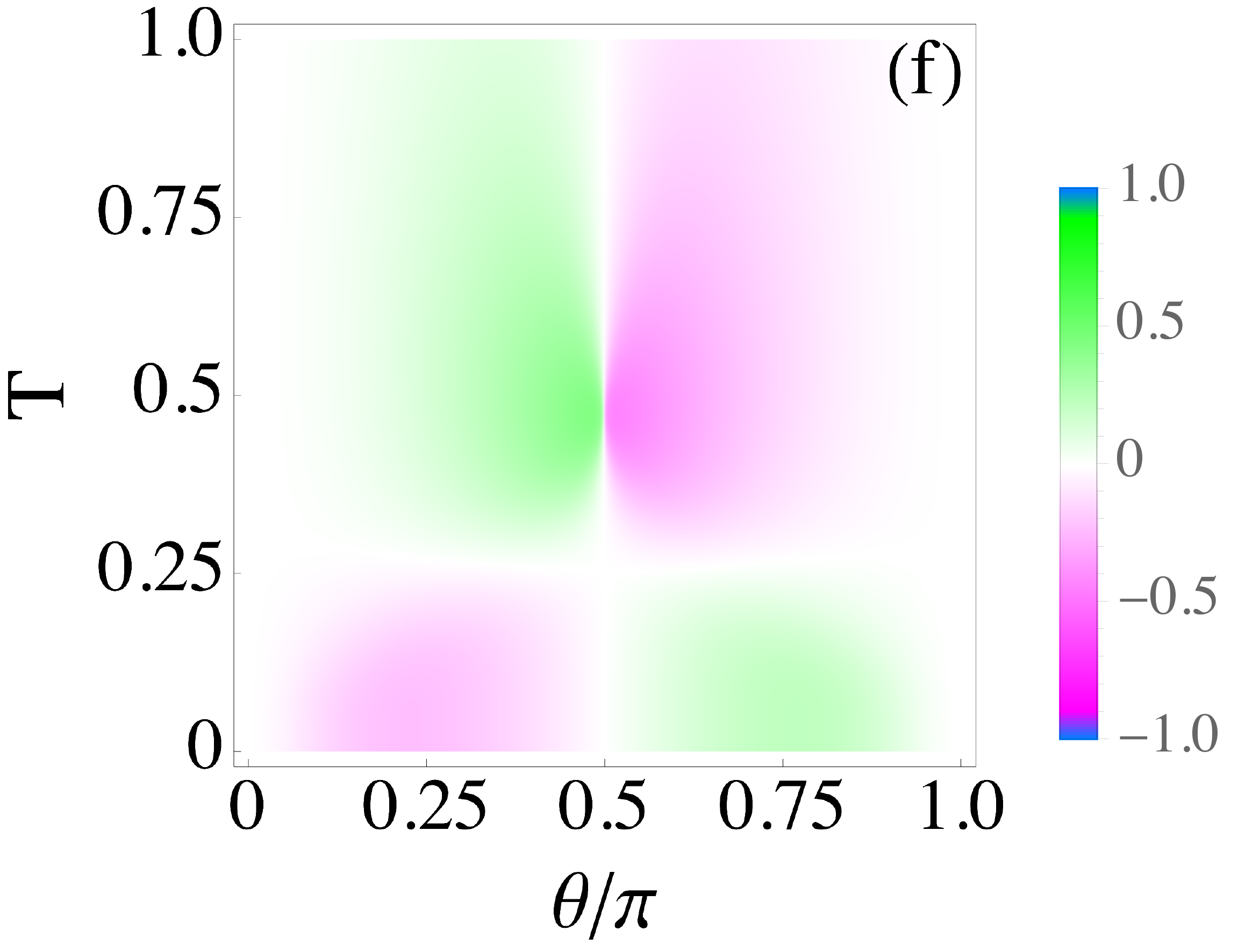}}
    \end{center}
    \caption{Color density maps of the Uhlmann phase of the composite system, $\Phi^{AB}$ [Eq.~(\ref{uhlman_bis})],  as a function of  $T$, and $\theta$, for different values of the coupling: (a) $g=0.02$, (b) $g=0.2$, (c) $g=0.4$, (d) $g=0.6$, (e) $g=0.8$ and (f) $g=0.9$. The vortex disappears at a critical value of the coupling.}
    \label{colormap1ABvsT}
\end{figure}

\begin{figure}[htbp] 
    \begin{center}
        \includegraphics[width=5.5cm]{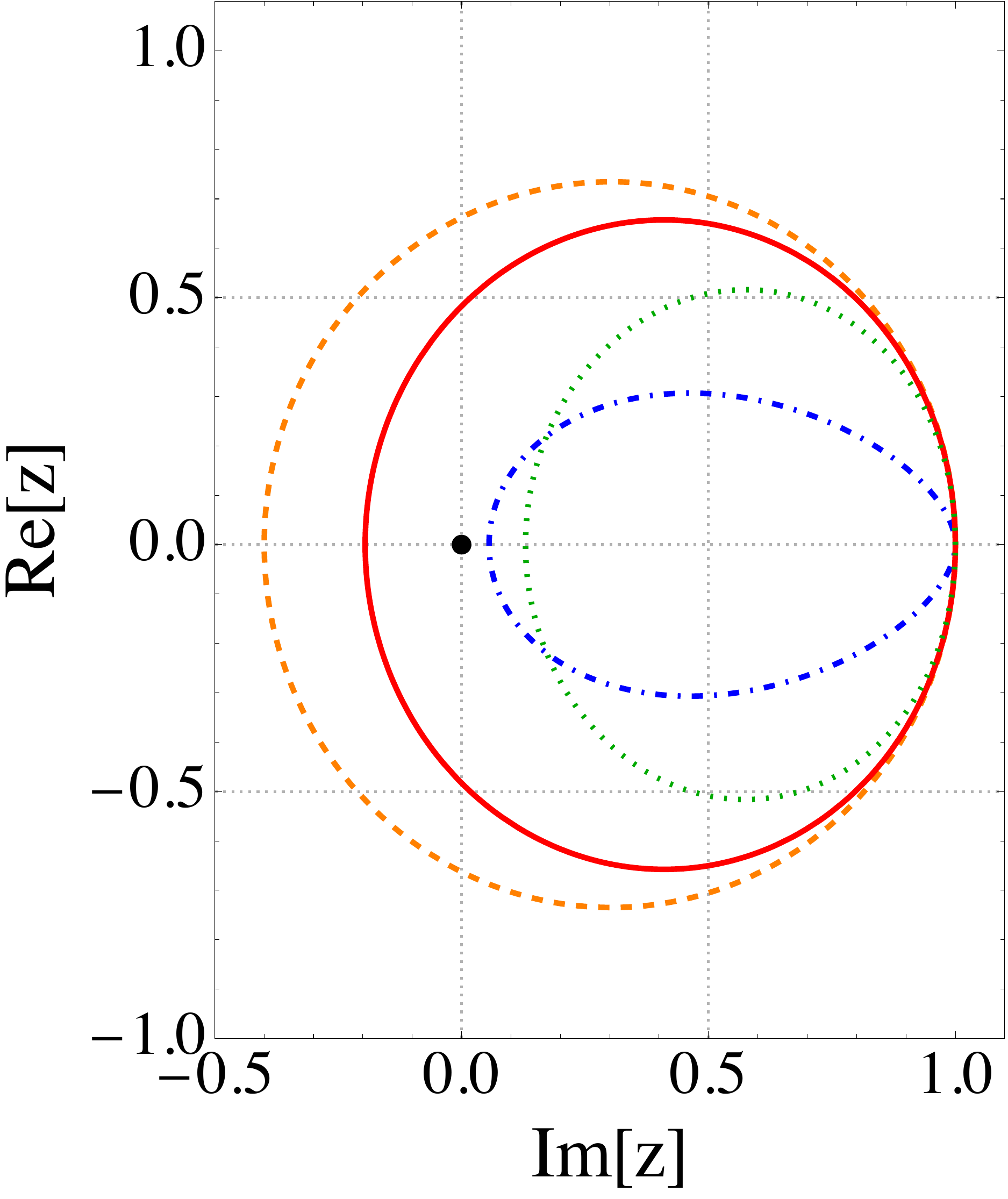}
    \end{center}
    \caption{ Argand diagram for $z(g,\theta,T)={\rm Tr}[\rho_{\phi_0}\,\hat{V}(\phi,\phi_0)]$ of system $AB$ in one-cycle evolution for several values of the temperature: $T=0.23$ (blue dashed dotted line), $T=0.5$ (orange dashed line), $T=0.6$, (red solid line), and $T=0.75$ (green dotted line). We have chosen $g=0.6$ in the calculation corresponding to the case of Fig.~\ref{colormap1ABvsT}(d). The winding number of the parametric curves changes twice as the temperature of the system increases.}
    \label{ArgandAB}
\end{figure}

\begin{figure}[htbp]
    \begin{center}
        \includegraphics[width=7.5cm]{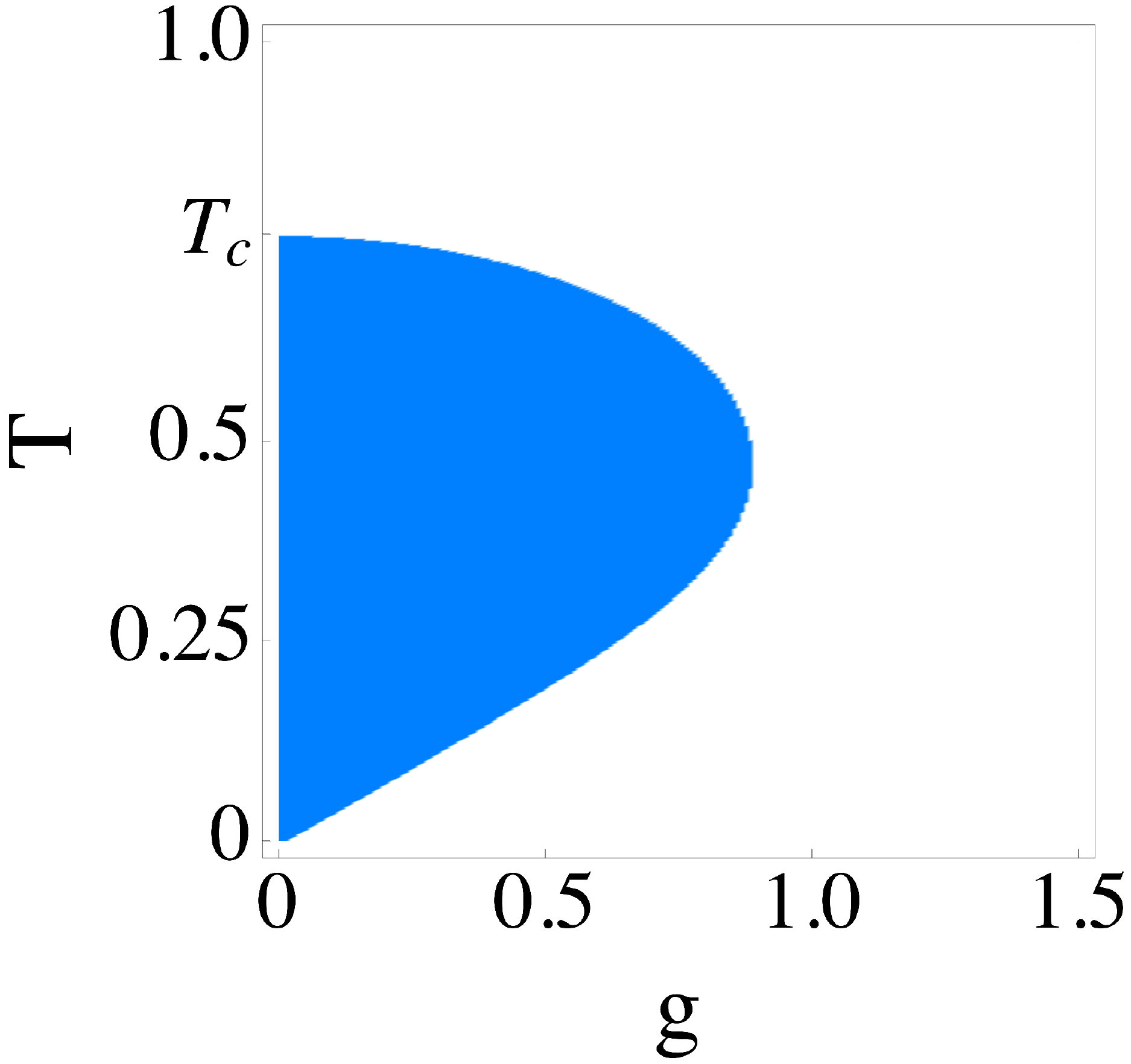}
    \end{center}
    \caption {Color density map for the Uhlmann phase $\Phi^{AB}$ as a function of $g$ and $T$ at a fixed direction $\theta=\pi/2$. The position of the vortex observed in Fig.~\ref{colormap1ABDepolarizing} corresponds to the set of points $(g,T)$ which define the Uhlmann phase boundary  where the phase changes abruptly  from $0$ to $\pi$ (blue). There are no phase transitions for temperatures higher than the critical value $T_c$.}
    \label{phaseABthetapie2}
\end{figure}

\begin{figure}[htbp] 
    \begin{center}
        \includegraphics[width=8.5cm]{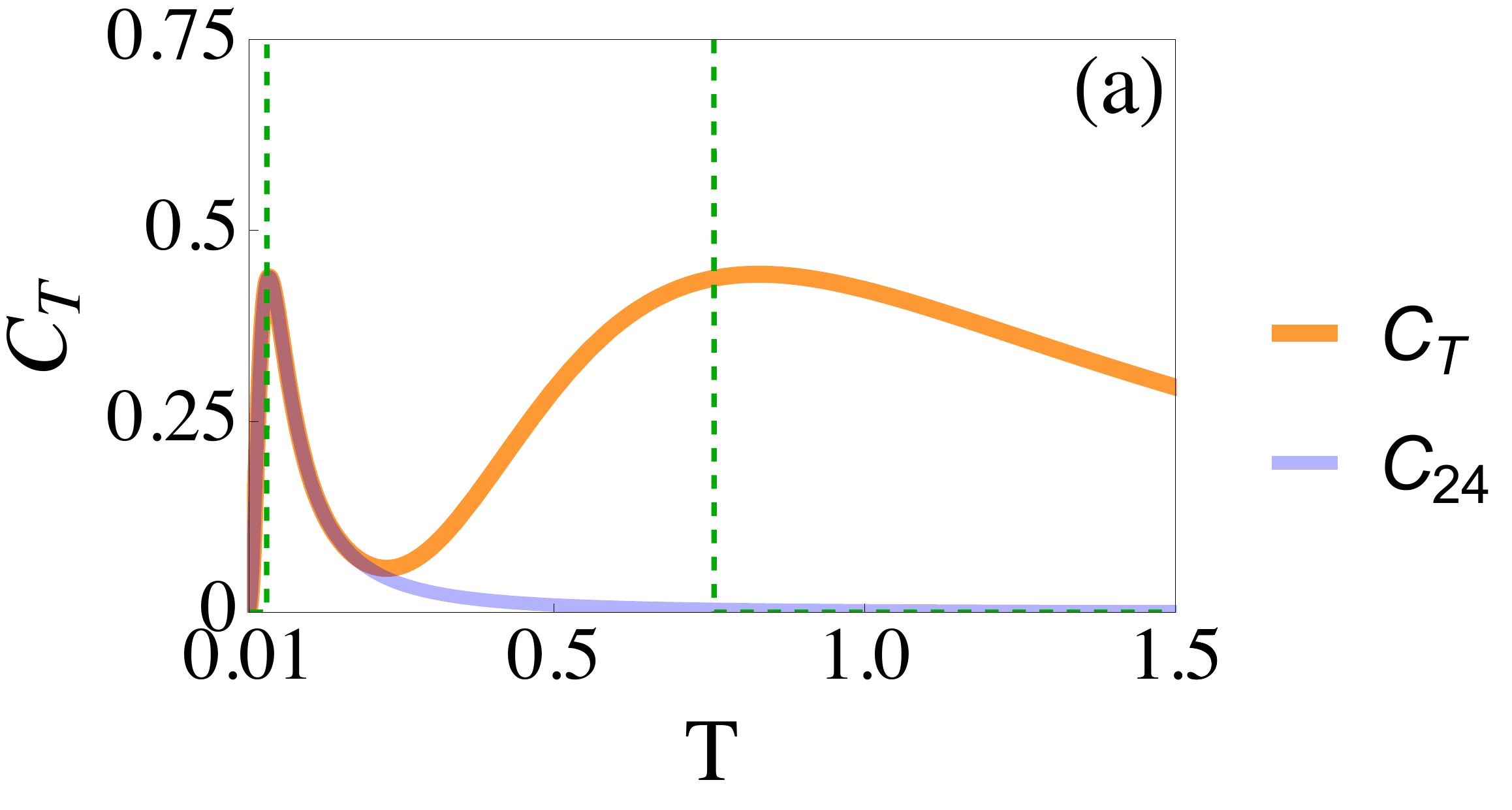}
        \includegraphics[width=8.5cm]{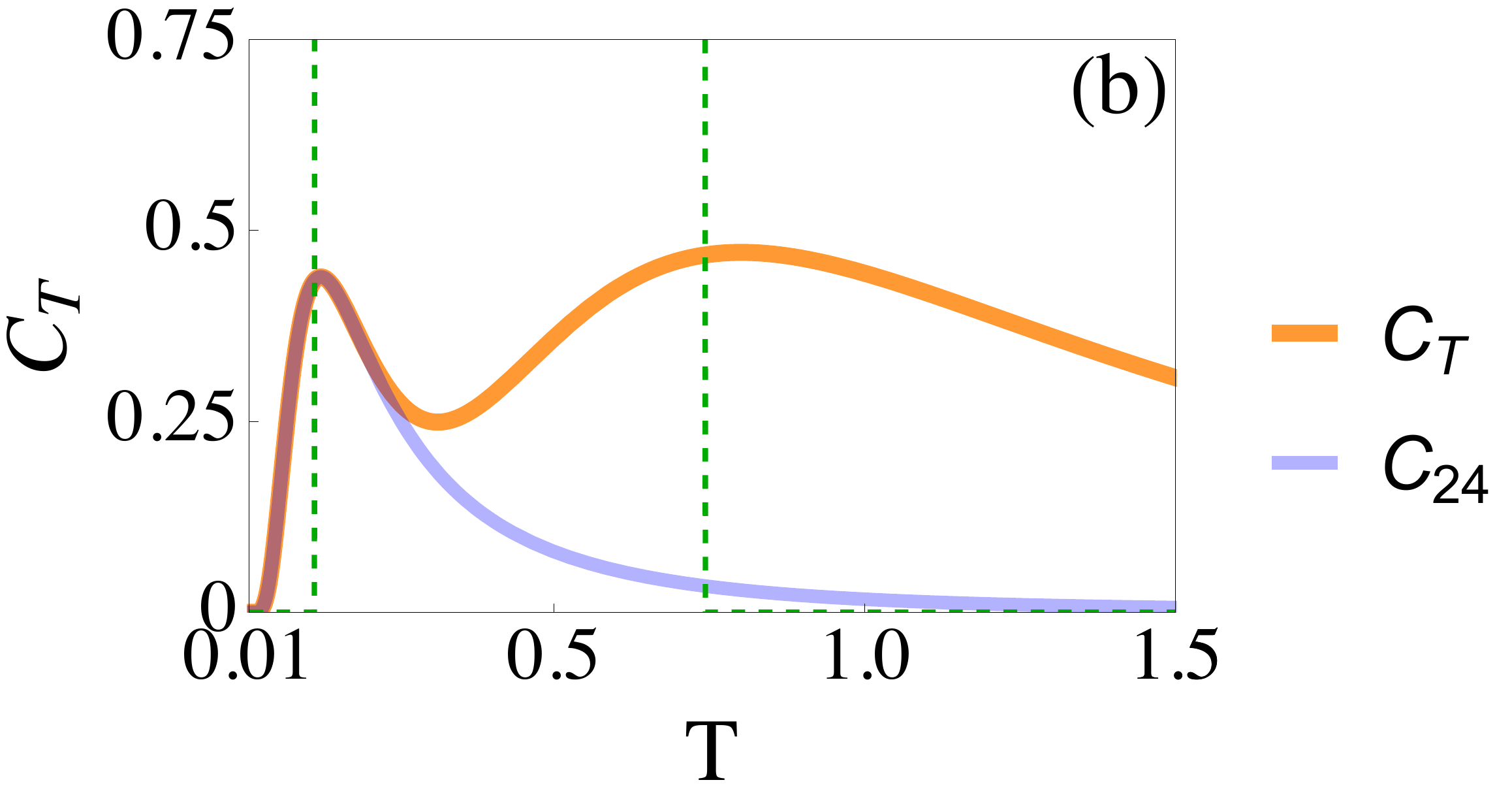}
        \includegraphics[width=8.5cm]{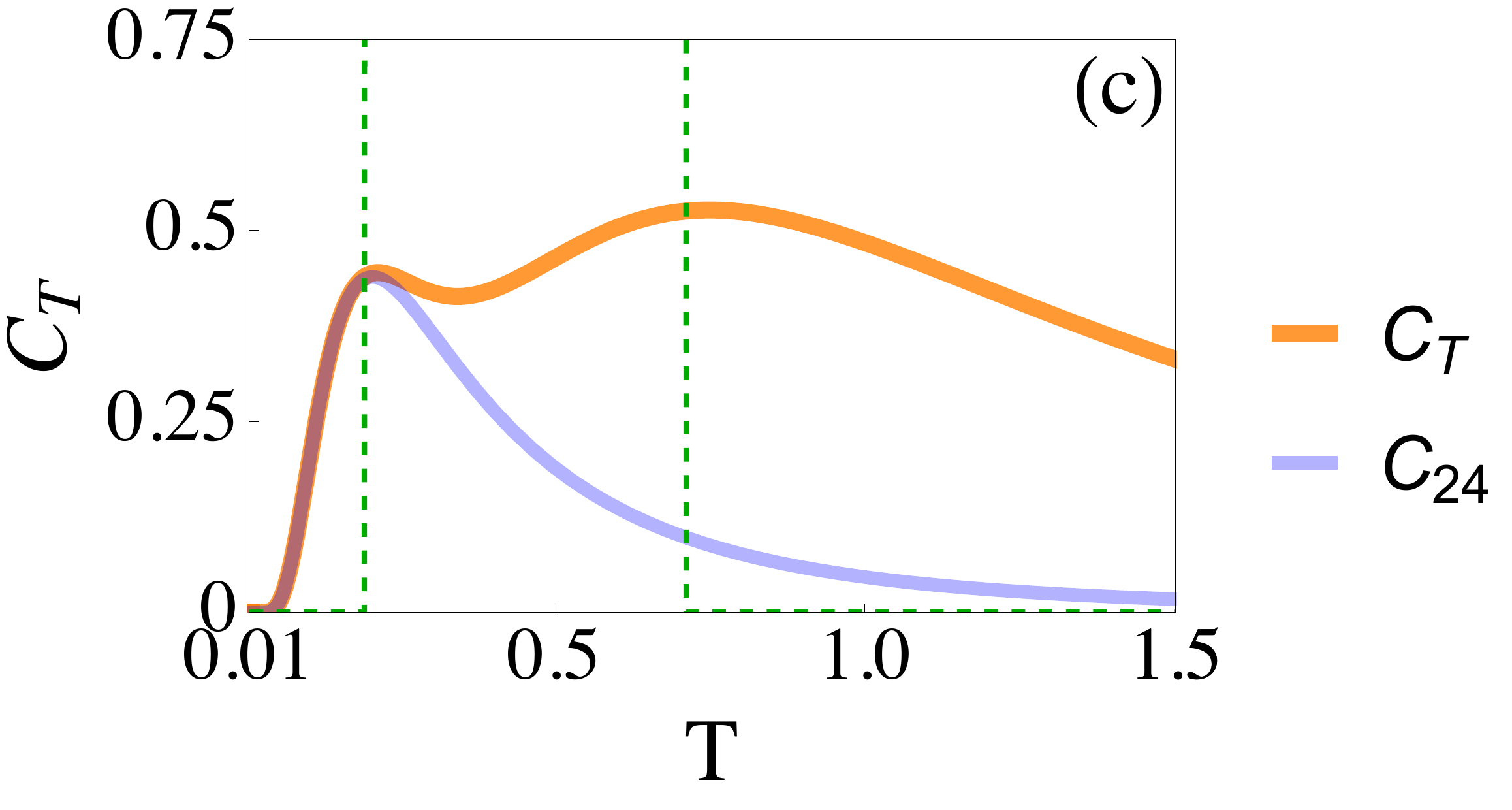}
        \includegraphics[width=8.5cm]{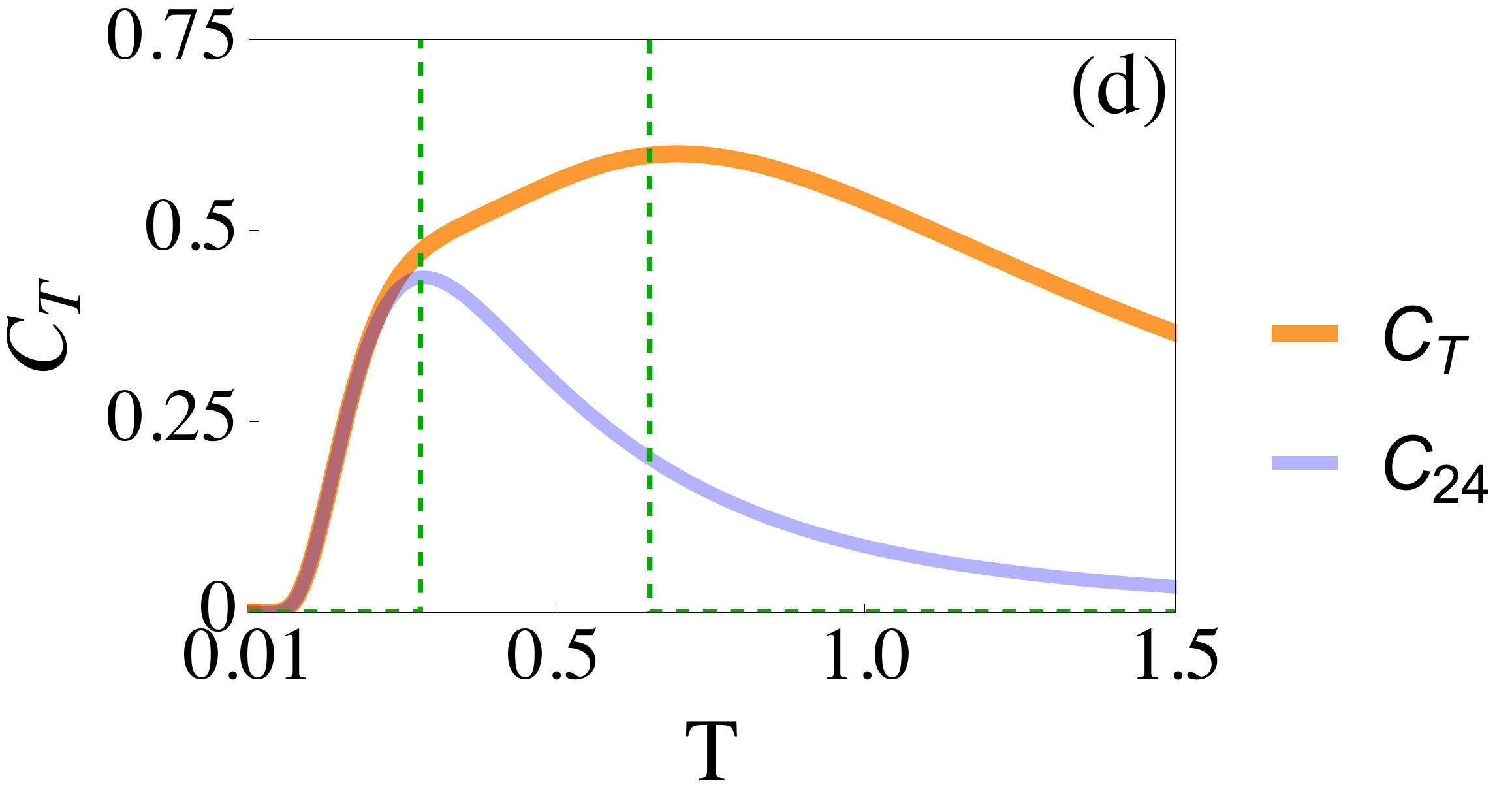}
    \end{center}
    \caption{(a) Heat capacity $C_T$ [Eq.~(\ref{CTraw})] (orange solid line) at $\theta=\pi/2$ for the couplings: (a) $g=0.1$, (b) $g=0.3$, (c) $g=0.5$, and (d) $g=0.7$. We show  that the width $\Delta T$ of the Uhlmann phase is roughly the separation of the $C_T$ peaks. We include the two-level contribution $C_T^{24}$ (blue solid line) responsible for the Schottky anomaly of $C_T$. For comparison we also include the  Uhlmann phase $\Phi^{AB}$ transitions (green dashed line). }
    \label{Ct_vs_Uhl}
\end{figure}

Next, we present in Fig.~\ref{colormap1ABvsT} the behavior of the Uhlmann phase $\Phi^{AB}$ as a function of temperature along all directions $\theta$ for different values of the coupling. In this case, we show the appearance of a double vortex in the system along the path $\theta=\pi/2$. The vortices disappear after we reach a particular critical value of the coupling.
We examine the phase transitions observed in Fig.~\ref{colormap1ABvsT}  from another perspective, by analyzing the Argand diagram of $z(g,\theta,T)={\rm Tr}[\rho_{\phi_0}\,\hat{V}(\phi,\phi_0)]$, using the Argument Principle of complex analysis \cite{VisualComplexFunctions}.
Figure~\ref{ArgandAB} shows the parametric plot $z(g,\theta, T)$ for different temperature values at a fixed coupling value.  
The zeros of ${\rm Tr}[\rho_{\phi_0}\,\hat{V}(\phi,\phi_0)]$ get mapped to the origin of $z$-plane (solid black dot), with parametric curves that wind (or not) around it. According to the Argument Principle, if the parametric curves wind once around the origin in the $z$-plane, that tells us that the corresponding curves in the complex plane of ${\rm Tr}[\rho_{\phi_0}\,\hat{V}(\phi,\phi_0)]$ must have had one zero inside it. Likewise, if the curve does not wind around the origin, there must have been no zeros. The change in winding numbers corresponds to the number of times that the Uhlmann phase $\Phi^{AB}$ [(Eq.~(\ref{uhlman_bis})] changes from 0 to $\pi$. In Fig.~\ref{colormap1ABvsT}, the winding number changes twice for a fixed value of $g$ for increasing temperature values.
%
In Fig.~\ref{phaseABthetapie2}, we emphasize the  behavior of the vortex, where we show a density plot of  $\Phi^{AB}$  vs  $g$ and $T$ for  $\theta=\pi/2$. The vortex position corresponds to all those sets of values $(g, T)$ that define the  Uhlmann phase boundary where $\Phi^{AB}$ changes abruptly from 0 to $\pi$.
%
%
 %
%
%
In Fig.~\ref{phaseABthetapie2}, we can see that in the limit  $g\rightarrow 0$, the temperature gap, $\Delta T$,  of the Uhlmann phase tends to the known result for spin-$\frac 1 2$ fermions in crystal momentum $k$-space \cite{viyuelaprl14}. Only one critical temperature $T_c$ is observed, which corresponds to a single vortex, as shown in Fig.~\ref{colormap1ABvsT}(a).
The gap $\Delta T$ begins to narrow for increasing values of  $g$,  revealing that there are two critical temperatures for a single value of the coupling.
To motivate the discussion about the observed behavior of $\Delta T$ exhibited by the Uhlmann phase for different values of the coupling, let us analyze the heat capacity of the system. The latter is defined as 
$C_T=\partial \braket{E}/\partial T=\beta^2\,\partial^2 (\operatorname{ln} Z)/\partial \beta^2$, where $\braket{E}$  is the thermal average of the energy, which leads us to the following expression for $\theta=\pi/2$, 
%
%
\begin{equation}
    C_T=\frac{g^2 \operatorname{sech}^2\left(g/2 T\right)+\left(g^2+4\right) \operatorname{sech}^2\left(\sqrt{g^2+4}/2 T\right)}{4 T^2}.
    \label{CTraw}
\end{equation}
In Fig.~\ref{Ct_vs_Uhl}(a)-(d), we show $C_T$ [Eq.~(\ref{CTraw})] as a function of temperature, which exhibits a structure characterized by a two-peak specific heat anomaly observed in multilevel models \cite{souza_specific_2016}.
%

%
For a wide range of coupling values, $g$, the first phase transition of $\Phi^{AB}$ corresponds to the first peak of $C_T$. The latter is well-defined for small values of $g$, while it becomes a shallow maximum for larger values of $g$. The second peak of $C_T$ occurs near the second phase transition of the Uhlmann phase, but this is different for larger values of $g$.
Since there is a correlation between the phase transitions of $\Phi_{AB}$ and the position of the peaks of $C_T$, we proceed to further investigate the physical quantities responsible for the emergence of the maxima in the heat capacity. We derive an alternative exact expression for the heat capacity $C_T=\sum _{i<j}\, C_T^{ij}$ that involves the contributions due to the energy gaps of the system spectrum,  where the $C_T^{ij}$ are the two-level type contributions to the heat capacity given by,  
%
%
%
\begin{equation}
    C_T^{ij}= \left(\beta/Z\right)^2  \operatorname{e}^{-2\beta E_i}\,\Delta_{ij}^2\,\operatorname{e}^{\beta \Delta_{ij}}=\beta^2\,p_i^2\,  \Delta_{ij}^2\, \,\operatorname{e}^{\beta \Delta_{ij}},
    \label{CTrawbis}
\end{equation}

and $\Delta_{ij}=E_i-E_j$ are the energy gaps with 
$E_1=-E_2=(g+\sqrt{g^2+4})/2$ and $E_3=-E_4=(-g+\sqrt{g^2+4})/2$.
%
Interestingly, the crossing of $\Delta_{13}$ and $\Delta_{34}$  occurs at $g=2/\sqrt{3}$, which will be of relevance  when we study phase transitions in the subsystems.
From Eq.~(\ref{CTrawbis}) we emphasize that 
the double-peaked structure of $C_T$ arises as an interplay of multiple two-level contributions. 
In particular, we demonstrate that the characteristic first peak in the low-temperature regime is governed by the two-level contribution $C_T^{24}$ \cite{souza_specific_2016} involving the ground state $E_2$ and first excited state $E_4$. 
In Figs.~\ref{Ct_vs_Uhl}(a)-(d), we show that the first critical temperature of the Uhlmann phase occurs at the maximum of $C_T^{24}$.  
Moreover, by taking the limit of Eq.~(\ref{CTrawbis}) in the low-temperature regime and using  
$Z\simeq \operatorname{e}^{- \beta E_2}(1+\operatorname{e}^{\beta \Delta_{24}})$, 
we show that
\begin{equation}
    C_T^{24}\simeq\frac{(\beta\, \Delta_{24})^2 \,\operatorname{e}^{\beta \Delta_{24}} }{\left( 1+ \operatorname{e}^{\beta \Delta_{24}}\right)^2},
    \label{ctschotkybis}
\end{equation}
which is the well-known formula for the \textit{Schottky anomaly} of the heat capacity.
In the case of the second critical temperature of the Uhlmann phase, which is very close to the second maximum of the $C_T$, the situation is more involved because the  $C_T^{ij}$'s have different weights for the other coupling and temperature regimes.
These results show how the topological  phase transitions of an abstract quantity, such as the Uhlmann geometric phase, can be related to a measurable physical quantity in solid-state physics, such as the heat capacity of the system. 
%

%
%
%

We have obtained two results  that we want to highlight: 
the first  is that the Uhlmann topological phase transition disappears for temperature values $T\geq T_c$, where $T_c$ is the critical temperature. The latter is a characteristic parameter of fermionic systems,  reported by Viyuela \textit{et al.} \cite{viyuelaprl14}. 
Interestingly, we did not observe these types of transitions in our previous work involving a composite system with mixed states induced by noisy channels \cite{pravillaeta21}, even using alternative definitions to describe geometric phases such as interferometric phases. 
The second is the surprising finding that the Uhlmann topological phase transitions induced by thermal effects are related to the heat capacity of the system. In particular,  we show evidence of a non-trivial correspondence between the Schottky anomaly of the heat capacity and a topological phase transition.


%
%
From our previous results, we expect a more elaborate structure of the Uhlmann phase in the subsystems, which we will explore in the next section.

\section{Thermal effects on Uhlmann phase for the subsystems.}\label{sec:subsysAandB}

We study the geometric phase of subsystems $\mathcal{H}^A$ (driven fermion) and $\mathcal{H}^B$ (undriven fermion) derived from the composite state, $\rho$,  and investigate the main features of the Uhlmann phase for different temperature values.
We obtain the density matrices for the subsystems $A$ ($B$) by computing the trace of $\rho$ over $B$ ($A$), given by $\rho^A={\rm Tr}_B[\rho]$, and  $\rho^B={\rm Tr}_A[\rho]$, respectively.
The $\rho^s$ are represented by general $2 \times 2$ matrices,
\begin{equation}
    \rho^s=
     \begin{pmatrix}
     a_{s} & c_{s} \,e^{-i \phi}\\ 
     c_{s}  \,e^{+i \phi}  & 1-a_{s}
    \end{pmatrix}, 	
    \label{bastardnotation}
\end{equation}
where the real coefficients $a_{s}$, and $c_{s}$ ($s=A,B$) for each eigenstate,  depend on the direction 
$\theta$, the coupling parameter $g$, and the temperature $T$, and  are independent of $\phi$:
\begin{eqnarray}
    a_{A}(\theta,g,T)&=&\sum_{j=1}^4{\cal N}_j^{-1}\left[ \left(u^{(1)}_j\right)^2+\left(u^{(2)}_j\right)^2\right]\,p_j; \nonumber \\ 
    c_{A}(\theta,g,T)&=&\sum_{j=1}^4{\cal N}_j^{-1}\left[ u^{(1)}_j\,u^{(3)}_j+ u^{(2)}_j\,u^{(4)}_j\right]\,p_j,
    \label{coefA}
\end{eqnarray}
and
\begin{eqnarray}
    a_{B}(\theta,g,T)&=&\sum_{j=1}^4{\cal N}_j^{-1}\left[ \left(u^{(1)}_j\right)^2+\left(u^{(3)}_j\right)^2\right]\,p_j; \nonumber \\ 
    c_{B}(\theta,g,T)&=&\sum_{j=1}^4{\cal N}_j^{-1}\left[ u^{(1)}_j\,u^{(2)}_j+ u^{(3)}_j\,u^{(4)}_j\right]\,p_j.
\label{coefB}
\end{eqnarray}
The eigenvalues of $\rho^s$  are
\begin{eqnarray}
    p_{s,1}&=&\left[1-\sqrt{(1-2a_{s})^2+4c_{s}^2}\, \right]/2; \\
    p_{s,2}&=& \left[1+\sqrt{(1-2a_{s})^2+4c_{s}^2}\, \right]/2,
\label{eigenvalspin}    
\end{eqnarray}
which satisfy the conditions  $p_{s,1}+p_{s,2}=1$, and $p_{s,1}\, p_{s,2}={\rm det}[\rho^s]=a_{s}(1-a_{s})-c_{s}^2$. The corresponding eigenvectors are,
\begin{equation}
    \ket{v_{s,l}}=\frac{1}{\sqrt{N_{s,l}}}
    \begin{bmatrix}
    \beta_{s,l}  \,e^{-i \phi}\\ 1 
    \end{bmatrix}, \\
\label{eigenvecspin}    
\end{equation}
where $l=1,2$,  $N_{s,l}=\beta_{s,l}^2+1$,
with $\beta_{s,l}=c_{s}/(p_{s,l}-a_{s})$.
The Uhlmann connection  can be computed from Eq.~(\ref{uhlAUb}), by considering the variation of the parameter $\phi$,
which leads us to $A^s(\phi)=-2i\Delta p_{s}\,(\bm{n}_{\delta_s}\cdot \bm{\sigma})\, d\phi$,  with $\bm{n}_{\delta_s}=(-\delta_s\cos\phi,-\delta_s\sin\phi,1)$, where
%
$\Delta p_{s}=[1-2\sqrt{{\rm det}[\rho^s]}]/N_{s,1}N_{s,2}$,
and the parameter $\delta_s=(2a_{s}-1)/2c_{s}$.
%
%
%
We derive an exact analytical solution for the \textit{Uhlmann phase}, $\Phi^s$, of  subsystem $s$, 
by following a procedure that involves the explicit calculation of the evolution operator in a rotating frame \cite{Bohm}. The procedure yields the following Uhlmann phase of the subsystems $A$ and $B$,
\begin{equation}
    \Phi^s(\theta,g,T)={\rm Arg}\left\{-\cos(\pi r_s)-i\, \left[\bar{\gamma}^s-\pi\right]\, \frac{\sin(\pi r_s)}{\pi\,r_s} \right\},
\label{analiticalUhlmann}
\end{equation}
with $r_s=r_s(\theta,g,T)$  defined as,
\begin{equation}
    r_s(\theta,g,T)= \left(1-\gamma^{s,1}\, \gamma^{s,2}\,\left[1-4\,{\rm det}[\rho^s]\right]\,/\pi^2 \right)^{1/2},
\label{lars}
\end{equation}
which is written in terms of the Berry phases $\gamma^{s,l}$ of the eigenstates of the subsystem $s$ 
\begin{equation}
    \gamma^{s,l}(\theta,g,T)=\int_0^{2\pi}\, d\phi \braket{v_{s,l}|i\partial_{\phi}v_{s,l}}=2\pi\, \left( \beta^2_{s,l}/N_{s,l}\right).
\label{BerrySubsystems2}    
\end{equation}
%
%
The result (\ref{analiticalUhlmann}) involves also the composed phase $\bar{\gamma}^s=\sum_{l=1}^{2} p_{s,l}\,\gamma^{s,l}$, for which 
it is verified that $\bar{\gamma}^A+\bar{\gamma}^B-2\pi= \bar{\gamma}^{AB}$, where $\bar{\gamma}^{AB}=\sum_{j=1}^{4} p_{j}\,\gamma_j$, as defined in Ref.~\onlinecite{yiprl04}. Although the latter is not the appropriate phase for mixed states, we note that it occurs naturally in the Uhlmann phase (\ref{analiticalUhlmann}).

\begin{figure}[htbp] 
    \begin{center}
        \subfloat{\includegraphics[width=4.5cm]{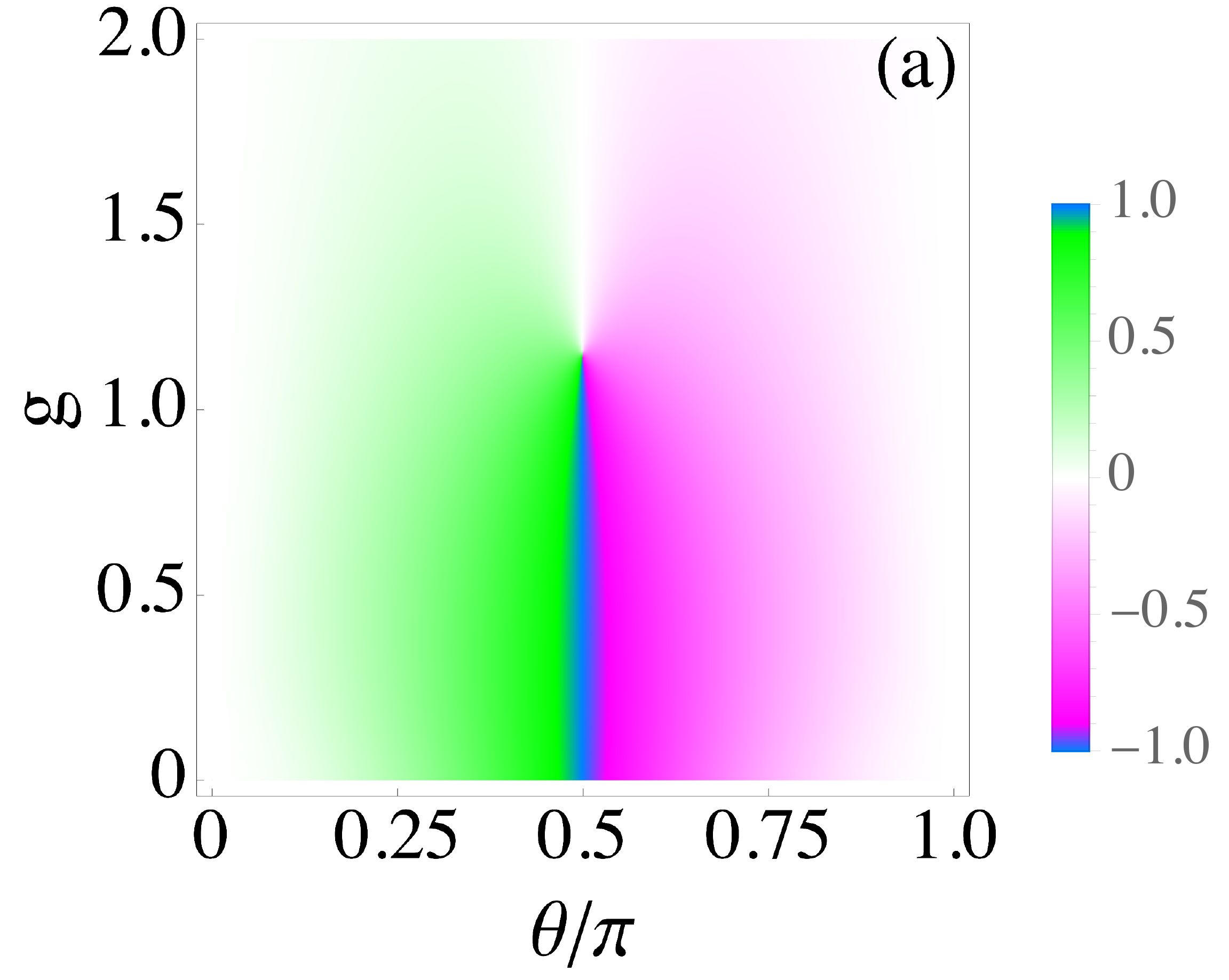}}
        \subfloat{\includegraphics[width=4.5cm]{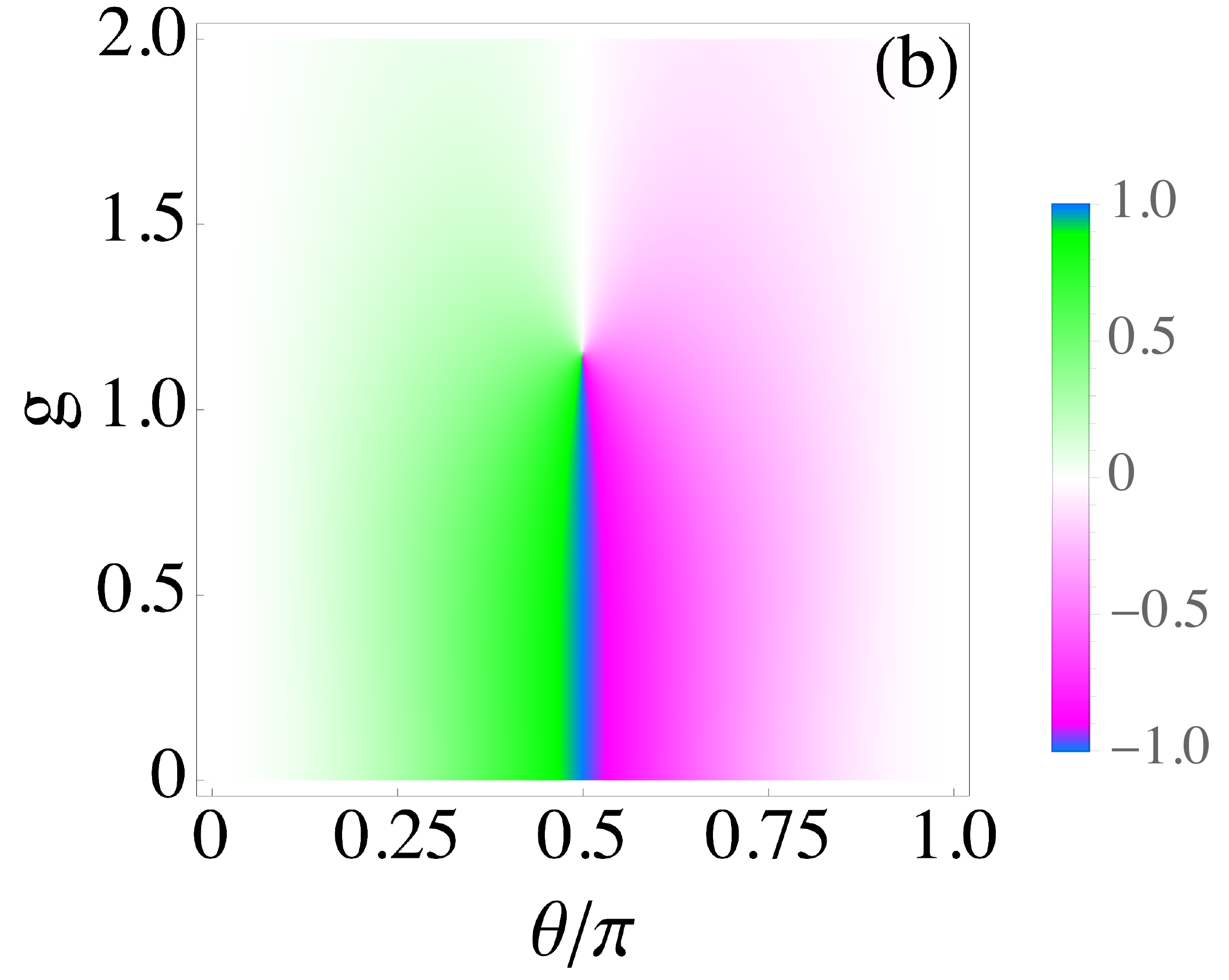}}
        \hspace{0.25cm}
        \subfloat{\includegraphics[width=4.5cm]{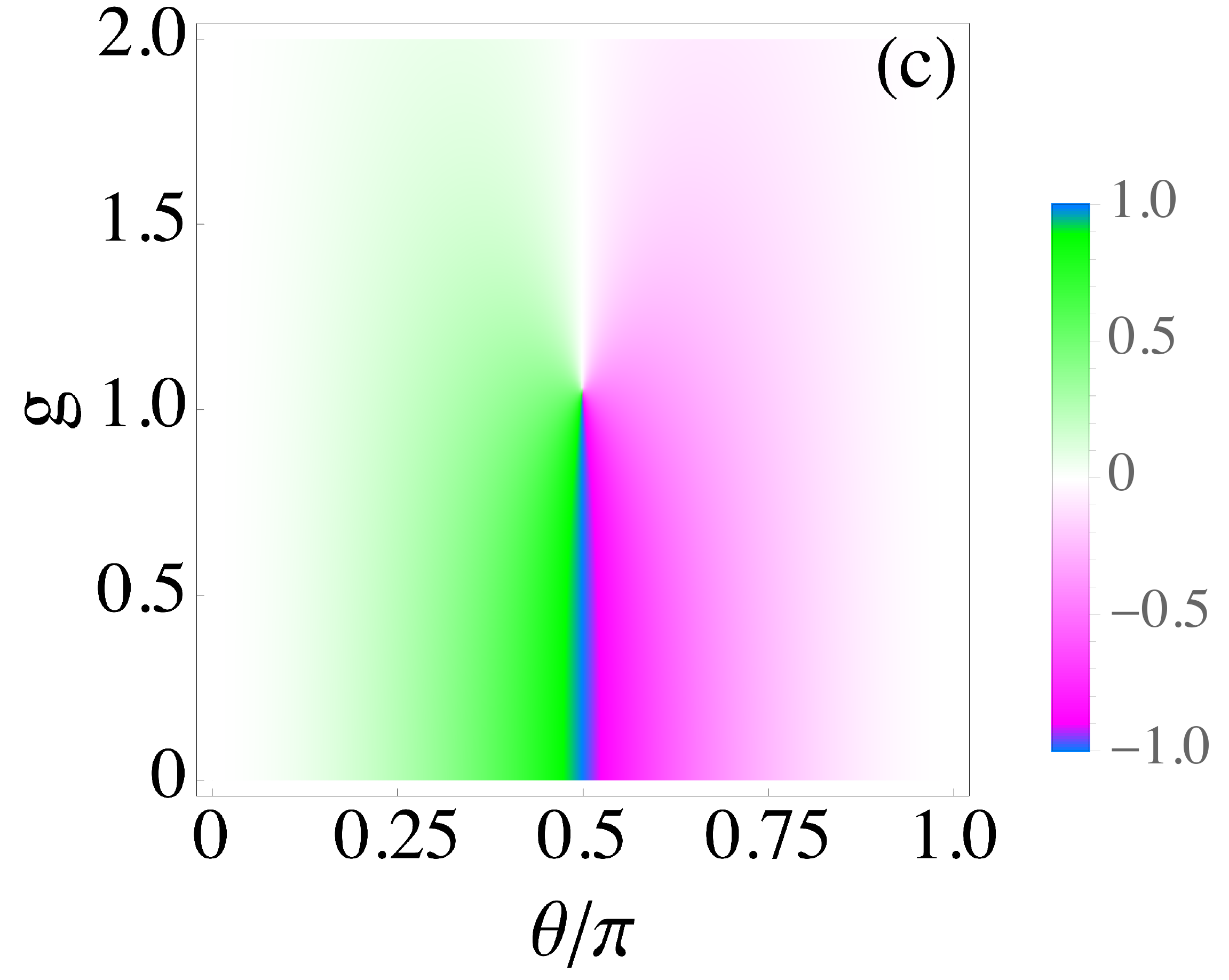}}
        \subfloat{\includegraphics[width=4.5cm]{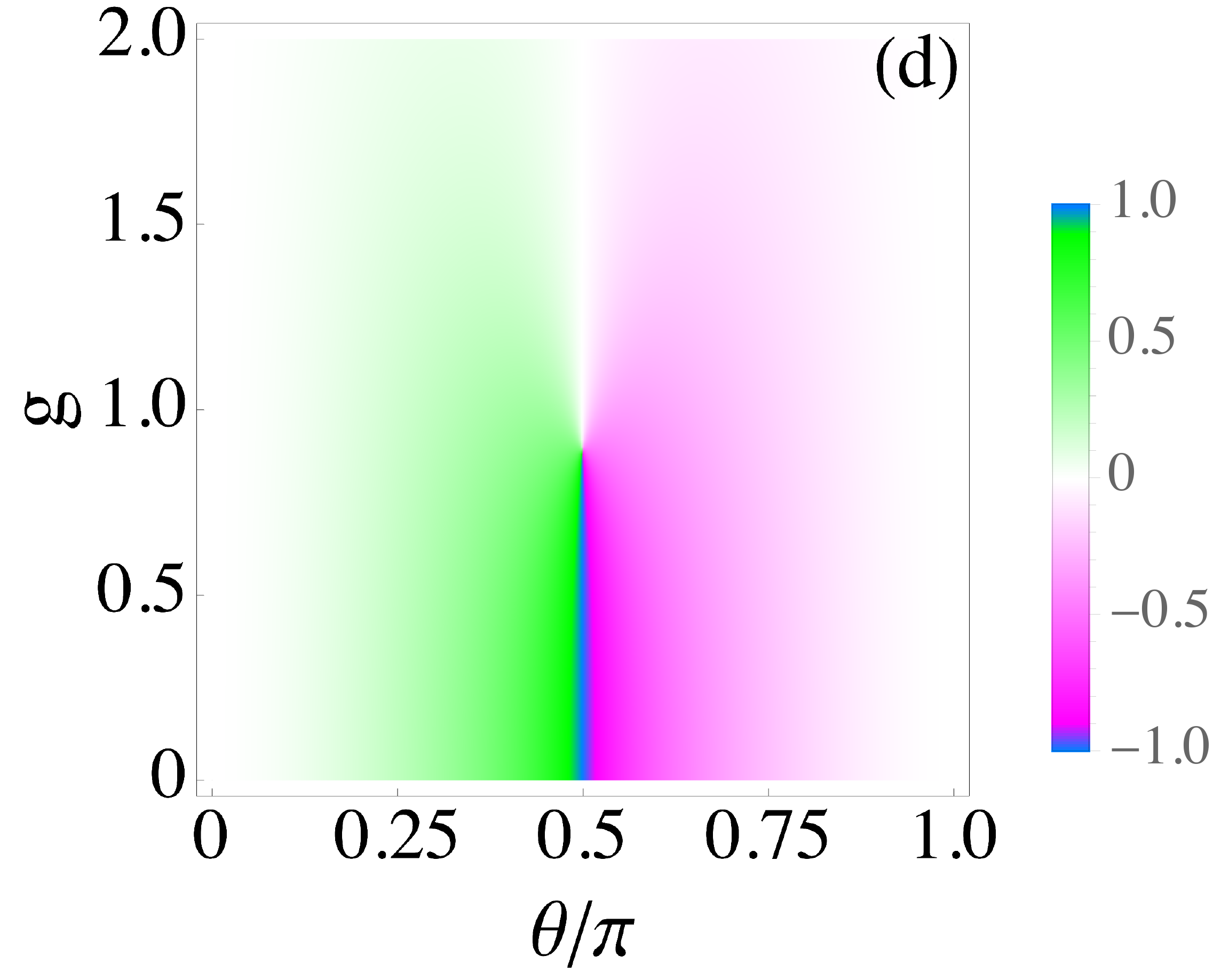}}
        \hspace{0.25cm}
        \subfloat{\includegraphics[width=4.5cm]{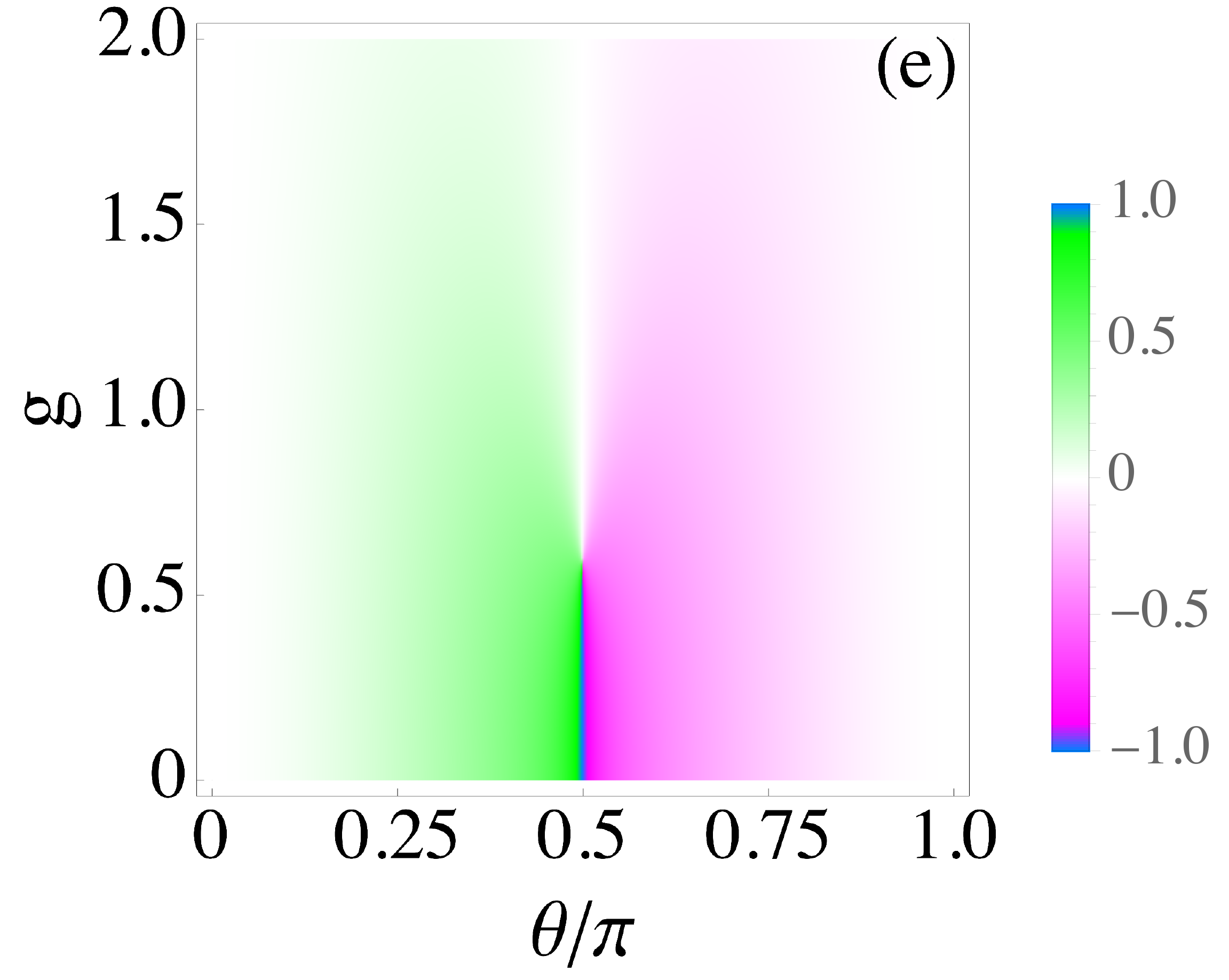}}
        \subfloat{\includegraphics[width=4.5cm]{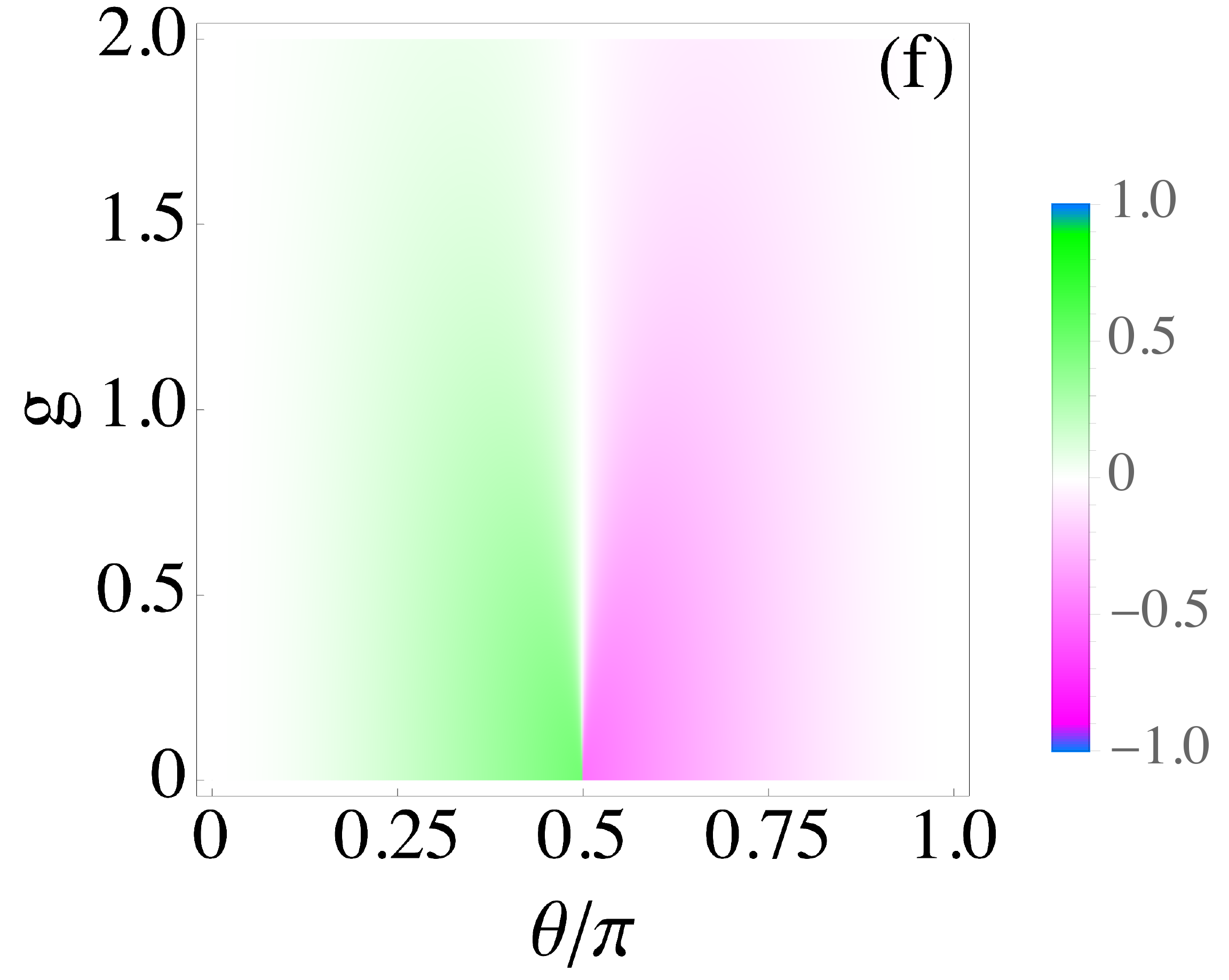}}
    \end{center}
    \caption{Color density maps of the Uhlmann phase for the subsystem $A$, $\Phi^A$ [Eq.~(\ref{analiticalUhlmann})] as a function of the coupling parameter $g$, and $\theta$, for different values of the temperature: (a) $T=0.02$, (b) $T=0.2$, (c) $T=0.5$, (d) $T=0.6$, (e) $T=0.7$, and (f) $T=T_c$. In all cases, we emphasize the presence of a vortex profile along $\theta=\pi/2$, occurring at a critical value of $g$. The vortex disappears at a critical temperature $T_c$.}
    \label{colormapUhlsubA}
\end{figure}

\begin{figure}[htbp] 
    \begin{center}
        \subfloat{\includegraphics[width=4.5cm]{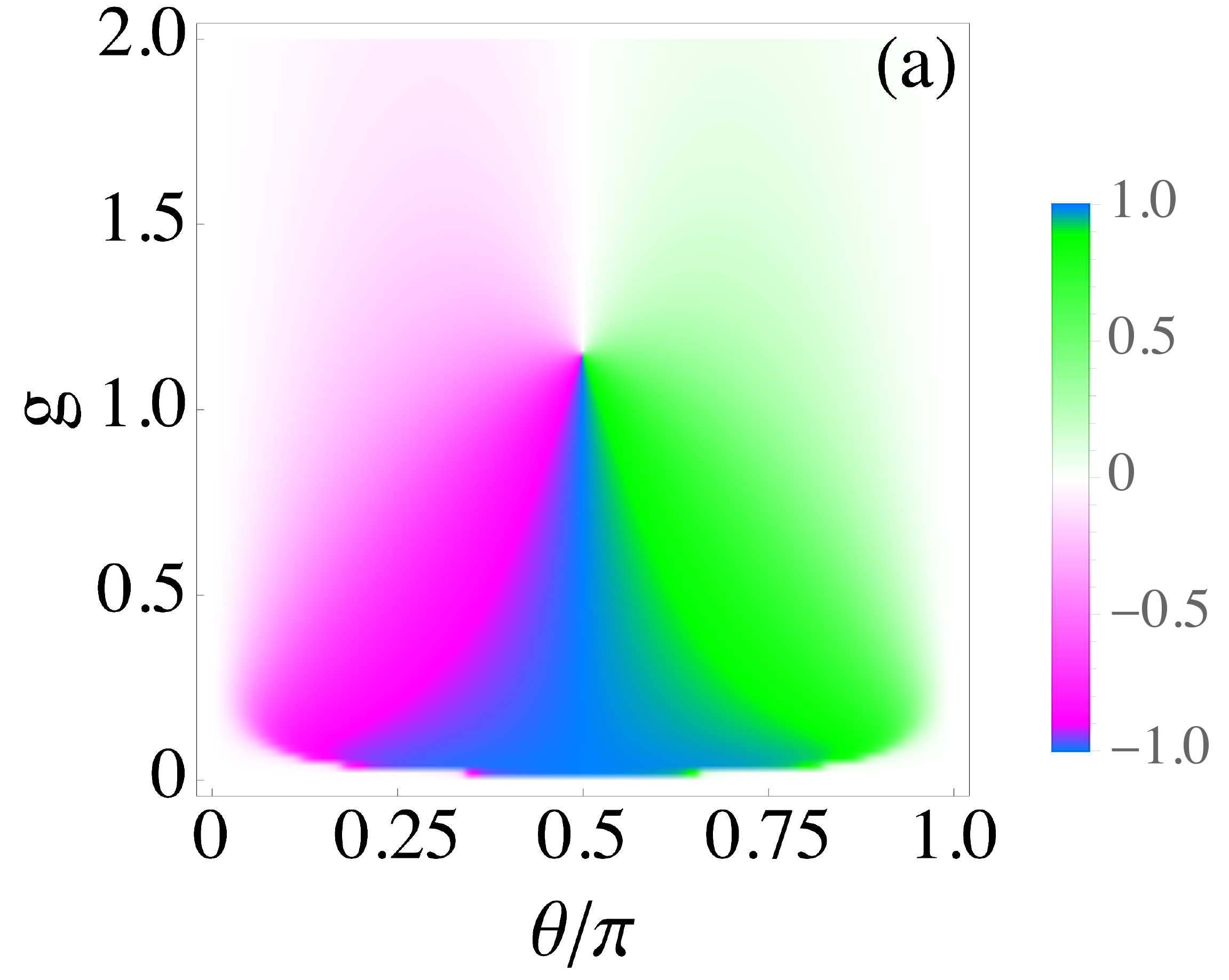}}
        \subfloat{\includegraphics[width=4.5cm]{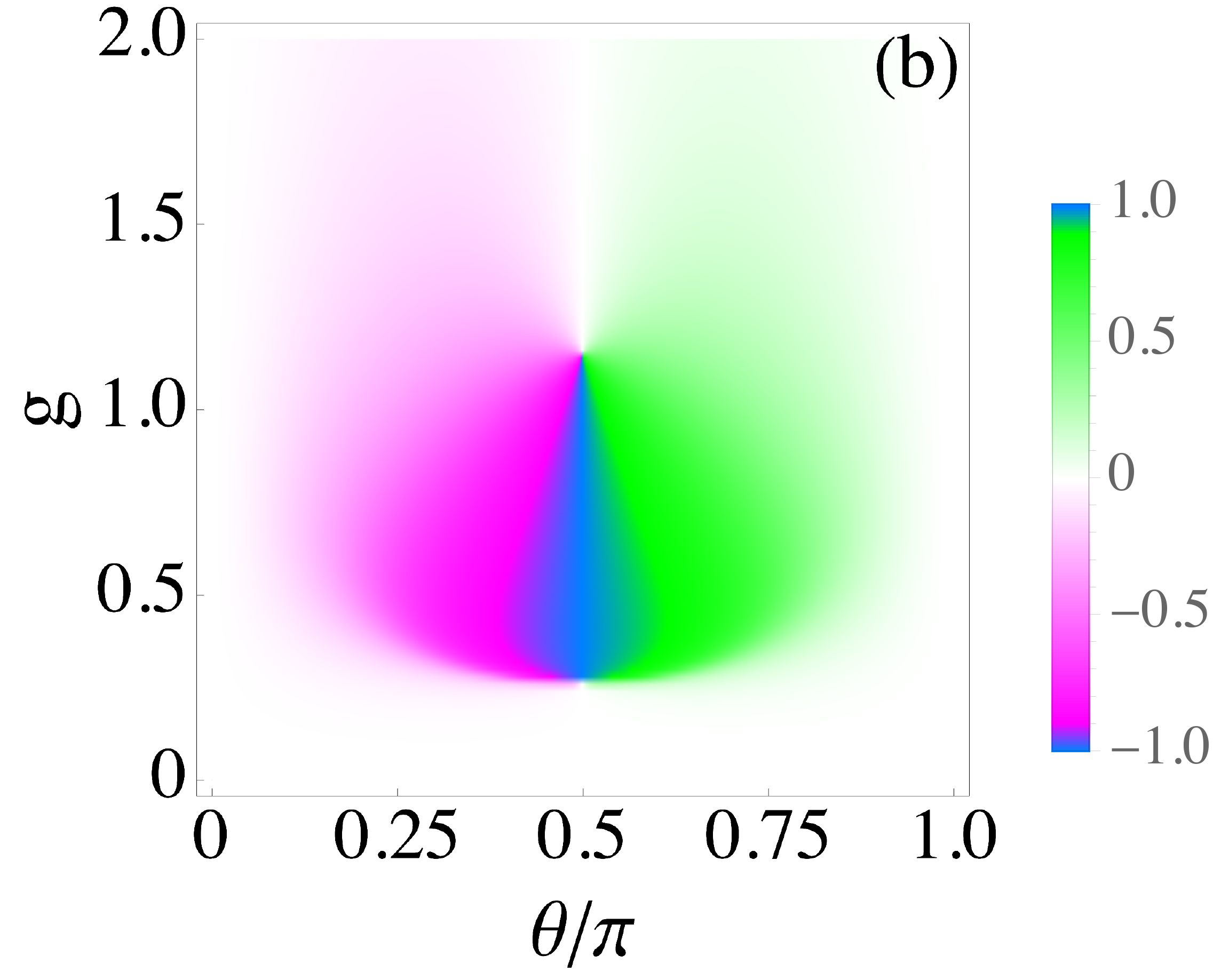}}
        \hspace{0.25cm}
        \subfloat{\includegraphics[width=4.5cm]{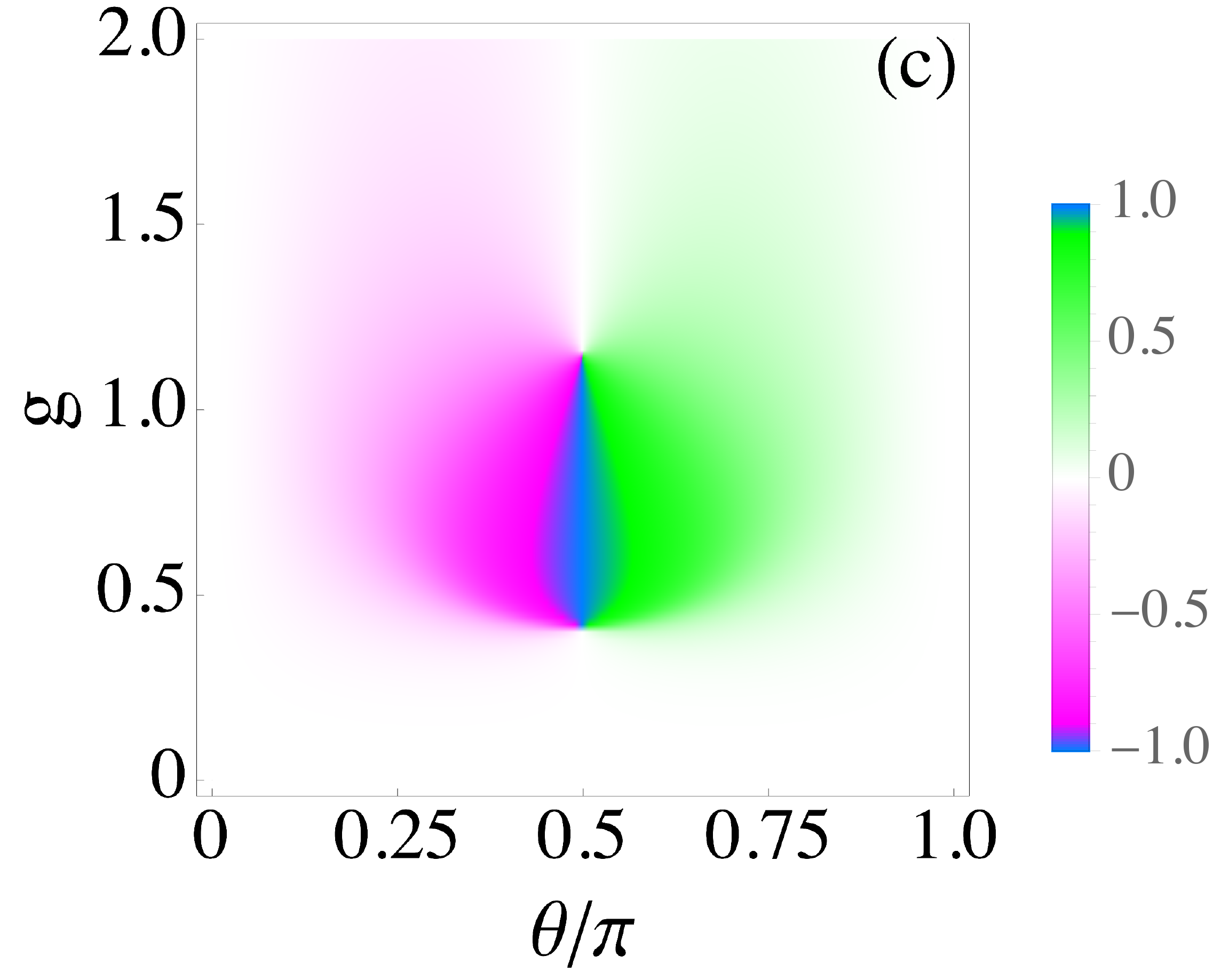}}
        \subfloat{\includegraphics[width=4.5cm]{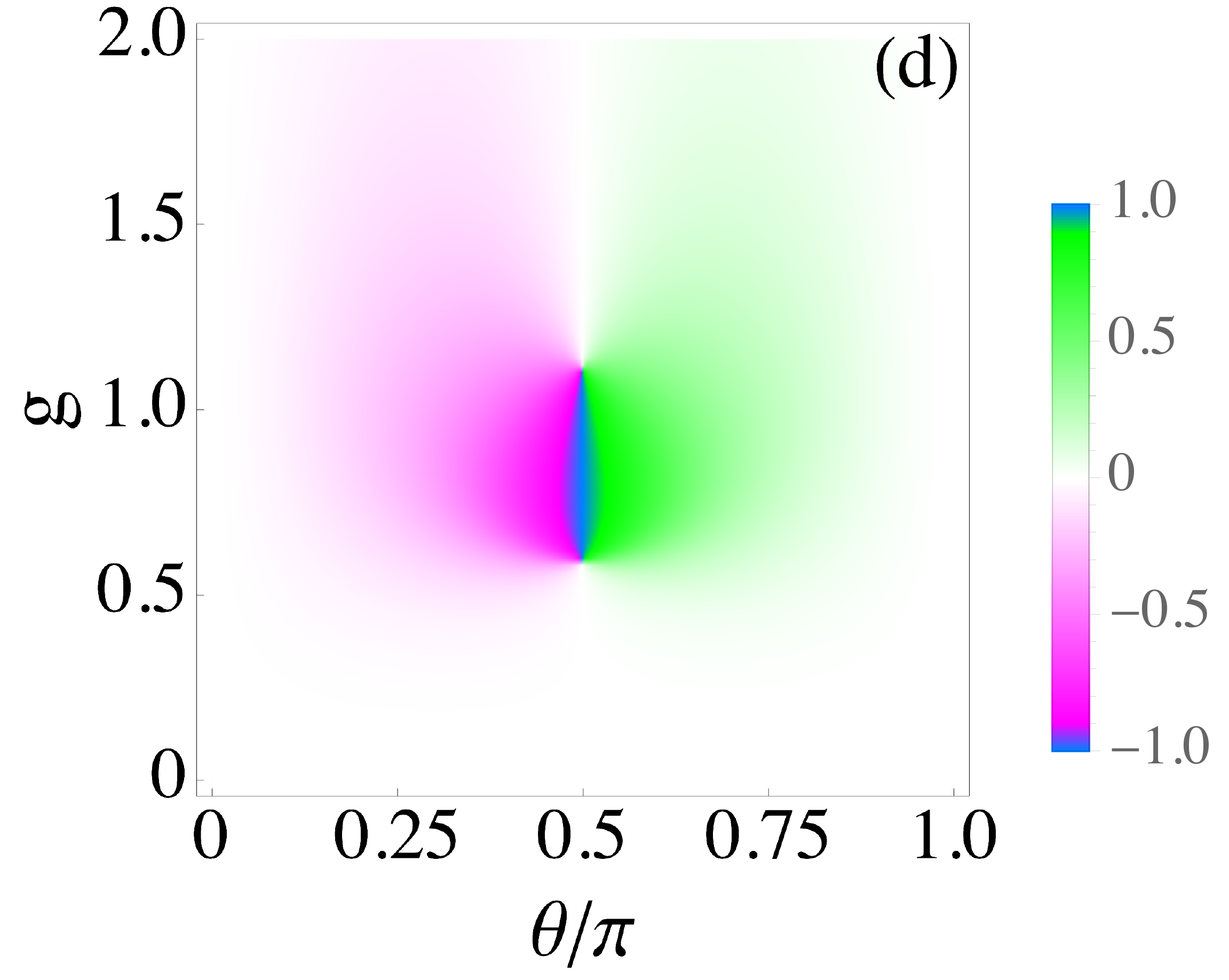}}
        \hspace{0.25cm}
        \subfloat{\includegraphics[width=4.5cm]{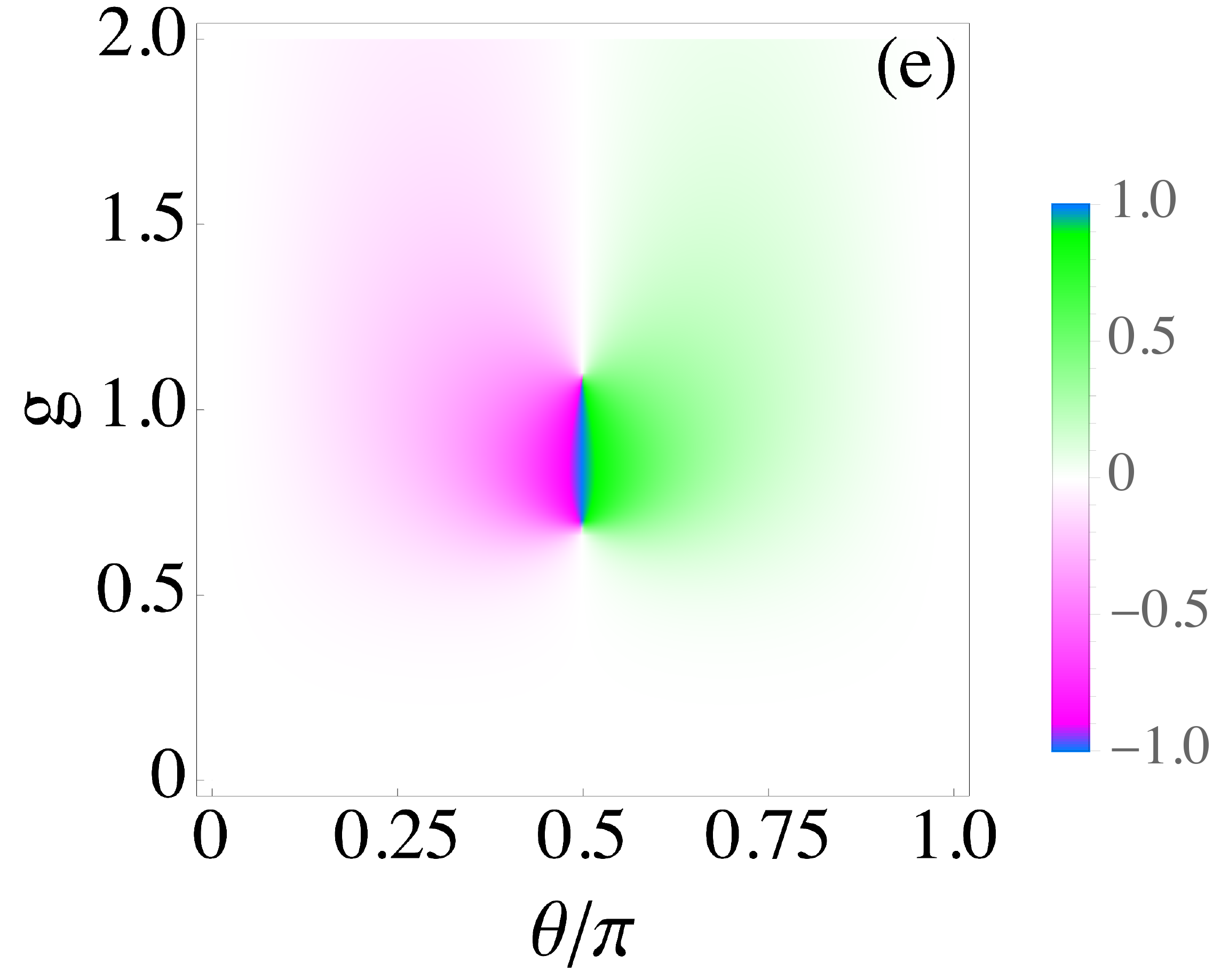}}
        \subfloat{\includegraphics[width=4.5cm]{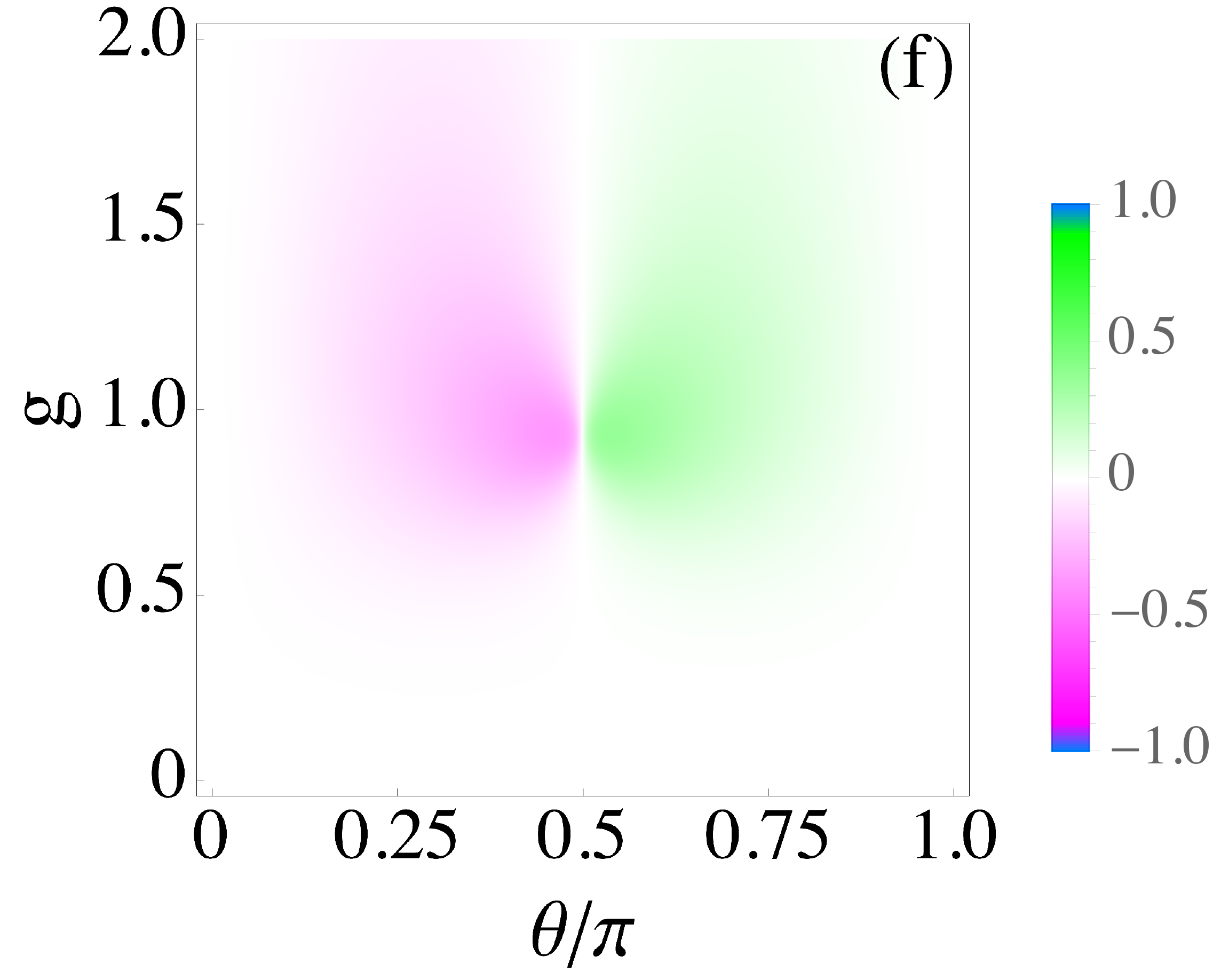}}
    \end{center}
    \caption{Color density maps of the Uhlmann phase for subsystems $B$, $\Phi^B$ [Eq.~(\ref{analiticalUhlmann})], as a function of the coupling parameter $g$, and $\theta$, for different values of the temperature: (a) $T=0.01$, (b) $T=0.1$, (c) $T=0.15$, (d) $T=0.2$, (e) $T=0.22$, and (f) $T=0.25$. In all cases, we emphasize the presence of two vortices along $\theta=\pi/2$, occurring at two critical values of $g$. }
    \label{colormapUhlsubB}
\end{figure}

We explore the Uhlmann phase [Eq.~(\ref{analiticalUhlmann})] for the subsystems $A$ and $B$ 
to show its dependence on the coupling, $g$, in all directions of the field for increasing temperature values, $T$. In Fig.~\ref{colormapUhlsubA}, we present color density maps of  $\Phi^A$   [Eq.~(\ref{analiticalUhlmann})], where we show that the Uhlmann phase exhibits a vortex at $\theta=\pi/2$.
The position of the vortex observed in Figs.~\ref{colormapUhlsubA}(a)-(b) appears to be fixed at a particular value of $g$ in the regime of small temperature values. However, in the sequence of Figs.~\ref{colormapUhlsubA}(c)-(e), we show that as the temperature increases, 
the vortex occurs for smaller values of $g$. In Fig.~\ref{colormapUhlsubA}(f), we show that the vortex disappears once we reach the critical temperature $T_c$.
In Fig.~\ref{colormapUhlsubB}, we present color density maps of $\Phi^B$  [Eq.~(\ref{analiticalUhlmann})], where we show that the behavior of the vortices as a function of temperature  is more dramatic than in system $A$. While in the latter, we have a single vortex whose position in $g$ decreases as the temperature increases, in $B$, we have completely different behavior: the appearance of two vortices that define two critical values of the coupling $g$ for the same temperature. 
In our recent study regarding the effects of a depolarizing channel in two-coupled fermions  Ref.~\onlinecite{pravillaeta21}, we observed no such behavior in the subsystems.
%
%
%
%
\begin{figure}[htbp] 
    \begin{center}
        \subfloat{\includegraphics[width=4.5cm]{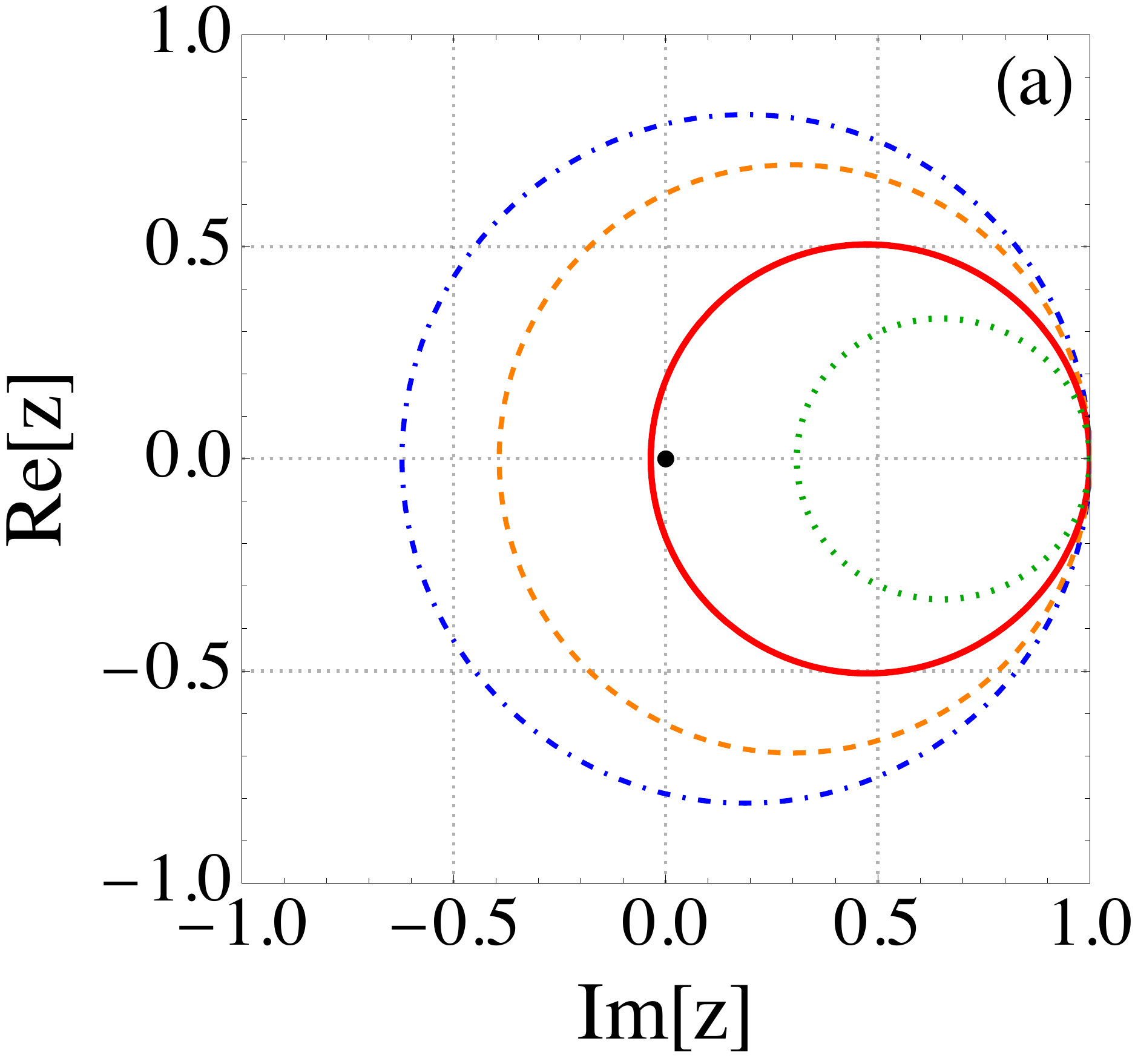}}
        \subfloat{\includegraphics[width=4.5cm]{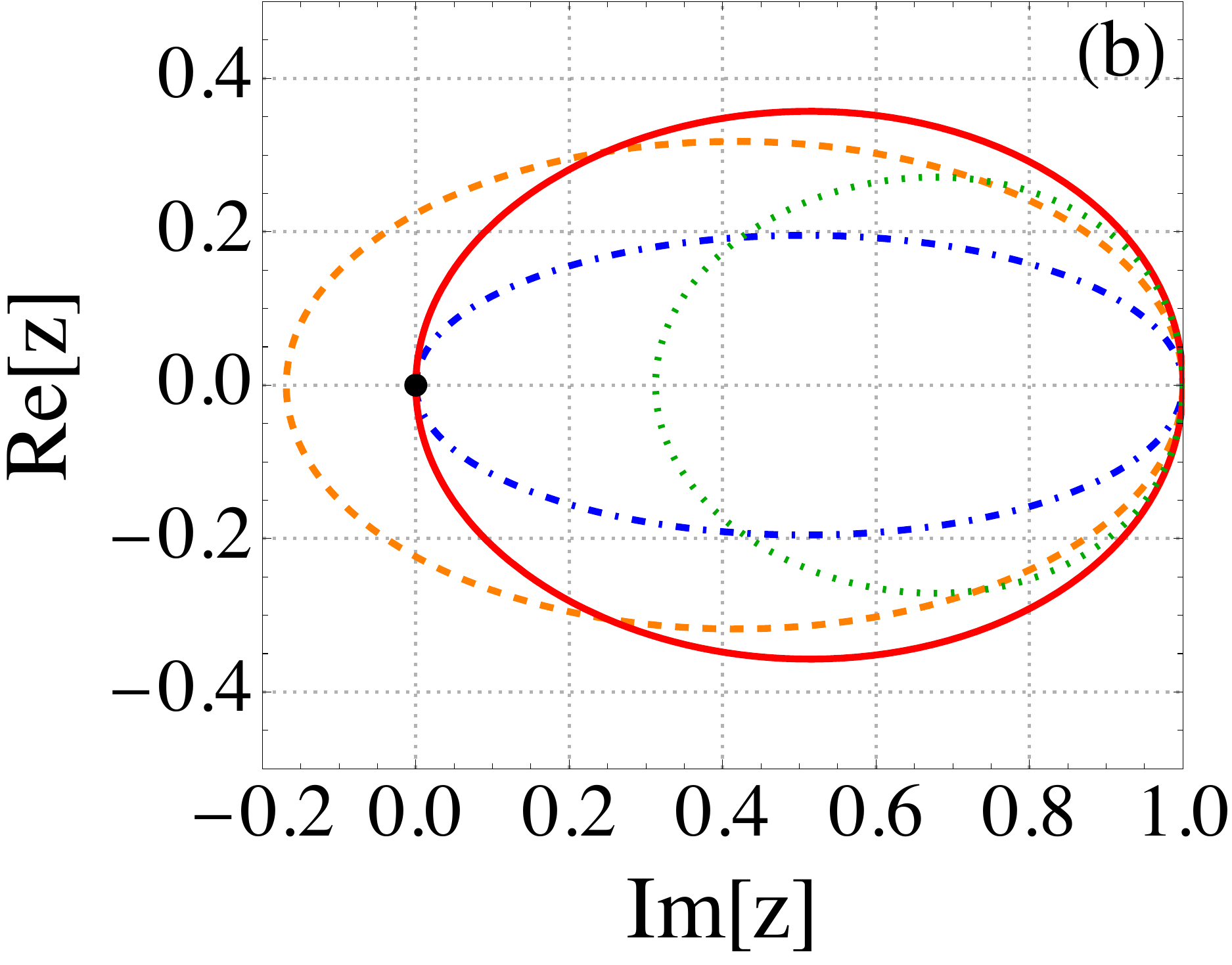}}
    \end{center}
    \caption[]{Argand diagrams for (a) $z^A(\theta)$ and (b) $z^B(\theta)$, at several values of the coupling strength: $g=0.597$ (blue dashed dotted line), $g=0.8$ (orange dashed line), $g=1.12$, (red solid line), and $g=1.5$ (green dotted line). We have chosen T=0.2 in the calculation. In case (b) there is a double change in the winding number  occurring at the same value of $T$. }
\label{Argand}
\end{figure}
The phase change observed in Figs.~\ref{colormapUhlsubA} and \ref{colormapUhlsubB} can be characterized by a change of a winding number.
To perform this task, we write the Uhlmann phase (\ref{analiticalUhlmann}) in the form
$\Phi^s={\rm Arg}\{-U_1(z^s(\theta))\}$ where  $U_1(z)=2z$ is the second-kind Chebyshev polynomial of order one, with argument $z^s(\theta)=\{\cos(\pi r_s)+i\, [(\bar{\gamma}^s-\pi)] \sin(\pi r_s)/\pi r_s\}/2$.
In Fig.~\ref{Argand} we  plot the curve $z^s(\theta)$ for several values of the coupling strength $g$.
%
In Fig.~\ref{Argand}(b), we show that in the system $B$, the parametric curve cross the zero twice, corresponding to a double phase transition of $\Phi^{B}$. The latter is not the case for system $A$ [Fig.~\ref{Argand}(a)], where we observe only one crossing of the zero, which corresponds to a single phase transition in the Uhlmann phase.   

We can gain more insight into the observed behavior of the Uhlmann phase by using the Bloch representation of the density matrices, $\rho^s$.
The latter can be written as $\rho^s=\frac{1}{2}( \mathbb{1}+\boldsymbol{n}_s\cdot \boldsymbol{\sigma})$, where $\boldsymbol{n}_s=(2 c_s \operatorname{cos}\phi, 2 c_s \operatorname{sin}\phi,2 a_s-1)$. 
For the case $\theta=\pi/2$, we have $a_s=1/2$, thus $\boldsymbol{n}_s=2 c_s ( \operatorname{cos}\phi, \operatorname{sin}\phi,0)$, which describes a circumference in the $n_x n_y$ plane, of radius $R_s=|2 c_s|$. 
In what follows, we show that the zeros of the Uhlmann phase  ($\Phi^s=0$) correspond to a critical value of $R_s$. 
By using the fact that $\bar{\gamma}^s=\pi$, the zeros are  calculated from $\Phi^s(\pi/2,g,T)={\rm Arg}\{-\cos[\pi r_s(\pi/2,g,T)]\}$, and these correspond to $r_s(\pi/2,g,T)$=1/2.
We evaluate $r_s(\pi/2,g,T)$ from  Eq.~(\ref{lars}) by substituting   $\gamma^{s,1}=\gamma^{s,2}=\pi$, which leads us to $r_s(\pi/2,g,T)=2\,{\rm det}[\rho^s]^{1/2}$. 
We evaluate ${\rm det}[\rho^s]$ from Eq.~(\ref{bastardnotation}) to obtain the following result: $r_s(\pi/2,g,T)=(1-R_s^2)^{1/2}=1/2$. That is, the condition $\Phi^s=0$ corresponds to a critical radius $R_{c,s}=\sqrt{3}/2$, which is the same value as reported by Viyuela \textit{et al.} in a quantum simulator model based on superconducting qubits \cite{viyuelanpj18}.
%
%
The above allows us to calculate the critical values of $(g,T)$ in each subsystem for which $R_s=R_{c,s}$. 
%

\begin{figure}[H]
    \begin{center}
        \includegraphics[width=6.cm]{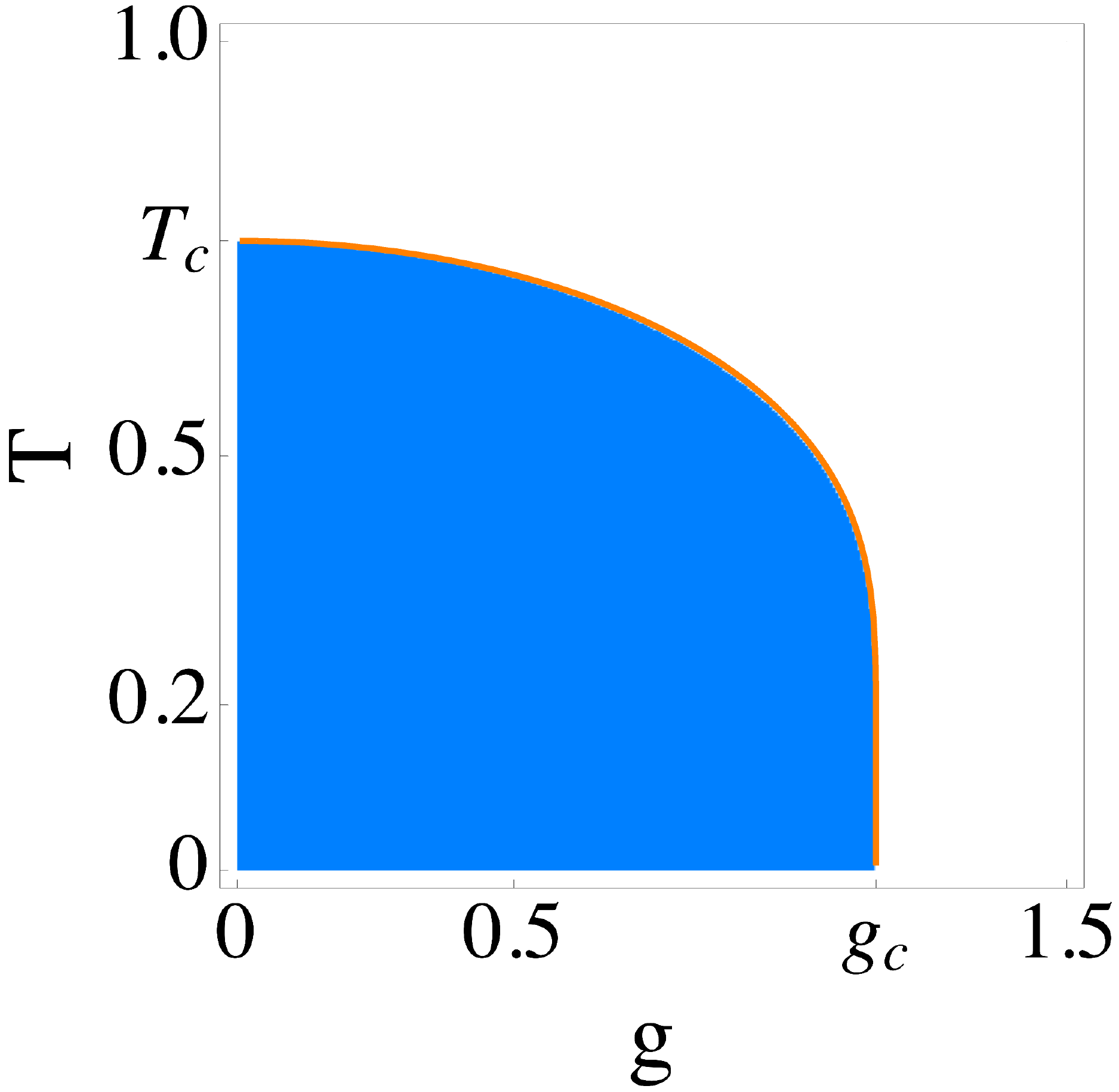}
    \end{center}
    \caption{Color density map for the Uhlmann phase $\Phi^{A}$ as a function of $g$ and $T$ at a fixed direction $\theta=\pi/2$. The position of the vortex observed in Fig.~\ref{colormapUhlsubA} , for subsystem $A$, corresponds to the set of points $(g,T)$ which define the Uhlmann phase boundary  where the phase changes abruptly from $0$ to $\pi$ (blue). We also include the roots of $r_A(g,\pi/2,T)-1/2=0$  (orange solid line) corresponding to the zeros of $\Phi^A$.}
\label{RootsSubsA}
\end{figure}

We analyze the critical values of temperature and coupling for subsystems $A$ and $B$, where the values that meet the critical radius condition are given and determine the positions of the vortices observed in Figs.~\ref{colormapUhlsubA} and \ref{colormapUhlsubB}.
In Fig.~\ref{RootsSubsA}, we show that for subsystem $A$ each temperature value corresponds to a single value of $g$, as long as the temperature does not exceed the critical value of 
$T_c$,
which occurs for minimal values of $g$. 
Interestingly, this critical temperature $T_c$ has been observed by Viyuela \textit{et al.} \cite{viyuelaprl14} in different one-dimensional fermionic models in crystal momentum $\boldsymbol{k}$-space, where the Uhlmann phase goes discontinuously and abruptly to zero when $T=T_c$.
In our case, for temperatures $T \geq T_c$, it is impossible to observe vortices, \textit{i.e.}, there are no phase transitions.
We also observe that in subsystem $A$, a critical value of $g$ is given by $g\equiv g_c=2/\sqrt{3}$, where for temperatures $T<0.2$, the vortex position remains almost fixed around this coupling value.

\begin{figure}[H]
    \begin{center}
        \includegraphics[width=6.cm]{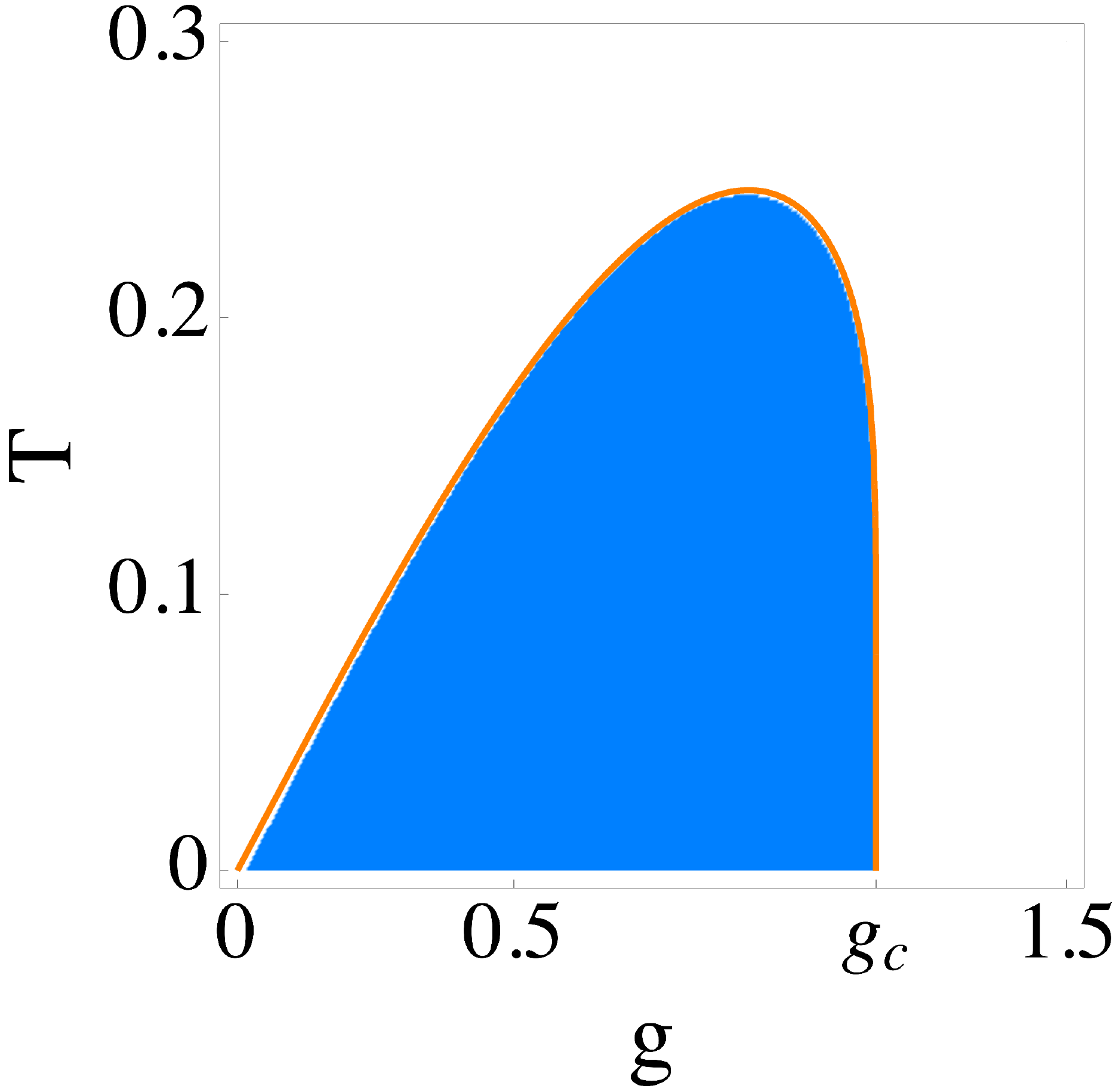}
    \end{center}
    \caption{Color density map for the Uhlmann phase $\Phi^{B}$ as a function of $g$ and $T$ at a fixed direction $\theta=\pi/2$. The position of the vortex observed in Fig.~\ref{colormapUhlsubB}, for subsystem $B$, corresponds to the set of points $(g,T)$ which define the Uhlmann phase boundary where the phase changes abruptly from $0$ to $\pi$ (blue). We also include the roots of $r_B(g,\pi/2,T)-1/2=0$  (orange solid line) corresponding to the zeros of $\Phi^B$. Notice that there is a double zero occurring at the same value of $T$.}
    \label{RootsSubsB}
\end{figure}

In Fig.~\ref{RootsSubsB} we present the case of subsystem $B$, and we observe a completely different behavior from $A$'s. In this case, we show that each temperature value corresponds to two values of $g$, as long as the temperature does not exceed the critical value given by the curve's maximum, which occurs at $(g, T)=(0.94,0.25)$. We also show that in the low-temperature regime, one of the vortices occurs at a minimal $g$. At the same time, we observe that the second vortex remains fixed around the critical value
$g_c$,
consistent with the observed behavior in Fig.~\ref{colormapUhlsubB}.

\begin{figure}[htbp!] 
    \begin{center}
        \subfloat{\includegraphics[width=4.5cm]{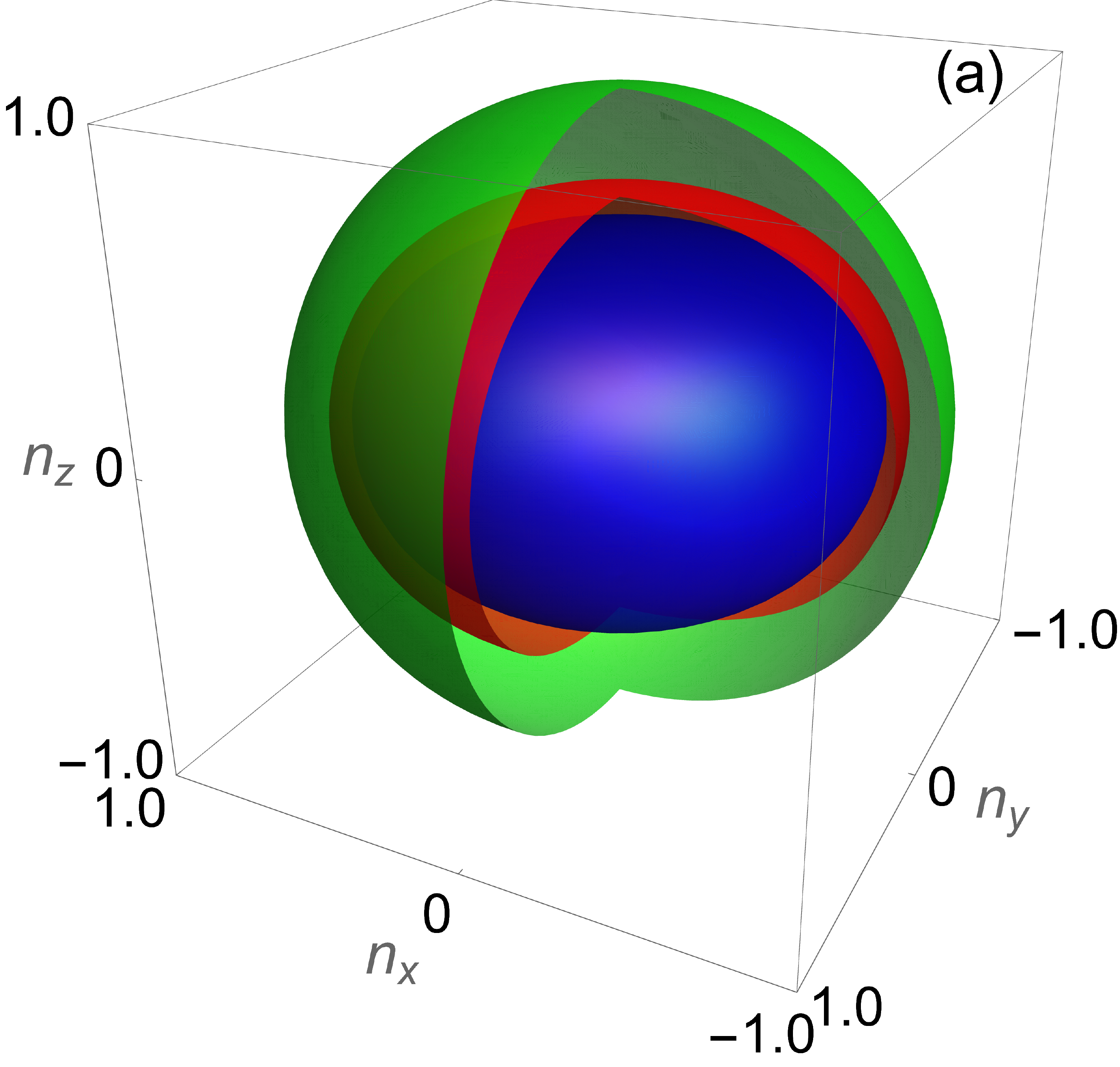}}
        \subfloat{\includegraphics[width=4.5cm]{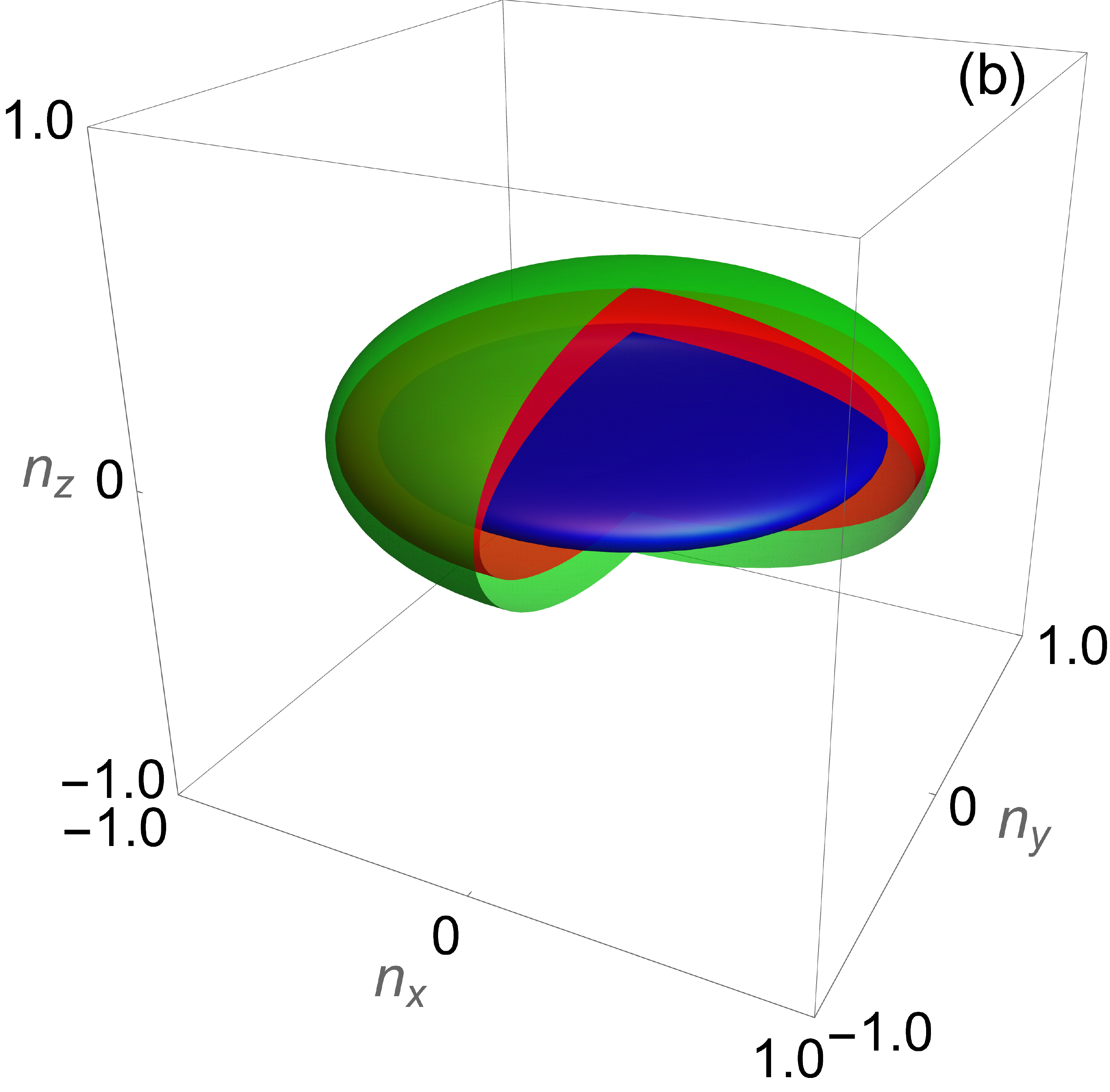}}
        \hspace{0.25cm}
        \subfloat{\includegraphics[width=4.5cm]{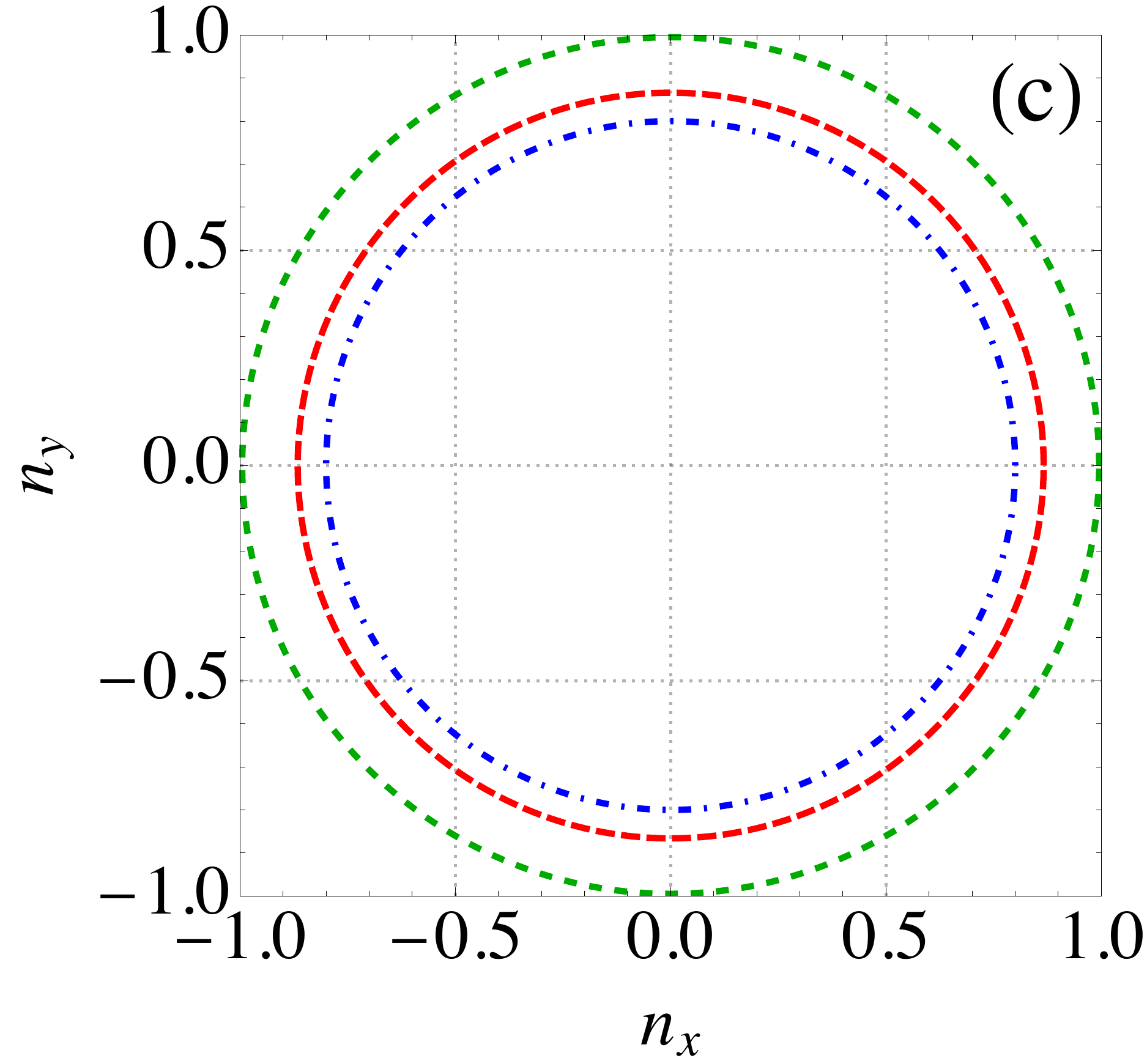}}
        \subfloat{\includegraphics[width=4.5cm]{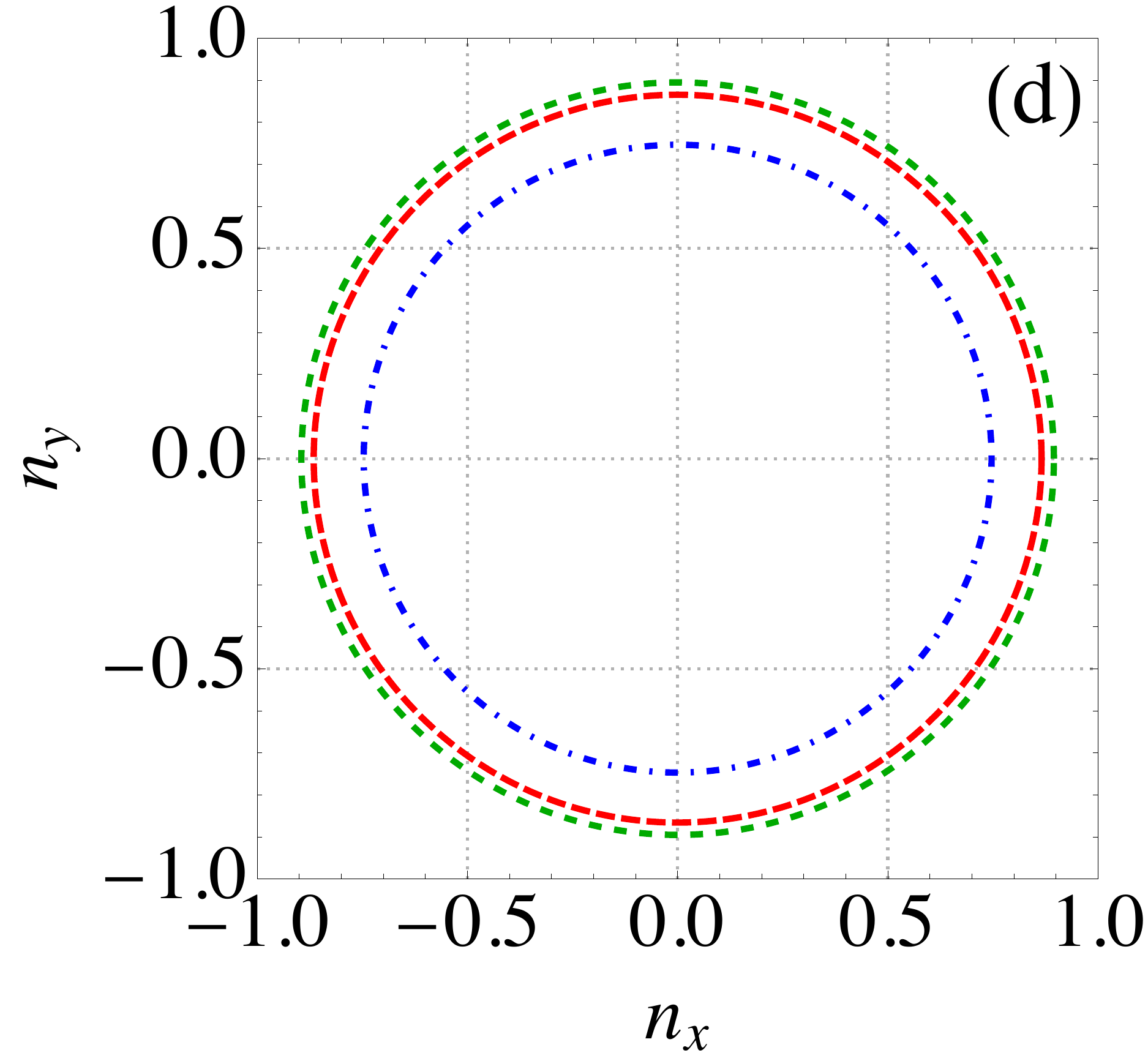}}
    \end{center}
    \caption{(a)-(b) Bloch representation for the subsystems for the cases with $T=0.2$ shown in Figs.~\ref{colormapUhlsubA}(b) and \ref{colormapUhlsubB}(d), respectively. In case (a) for the subsystem, $A$ the ellipsoids correspond to the couplings: $g=0.2$ (green),  $g=1.1546\simeq g_c$ (red) (critical spheroid), and  $g=1.5$ (blue). The radii $R_A$ of the ellipsoids decrease as $g$ increases. In case (b) for subsystem $B$, the ellipsoids correspond to the couplings:  $g=0.4$ (blue), $g=0.596$ (red) (first critical spheroid), and $g=0.8$ (green). The radii $R_B$ of the ellipsoids increase with $g$. We also include the circular cross-sections at  $\theta=\pi/2$ as parametric plots for subsystems (c) $A$, and (d) $B$  showing in both cases the crossing of the critical spheroids (red line).}
    \label{firstcrosing}
\end{figure}

%
%
%
%
In Figs.~\ref{firstcrosing}(a)-(b), we show the Bloch representation for the subsystems, for the fixed temperature $T=0.2$ used in Figs.~\ref{colormapUhlsubA}(b) and \ref{colormapUhlsubB}(d),  for different values of the coupling $g$. 
%
\begin{figure}[htbp!] 
    \begin{center}
        \subfloat{\includegraphics[width=4.5cm]{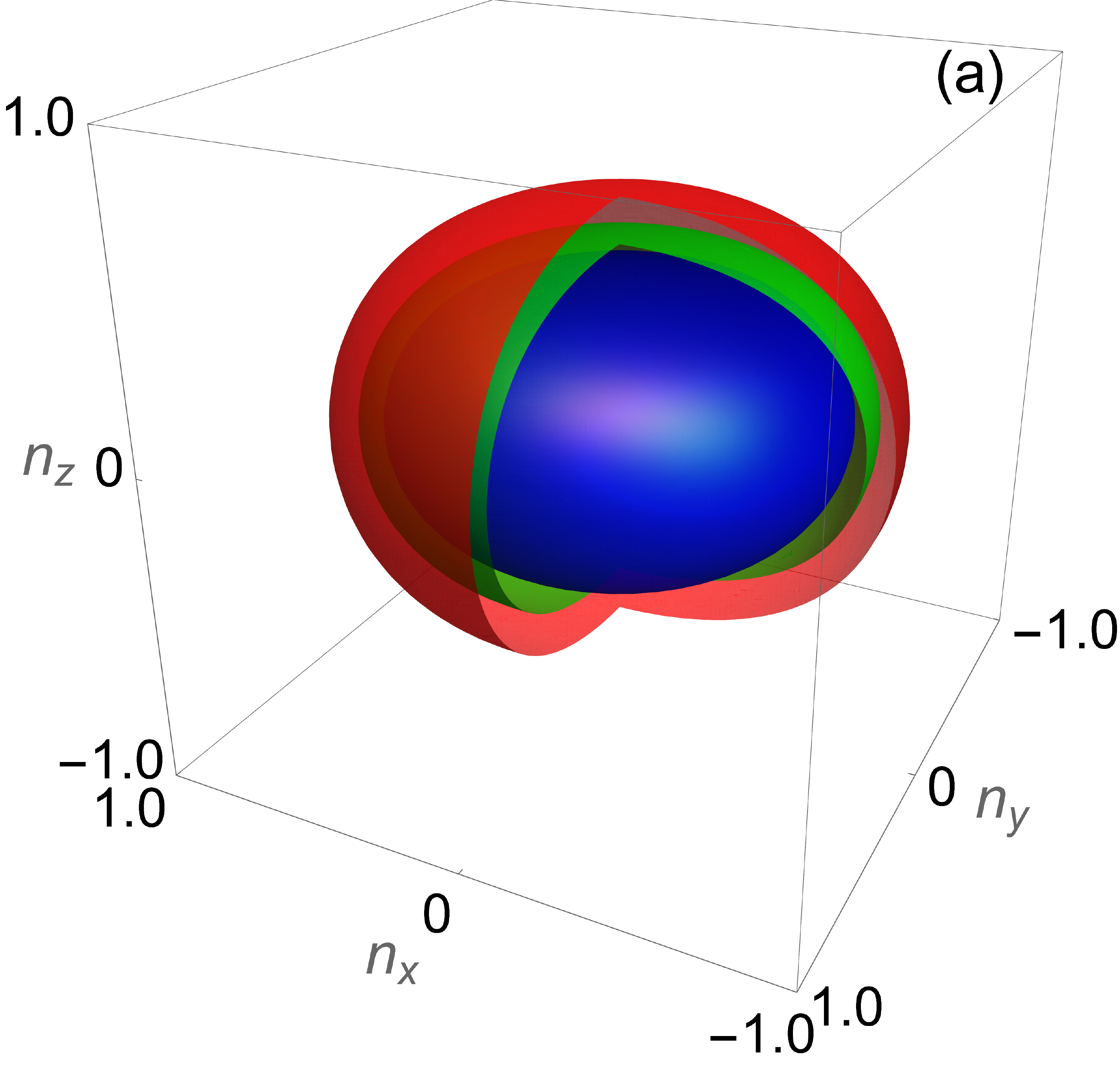}}
        \subfloat{\includegraphics[width=4.5cm]{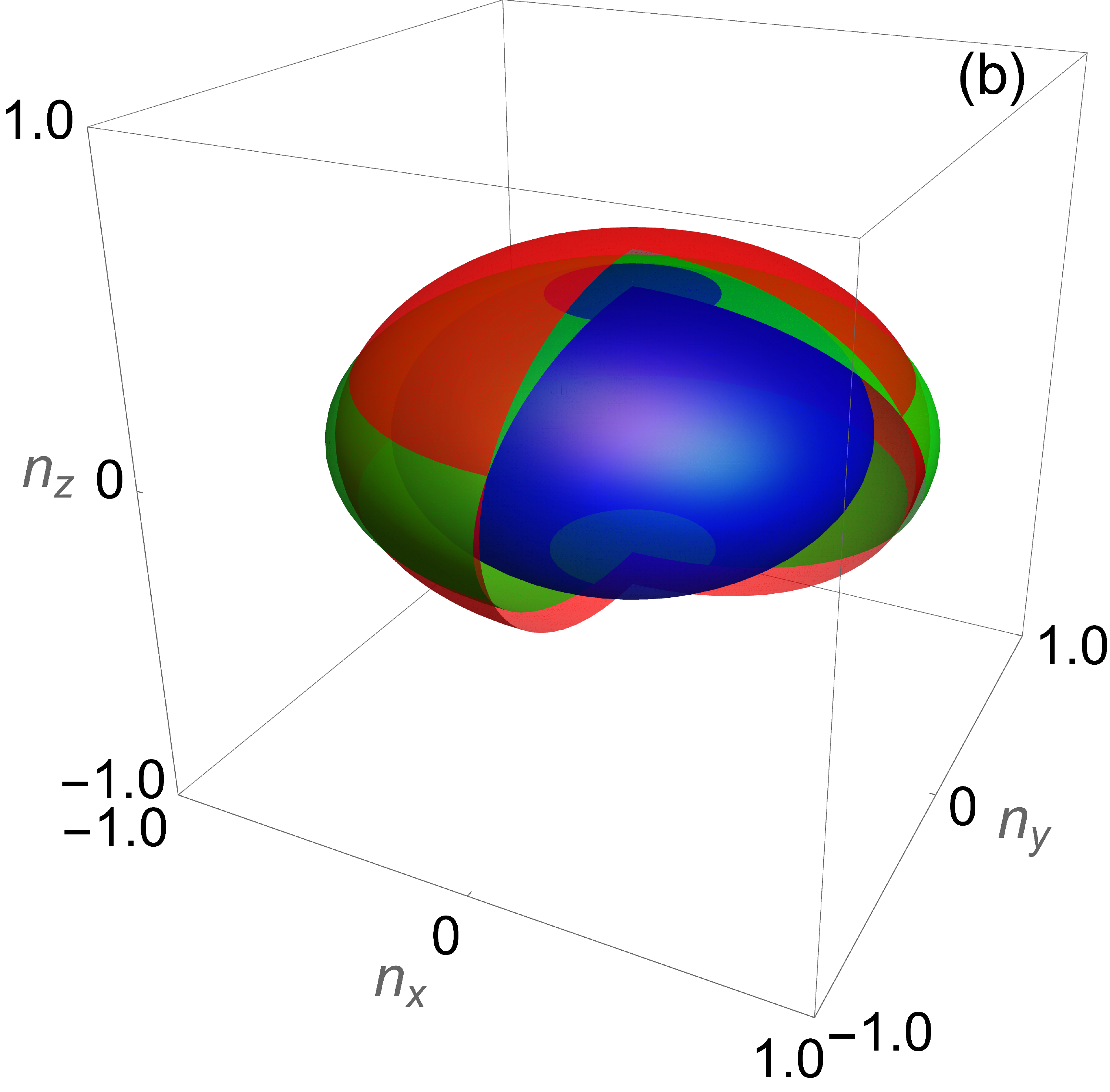}}
        \hspace{0.25cm}
        \subfloat{\includegraphics[width=4.5cm]{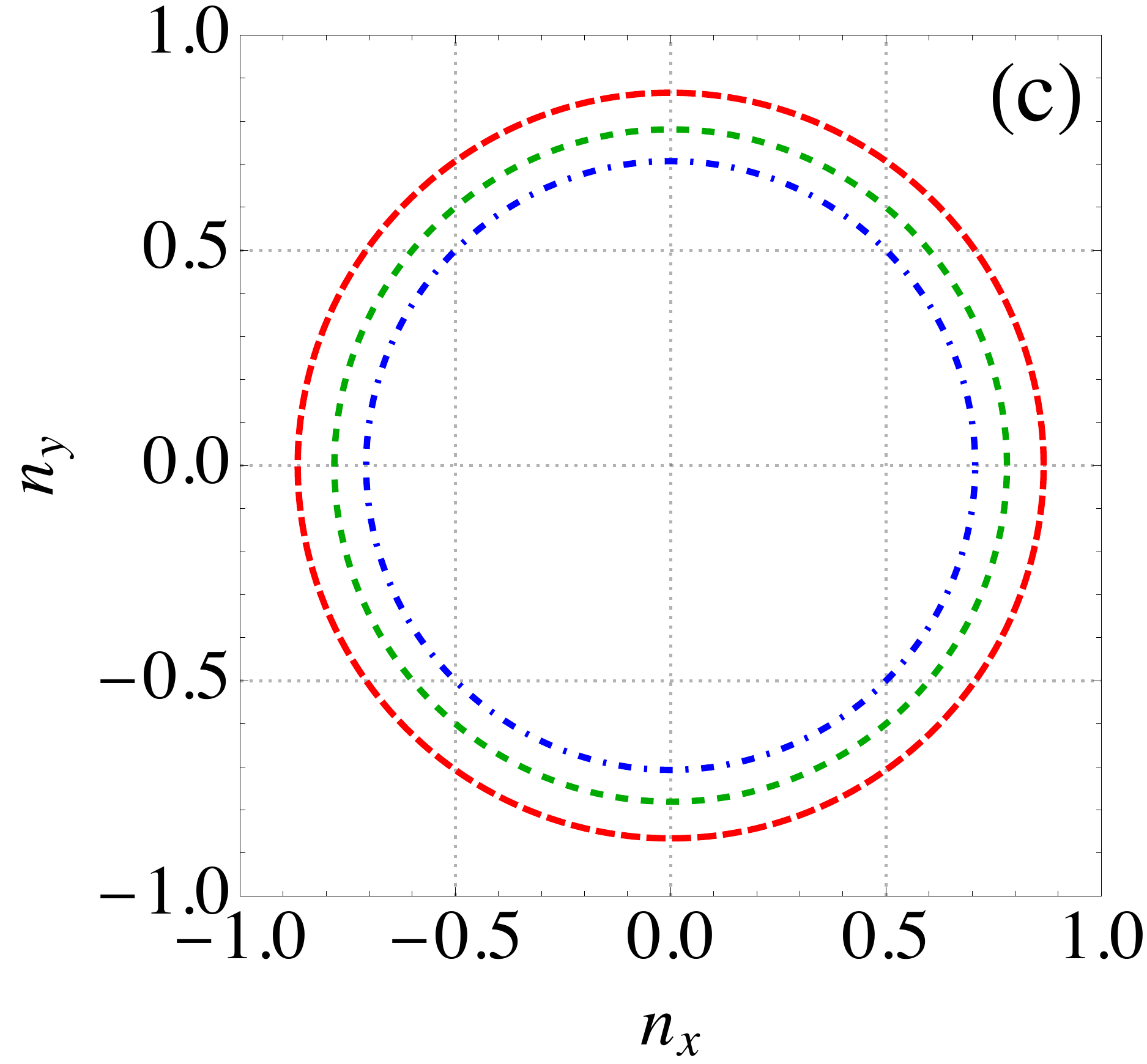}}
        \subfloat{\includegraphics[width=4.5cm]{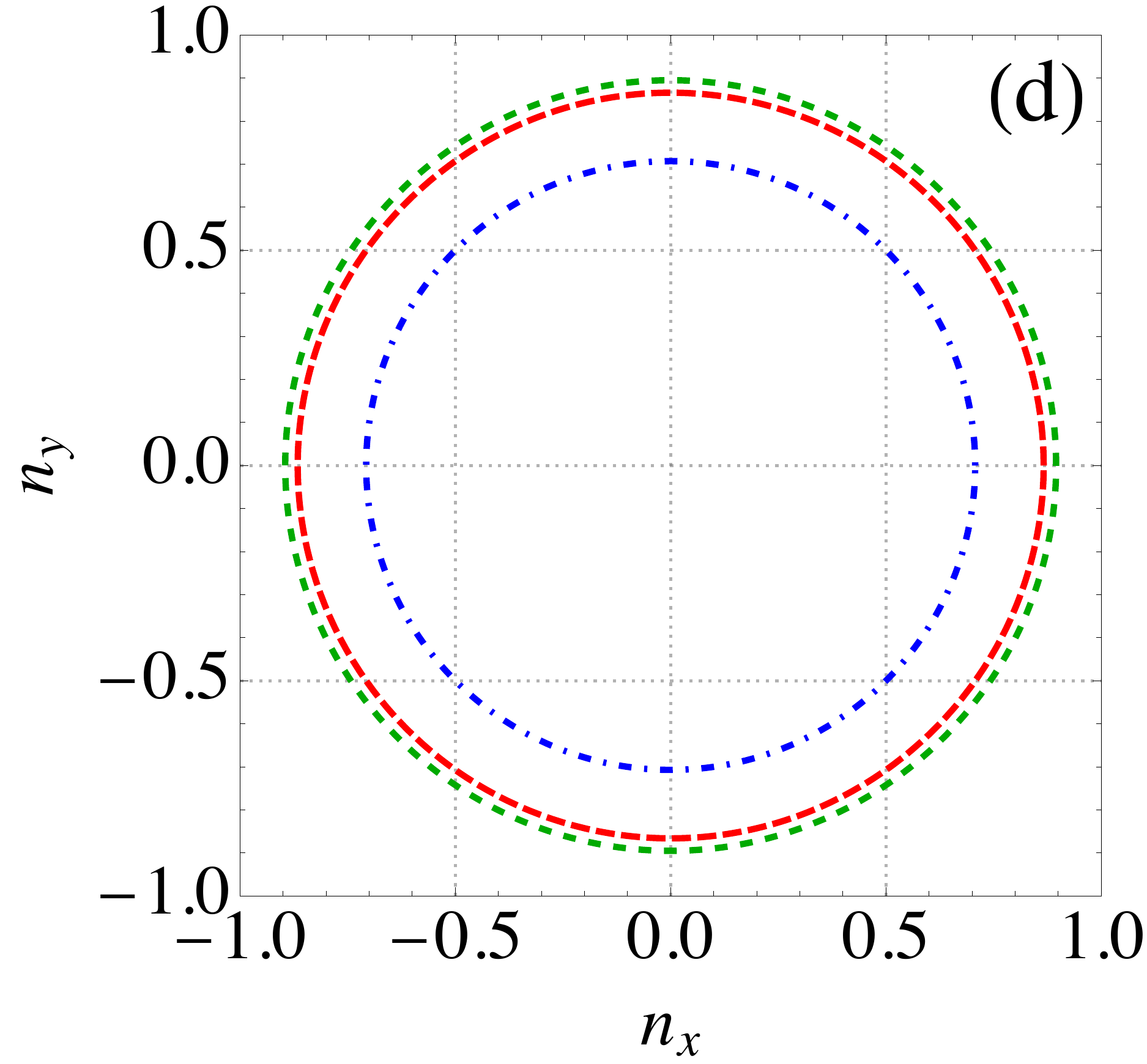}}
    \end{center}
    \caption{(a)-(b) Bloch representation for the subsystems for the cases with $T=0.2$ shown in Figs.~\ref{colormapUhlsubA}(b) and \ref{colormapUhlsubB}(d), respectively. In case (a) for the subsystem, $A$ the ellipsoids correspond to the couplings:  $g=1.1546\simeq g_c$ (red) (critical spheroid), $g=1.6$ (green), and $g=2.0$ (blue). The radii of the ellipsoids keeps decreasing as $g$ increases.  In case (b) for system $B$ we increase the coupling: $g=0.8$ (green), $g=1.12$ (red) (second critical spheroid), and $g=2.0$ (blue). We show that the Bloch radius crosses for a second time its critical value. We also include the circular cross-sections at  $\theta=\pi/2$ as parametric plots for subsystems (c) $A$, and (d) $B$  showing that the latter crosses the critical spheroid (red line) for a second time.}
    \label{secondcrossingB}
\end{figure}
Figures~\ref{firstcrosing}(a)-(b) show that the main effect of the temperature is to shrink the Bloch ball of the subsystems into an \textit{oblate spheroid} about the $n_z$  axis, with a circular cross-section of radius $R_s=|2 c_s|$ in the $n_x\,n_y$  plane. The corresponding circular cross-sections occurring a $\theta=\pi/2$ are also shown in  Figs.~\ref{firstcrosing}(c)-(d), for systems $A$ and $B$, respectively. 
For the chosen value of $T$, we show in Figs.~\ref{firstcrosing}(a) that the spheroids are contracted as $g$ increases, crossing once the critical spheroid of radius $R_{c,A}$ as their radius $R_A$ diminishes. The crossing of the critical spheroid corresponds to the solitary vortex observed in Fig.~\ref{colormapUhlsubA}(b). 
In  Fig.~\ref{firstcrosing}(b), we show that the effect is more dramatic in subsystem $B$ since the ellipsoids experiment a noticeable contraction along the $n_z$ axis. 
The radii of the ellipsoids $R_B$ increase with $g$, contrary to the observed behavior in (a). Also, as the radii of the ellipsoids increase, they cross the critical ellipsoid (red) with radius $R_{c,B}$. 
The latter corresponds to the appearance of the lower vortex in Fig.~\ref{colormapUhlsubB}(d). 
Although the behavior observed in the Bloch representation exhibits some aspects resembling the action of depolarizing or phase-damping channels \cite{nielsen11,jagadish_invitation_2018} (or possible combinations of both), the behavior is not trivial.

%
%
In Figs.~\ref{secondcrossingB}(a)-(b), we show the Bloch representation for the subsystems, for the fixed temperature used in Figs.~\ref{colormapUhlsubA}(b) and \ref{colormapUhlsubB}(d),  for larger values of the coupling $g$.
In Fig.~\ref{secondcrossingB}(a), we demonstrate that spheroids contract along all the axes as $g$ increases, and in the process, no further crossings of the critical spheroid (red) occur. See also their corresponding cross-sections in Fig.~\ref{secondcrossingB}(c). However, in Fig.~\ref{secondcrossingB}(b), we observe a peculiar behavior of the ellipsoids since they begin to shrink, and in the processes, their radii cross for a second time the critical spheroid (red) of radius $R_{c,B}$. See their corresponding cross-sections in Fig.~\ref{secondcrossingB}(d). 
The observed behavior is consistent with the appearance of the top vortex in Figs.~\ref{colormapUhlsubB}(d). 
Notice also that although the critical ellipsoids in cases of Fig.~\ref{secondcrossingB}(b) and Fig.~\ref{firstcrosing}(b) have the same radius, they exhibit a different elongation about the $n_z$ axis.

%
%

%

From the Bloch representations of the density matrix 
we show the various crossings of the critical ellipsoid by analyzing their respective sections in the equatorial plane. Even though all these cross-sections are very alike, we must highlight that this is not the case for their corresponding surfaces showing the dramatic effects of the driving field and the temperature on each subsystem.

\section{Conclusions}\label{sec:conclusions}

In this work, we study the effects of temperature on the Uhlmann phase in a system of two coupled spin-$\frac 1 2$ fermions where one of the fermions is driven by a magnetic field.
We derive analytical expressions involving unitary transformations  of the Uhlmann holonomy and show that the corresponding  phase for the composite system  exhibits two critical temperatures (vortices) 
that define a gap $\Delta T$ where the Uhlmann phase is not-trivial in all field directions with a fixed latitude $\theta=\pi/2$.
We show that for small couplings, $\Delta T\sim T_c$, 
the critical temperature of one-dimensional fermion systems described by two-band Hamiltonians in crystal momentum-space.
We also demonstrate that the first transition of the Uhlmann phase occurring in the low-temperature regime corresponds to the peak of the Schottky anomaly of the heat capacity $C_T$ characteristic of a two-level system involving the ground and first excited states. 
The second phase transition occurs at temperatures very close to the second maximum of  $C_T$ associated with a multilevel system.

We derive exact analytical expressions for the thermal Uhlmann phase for the subsystems $A$  (driven fermion) and $B$  (undriven fermion) and show that the temperature induces unexpected effects on the phase transitions.
In the case of the subsystem $A$,  we demonstrate that for small coupling values $g$, a topological phase transition of $\Phi^A$ appears at  $T_c$. 
For larger values of $g$, the transitions occur at lower temperature values and vanish when the coupling reaches the critical value $g_c=2/\sqrt{3}$.
We also find that the phase transition of $\Phi^B$  behaves very differently from that of $A$ and exhibits a peculiar behavior. 
We demonstrate that at low temperatures, there is only one phase transition at $g_c$. However, as the temperature increases,  we show the emergence of phase transitions corresponding to two different couplings separated by
$\Delta g$,
occurring at the same value of $T$. 
As the temperature increases, we show that the Uhlmann phase transitions (vortices) vanish as  
we reach a critical value of the temperature.

Alternatively, using the Bloch representation, we show that oblate spheroids describe the states of the subsystems with circular sections in the equatorial plane.
These ellipsoids are contracted along the polar axis by the effects of the temperature. We demonstrate that the phase transitions in the subsystems appear when the radii of these ellipsoids (in the equatorial plane) cross the critical ellipsoid of radius $R_{ c,s}= g_c^{-1}$, which occurs once in $A$ and twice in $B$, for a fixed value of $T$.

Our results show that although specific critical values of the coupling (or critical temperatures) typical to other fermionic systems underlie the structure of the geometric phases, the choice of the mechanism that generates the mixed states in the system can cause non-trivial effects on the behavior in their topological phase transitions.
For example, the behavior observed in other systems based on the same spin-coupled model, where the mixed states caused by noisy channels \cite{pravillaeta21} do not exhibit phase transitions in the bipartite system.

Finally, we remark that inducing a thermal mixing of the states allows us to correlate the phase transitions of an abstract quantity, such as the Uhlmann geometric phase, with an effect observed in solid-state physics, such as the Schottky anomaly of the heat capacity of the system.
For pure states, there are known connections between geometric quantities like the Berry curvature or the quantum metric and associated dipoles, and linear and nonlinear induced physical observables \cite{RevModPhys.82.1959,resta_insulating_2011,gianfrate_measurement_2020,PhysRevLett.122.227402,PhysRevResearch.2.033100,ahn_riemannian_2022,PhysRevLett.129.227401}.
In the same spirit, here
we find a similar relation but involving thermally induced mixed states.
We hope our work will stimulate future research to explore the fundamental aspects of geometric phases and their possible connection with physical observables.

\section{Acknowledgements}\label{acknowledgements}

The authors acknowledge partial financial support of DGAPA-UNAM-PAPIIT Grant No. IN111122, M\'exico. DMG and FNG acknowledge support from CONACYT (M\'exico). JV also thanks CNyN--UNAM for their kind hospitality during short stays, and I. Maldonado for fruitful discussions.

%

%

\end{document}